\documentclass[twocolumn,superscriptaddress,prx,amsmath,amstex,amssymb,citeautoscript,nofootinbib]{revtex4-1}
\pdfoutput=1
\usepackage[english]{babel}
\usepackage{letltxmacro}
\usepackage{latexsym}
\usepackage[utf8]{inputenc}
\LetLtxMacro{\ORIGselectlanguage}{\selectlanguage}
\makeatletter
\DeclareRobustCommand{\selectlanguage}[1]{%
  \@ifundefined{alias@\string#1}
    {\ORIGselectlanguage{#1}}
    {\begingroup\edef\x{\endgroup
       \noexpand\ORIGselectlanguage{\@nameuse{alias@#1}}}\x}%
}
\newcommand{\definelanguagealias}[2]{%
  \@namedef{alias@#1}{#2}%
}
\makeatother
\definelanguagealias{en}{english}
\definelanguagealias{English}{english}
\usepackage{graphicx}
\usepackage{amsmath}
\usepackage{amsfonts}
\usepackage{amsthm}   %
\usepackage{amssymb}
\usepackage{bm}
\usepackage{color}
\usepackage[percent]{overpic}
\usepackage{soul} 
\usepackage{amssymb}
\usepackage{wasysym}
\usepackage{dsfont}
\usepackage{float}
\usepackage{mathtools}
\usepackage{enumitem}
\usepackage{dsfont}
\usepackage{multirow}
\usepackage{colortbl}
\usepackage{mathrsfs}
\usepackage{makecell}

\usepackage{hyperref}
\hypersetup{
    bookmarks=false,         %
    unicode=false,          %
    pdftoolbar=false,        %
    pdfmenubar=true,        %
    pdffitwindow=false,     %
    pdfstartview={FitH},    %
    pdftitle={},    %
    pdfauthor={Authors},     %
    pdfsubject={},   %
    pdfcreator={},   %
    pdfproducer={}, %
    pdfkeywords={many-body localization} {matrix elements} {disordered systems}, %
    pdfnewwindow=true,      %
    colorlinks=true,       %
    linkcolor=black,          %
    citecolor=blue,        %
    filecolor=magenta,      %
    urlcolor=black           %
}

\newcommand{\be}{\begin{equation}}
\newcommand{\ee}{\end{equation}}
\newcommand{\bea}{\begin{eqnarray}}
\newcommand{\eea}{\end{eqnarray}}

\newcommand{\mI}{{\mathcal I}}

\newcommand{\ua}{\uparrow}
\newcommand{\da}{\downarrow}

\newcommand{\ri}{\mathrm{i}}
\newcommand{\rd}{\mathrm{d}}
\newcommand{\eff}{_\text{eff}}

\usepackage{graphicx}
\usepackage[colorinlistoftodos]{todonotes}
\usepackage{verbatim}

\newcommand{\bbrakket}[2]{\mbox{$ \langle\langle #1 | #2 \rangle\rangle $}}

\newcommand{\ket}[1]{\mbox{$| #1 \rangle$}}
\newcommand{\kket}[1]{\mbox{$| #1 \rangle\rangle$}}
\newcommand{\bra}[1]{\mbox{$\langle #1 |$}}
\newcommand{\bbra}[1]{\mbox{$\langle\langle #1 |$}}
\newcommand{\abs}[1]{\lvert#1\rvert}

\newcommand{\tr}{\mathrm{tr}}

\usepackage{soul}

\newtheorem*{lemma*}{Lemma}

\begin{document}

\title{Symmetry enriched phases of quantum circuits}
\author{Yimu Bao}
\affiliation{Department of Physics, University of California, Berkeley, California 94720, USA}
\author{Soonwon Choi}
\affiliation{Department of Physics, University of California, Berkeley, California 94720, USA}
\affiliation{Center for Theoretical Physics, Massachusetts Institute of Technology, Cambridge, MA 02139, USA}
\author{Ehud Altman}
\affiliation{Department of Physics, University of California, Berkeley, California 94720, USA}
\affiliation{Materials Science Division, Lawrence Berkeley National Laboratory, Berkeley, CA 94720, USA}

\begin{abstract}
Quantum circuits consisting of random unitary gates and subject to local measurements have been shown to undergo a phase transition, tuned by the rate of measurement, from a state with volume-law entanglement to an area-law state. From a broader perspective, these circuits generate a novel ensemble of quantum many-body states at their output. 
In this paper, we characterize this ensemble and classify the phases that can be established as steady states. Symmetry plays a nonstandard role in that the physical symmetry imposed on the circuit elements does not on its own dictate the possible phases. Instead, it is extended by dynamical symmetries associated with this ensemble to form an enlarged symmetry. 
Thus, we predict phases that have no equilibrium counterpart and could not have been supported by the physical circuit symmetry alone. We give the following examples.
First, we classify the phases of a circuit operating on qubit chains with $\mathbb{Z}_2$ symmetry. One striking prediction, corroborated with numerical simulation, is the existence of distinct volume-law phases in one dimension, which nonetheless support true long-range order. We furthermore argue that owing to the enlarged symmetry, this system can in principle support a topological area-law phase, protected by the combination of the circuit symmetry and a dynamical permutation symmetry. Second, we consider a Gaussian fermionic circuit that only conserves fermion parity. Here the enlarged symmetry gives rise to a $U(1)$ critical phase at moderate measurement rates and a Kosterlitz-Thouless transition to area-law phases. We comment on the interpretation of the different phases in terms of the capacity to encode quantum information. We discuss close analogies to the theory of spin glasses pioneered by Edwards and Anderson as well as crucial differences that stem from the quantum nature of the circuit ensemble.

\end{abstract}

\maketitle
\section{Introduction}

Experiments with quantum circuits designed as platforms for quantum information processing have recently shown remarkable advances and present a new class of many body systems~\cite{landsman2019verified,arute2019quantum,zhong2020quantum}.  
Generic quantum circuits consist of two distinct components: 
unitary evolution that generates entanglement and scrambles information
and measurements or couplings to a noisy environment that irreversibly reveal or destroy the encoded information.
The interplay of these components can lead to novel types of emergent phenomena, which bear on the circuit's capacity to offer a computational advantage over classical devices~\cite{arute2019quantum,zhong2020quantum}.
A case in point is the newly discovered phase transition in the dynamics of unitary circuits subject to local measurements.
As the rate of measurements exceeds a certain threshold, the steady state of the circuit changes from a highly entangled volume-law state to an area-law state~\cite{skinner2018measurement,li2018quantum,chan2018weak}. The establishment of a low entanglement state at high measurement rates  can be viewed in terms of the standard picture of quantum collapse of the wave function due to the repeated local measurements.
On the other hand, the essential mechanism that protects the wave function from collapse at low measurement rates is understood from the viewpoint of quantum error correction: generic unitary gates scramble quantum information and encode it in nonlocal degrees freedom, thereby affording partial protection of information from the deleterious effects of measurements.
Increasing the measurement rate degrades this encoding and reduces the capacity of the dynamics to keep information, the so-called \emph{quantum channel capacity}, until it vanishes at the phase transition point~\cite{choi2019quantum,gullans2019dynamical,fan2020self,ippoliti2020entanglement}. 

In his famous essay ``More is Different", Anderson remarked that ``it is only slightly overstating the case to say that physics is the study of symmetry''~\cite{anderson1972more}.
So, it may not be too surprising that
the measurement-induced transition (MIPT) in the entanglement entropy has an appealing theoretical description in terms of spontaneously broken symmetry~\cite{bao2020theory,jian2020measurement,nahum2020measurement}. 
What is somewhat unconventional, however, is that the symmetry in question is not a physical symmetry of the circuit elements, but is rather a consequence of the ensemble of quantum states that the circuit generates at its output. 

Each member state in this circuit ensemble corresponds to a particular sequence, or history, of measurement outcomes and appears with probability assigned by the Born rules. The symmetry arises because distinct features of the ensemble states can be seen only in nonlinear moments of the density matrix, like the entanglement entropy or 
fluctuations of observables between the different histories, while simple averages over observables give a trivial result. The time dependence of these moments can be expressed through the evolution of $n>1$ identical copies of the density matrix. %
Such dynamics, generated by unitary gates and measurements, has a  $\mathcal{S}_n$ symmetry to permutation among $n$ forward and, separately, $n$ backward propagating branches (i.e. both ket and bra wave functions). 
This is the symmetry, which is spontaneously broken in the MIPT. 
We note that the independent left and right replica symmetries arise from the need to describe a quantum state using a density matrix rather than a probability distribution. This is a crucial distinction from the replica symmetry breaking in classical spin glasses. We make further remarks on this point in the discussion section.

A systematic description of the MIPT has been obtained by mapping the circuit with Haar random unitary gates and measurements to a statistical mechanics model with $(\mathcal{S}_n\times \mathcal{S}_n)\rtimes \mathbb{Z}_2$ symmetry~\cite{hayden2016holographic,nahum2018operator,vasseur2018entanglement,zhou2019emergent,bao2020theory,jian2020measurement,nahum2020measurement}.
The extra $\mathbb{Z}_2$ symmetry is inherited from a symmetry of the time evolution to hermitian conjugation of all copies of the density matrix.
In these studies, the random unitary gates were assumed to be uniformly distributed over the Haar measure. 
With the purpose of understanding more structured circuits, it is natural to ask what other phases may arise if physical symmetries or other constraints are imposed on it. %

In this paper we elucidate how physical symmetries of the circuit elements combine with the intrinsic dynamical symmetries discussed above to determine the classification of the steady state phases. In the time evolution of $n$ copies of the density matrix the physical circuit symmetry $G$ is replicated to the $n$ forward and $n$ backward propagating branches. Conjugating a group element acting in one copy by a permutation in $\mathcal{S}_n$ transforms it to the corresponding group element in another copy. Thus, the dynamics of $n$ copies has the enlarged symmetry   $\mathcal{G}^{(n)}=[(G^{\otimes n}\rtimes \mathcal{S}_n)\times (G^{\otimes n}\rtimes \mathcal{S}_n)]\rtimes {\mathbb{Z}}_2^{\mathbb{H}}$. Here $\mathbb{Z}_2^{\mathbb{H}}$ acts like an anti-unitary time-reversal symmetry.
We shall see that for broad classes of circuit architecture the actual effective symmetry that determines the phase structure can be simplified. To substantiate the classification of phases, we develop an exact map of the dynamics of replicated density matrices in a broad class of quantum circuits to the imaginary time evolution under an effective quantum Hamiltonian that inherits the symmetry $\mathcal{G}^{(n)}$ (see Section~\ref{sec:framework}).
The possible steady states of the circuit ensemble correspond to the ground state phases of this effective Hamiltonian, which thus transcend a naive classification by the physical circuit symmetries. Note that having a volume-law state in the circuit does not imply that the corresponding effective Hamiltonian has a ground state with volume-law entanglement. Rather the volume-law entanglement of the circuit state  translates to a certain boundary operator, which roughly speaking measures the free energy of a spacetime domain wall. 

We offer an information theoretic interpretation of the phases of the circuit ensemble as distinct patterns of information encoding in the circuit, which remains well-defined in the replica limit $n\to 1$.
Nonetheless, we argue and provide numerical evidence that for certain purposes a model with $n=2$ replicas represents the true effective symmetries of the system and thus gives qualitatively correct predictions.

We demonstrate and explore these ideas on two examples representing different classes of circuits. The first is a circuit with $\mathbb{Z}_2$ symmetry operating on a chain of qubits. The second example is a Gaussian fermionic circuit, which conserves only the $\mathbb{Z}_2$ fermion parity. For the rest of this introduction, we provide a brief overview of the main results and insights obtained from studying the two examples and then lay out the general organization of the paper.

\subsection{Overview: quantum  circuits with $\mathbb{Z}_2$ symmetry}
In Section~\ref{sec:Z2model}, we investigate the phases of quantum circuits invariant under a global $\mathbb{Z}_2$ symmetry generated by the parity operator $\hat{\pi}=\prod_j Z_j$.
Thus, the single-qubit measurements of $Z_i$ or two-qubit measurements of $X_i X_{i+1}$, which commute with $\hat{\pi}$ are allowed, while single-qubit measurements of $X_i$ are not.
Similarly, all unitary gates must also commute with $\hat{\pi}$. 

It was previously demonstrated, based on a specific one-dimensional model~\cite{sang2020measurement}, that circuits with $\mathbb{Z}_2$ symmetry can exhibit at least two distinct area-law entangled phases in which the $\mathbb{Z}_2$ symmetry is either preserved or spontaneously broken.
The former is stabilized by on-site measurements of $Z_i$, while the latter is driven by measurements of $X_i X_{i+1}$.
The existence of a broken symmetry state in this system agrees with the intuition that area-law entangled states are akin to quantum ground states, which may spontaneously break a physical $\mathbb{Z}_2$ symmetry in one dimension.

One of our main results, however, is that circuits with $\mathbb{Z}_2$ symmetry generally admit a much richer phase structure derived from the enlarged dynamical symmetry.
This includes multiple area-law and volume-law phases, characterized by distinct broken symmetry order parameters, which could not have been established in presence of the circuit symmetry alone. 
Especially unique to this far-from-equilibrium setting is the establishment of states with volume-law entanglement entropy, which nonetheless exhibit spontaneously broken circuit symmetries in a one-dimensional system.
In the same vein, we point out the possibility of establishing a new type of symmetry protected topological (SPT) phase~\cite{pollmann2010entanglement,chen2011classification} as a steady state of the quantum circuit evolution, where the protecting symmetry is the enlarged dynamical symmetry rather than the physical circuit symmetry.
In our case, the $\mathbb{Z}_2$ circuit symmetry alone would not be sufficient to support an SPT phase as a ground state. However, such a state could be protected by a $\mathbb{Z}_2\times\mathbb{Z}_2$ subgroup of the enlarged dynamical symmetry. Here the second $\mathbb{Z}_2$ symmetry originates from the permutation symmetry of forward branches.

We consider a number of physical probes that can distinguish between the symmetry enriched phases of the circuit.
A useful probe of broken symmetry in such a system, which was also used in Ref.~\cite{sang2020measurement}, is the long distance Edwards-Anderson (EA) correlation function
\be
\chi_{EA}(i,j)=\overline{\sum_m p_m \langle X_i X_j\rangle^2}.
\ee
Here the sum is over the quantum trajectories of the circuit wave function, corresponding to different sequences $m$ of measurement outcomes obtained with probabilities $p_m$. The overline represents averaging over different unitary gates from an ensemble of quantum circuits. 
In the conclusions, we will discuss the intriguing connection between this problem and the theory of spin glasses pioneered by Edwards and Anderson, who also introduced this order parameter in the 1970s~\cite{edwards1975theory}.

Another useful diagnostic of  symmetry is the subsystem parity variance 
\be
\Pi(A)=\overline{\sum_m p_m \langle \prod_{j\in A} Z_j\rangle^2},
\ee
expected to be nondecaying in the limit of a large subsystem size when the state of typical trajectories has a well-defined parity.
At a superficial level, this probe appears to be a ``disorder parameter'' dual to the EA correlations.
However, we shall see that this duality does not hold in the enlarged symmetry space, thus allowing phases with coexisting long-range EA correlations and nondecaying parity variance. 

Complementary to these physical observables, we also consider information theoretic quantities,  such as Fisher information obtained by measurement outcomes about specific  perturbations to the initial state.
Such information theoretic probes allow to characterize the phases in terms of the flow of information in the circuit rather than steady state equal time properties such as $\chi_{EA}$ or $\Pi(A)$.

A crucial point is that all of these diagnostics must be defined as second (or even higher) moments in the ensemble of trajectories. For example, a simple average of order parameter correlation functions, i.e. $\overline{\langle X_i X_j\rangle}$ vanishes because the broken symmetry state is characterized by random $X$ orientations in a given trajectory due to random measurement outcomes. Evaluating second moments necessarily involves the dynamics of two replica copies, which should therefore be viewed as fundamental to the physics of these systems and not merely an averaging trick. Thus, we devote much of our attention to classifying phases in a model with two replicas ($n=2$). 

We note, however, that to ensure proper averaging over the ensemble one needs to invoke auxiliary replicas and ultimately take the the replica limit $n\to 1$. While we do not explicitly construct models for $n>2$ to facilitate such an extrapolation, we show that, apart from the coexistence phases that feature both nonvanishing $\chi_{EA}$ and $\Pi(A)$, the phases predicted for $n=2$ have natural extensions to higher replicas. Furthermore, when two phases can be distinguished by physical diagnostics in the two-replica model, this distinction holds for $n\geq 2$ and is thus expected to extrapolate to the replica limit. 

 In Section~\ref{sec:z2_numerics}, we provide numerical evidence for the phase structure predicted by the two-replica effective theory. The model we consider consists of two chains where chain 1 is unrestricted by symmetry, while operations on chain 2 obey a global $\mathbb{Z}_2$ symmetry. In other words, only qubits of chain 2 carry a global  $\mathbb{Z}_2$ charge. We show that the volume-law phase in this system can host more than two distinct broken symmetry phases, confirming that the symmetry dictating the phase structure is larger than the $\mathbb{Z}_2$ circuit symmetry. The coexistence of long-range quantum orders with volume-law entanglement presents a sharp contrast with thermal states, which exclude such order in one-dimensional systems. 

The fact that the volume-law phase can exhibit long-range order naturally raises the question if topological states and the ensuing edge modes can also remain protected in volume-law entangled steady states of quantum circuits. The results of the two-chain model suggest an affirmative answer.
We could replace the qubits in chain 2 with Majorana fermions conserving the $\mathbb{Z}_2$ fermion parity (still coupled to the qubits in chain 1). Under a Jordan-Wigner transformation this model maps to the one we computed numerically. The broken symmetry phase of the qubit circuit translates to a topological phase with edge zero modes and volume-law entanglement in the fermionic model.  

\subsection{Overview: Gaussian fermionic circuits}
In Section~\ref{sec:fermion} we investigate Gaussian fermionic circuits with unitary gates and measurements that only conserve the $\mathbb{Z}_2$ fermion parity, but not fermion number. 
The elements in these Gaussian circuits preserve the gaussianity of the wave function~\cite{terhal2002classical}.
In particular, we consider unitary gates generated by a (time-dependent) Hamiltonian that is quadratic in fermion operators and measurements of quadratic observables associated with rank one projectors, such as measuring the local parity. 
These systems provide a second example of the role played by the enlarged dynamical symmetry in dictating the phase structure. 

The effect of local Gaussian measurements on free fermions was first discussed by Cao et al.~\cite{cao2018entanglement}, who argued that these systems cannot sustain a volume-law state for any nonvanishing measurement rate (see also~\cite{fidkowski2020dynamical}). This can be understood as resulting from the absence of scrambling in Gaussian circuits, which therefore cannot protect quantum information (or equivalently entanglement) from measurements through nonlocal encoding~\cite{choi2019quantum,gullans2019dynamical}.  
Recent numerical results on monitored free fermion dynamics indicated, however, a possible measurement-induced phase transition from a critical phase, with entanglement entropy scaling logarithmically in system sizes, to a strict area-law phase~\cite{alberton2020trajectory}. 
Such a transition is also found in a specific fermion model via a mapping to classical loop models~\cite{sang2020entanglement}.
We note that critical states of nonunitary fermion models (though not measurement circuits) were studied in Refs.~\cite{chen2020emergent,liu2020non,jian2020criticality} as well as in measurement-only models~\cite{nahum2020entanglement}.

The framework developed in this paper naturally captures the instability of the volume-law state and the emergence of a critical phase. Both properties stem from the effective symmetry associated with the dynamics of two copies of the system, which is enlarged with respect to the physical symmetry of the circuit.  

We consider circuits with a  $\mathbb{Z}_2$ fermion parity symmetry, described using two Majorana operators on each physical site $j$, $\gamma_{2j-1}$ and $\gamma_{2j}$. The local observables measured  in the circuit are the fermion parities on sites  $\hat{\pi}_{s,j}=-\ri\gamma_{2j-1}\gamma_{2j}$ and the fermion parity on bonds connecting two nearest neighbor sites  $\hat{\pi}_{b,j}=-\ri\gamma_{2j}\gamma_{2j+1}$. These operators are also the generators of all the unitary gates. 

Due to the Gaussian constraint and fermion commutation relations, the enlarged dynamical symmetry of this model is different from that of qubit circuits.
In the absence of measurements, the unitary evolution of $n$ replica copies of a density matrix exhibits a global $O(2n)\rtimes \mathbb{Z}_2^\mathbb{T}$ corresponding to Bogoliubov rotations of Majorana operators between the $2n$ branches. $\mathbb{Z}_2^\mathbb{T}$ is a time-reversal like symmetry that squares to $-1$.
This is the same symmetry that emerges in the symplectic class of fermion systems in a random potential \cite{Ryu2007,Fu2012}.
There are two important differences, however. First, the replica limit in this dynamical problem is $n\to 1$, while it is $n\to 0$ in the Anderson localization problem.
Second, the measurements break the symmetry down to $[O(n)\times O(n)]\rtimes \mathbb{Z}_2^\mathbb{T}$.

As in the qubit circuits, we focus on the dynamics of the minimal model with two replica copies, which is mapped to imaginary time evolution generated by a quantum Hamiltonian.
In absence of measurements, averaging over a {\it purely unitary} circuit of Gaussian fermion gates gives rise to an effective ferromagnetic spin-1 Hamiltonian with $O(3)$ symmetry. The quantum ferromagnet allows establishment of long-range order, which translates to volume-law entanglement, even in a one dimensional system. 

With non-vanishing rate of measurements the symmetry of the effective model is reduced to $U(1)\rtimes(\mathbb{Z}_2\times \mathbb{Z}_2^\mathbb{T})$, which cannot support a broken symmetry state. For moderate measurement rate this leads to the establishment of a critical ground state of $H_{\text{eff}}$. The sub-system purity $\exp(-S_A^{(2)})$ of the circuit state translates to a boundary correlation function, which decays as a power law in the critical phase of $H_{\text{eff}}$ with a decay exponent that decreases continuously with increasing measurement rate. Accordingly, the entanglement entropy scales as $\log|A|$ with a continuously varying pre-factor. 

The two-replica theory predicts that the critical phase ends with a Kosterlitz-Thouless transition that takes the system into one of two area-law phases. The circuits diagnostics that allow to distinguish between the two area-law phases are the subsystem site and bond parity variances:
\be
\Pi_{a,A}\equiv\overline{\sum_m p_m\langle\prod_{j\in A}\hat{\pi}_{a,j}\rangle_m^2}.
\ee
Here the subscript $a=s,b$ stands for site or bond.

In the area-law phase with dominant site measurements, the 
subsystem site parity variance is non-decaying, whereas the bond parity variance decays exponentially. If the bond measurements dominate, then the bond parity variance is non-decaying. 
In terms of the effective ground state description, these two states correspond to a trivial and a Haldane SPT phase, respectively.  
The transition between the two area-law phases with the same characteristics has been discussed in the context of measurement-only dynamics of a Majorana chain~\cite{nahum2020entanglement,lang2020entanglement}. 

The fact that all phases can be diagnosed by quantities that are second moments over the ensemble of quantum trajectories hints at that a model of two replica copies captures the correct emergent symmetries in the problem.
Nonetheless, appropriate averaging of $\Pi_{a,A}$ requires introducing auxiliary replicas  followed by extrapolation to the physical replica limit $n\to 1$. This may alter the critical behavior.  

In Section~\ref{sec:fermi_numerics}, we compare the predictions of the effective two-replica theory to exact numerical simulations of the dynamics in the Gaussian fermionic circuits, which can be done on relatively large systems. We find good agreements with the expected properties of the critical phase, the two area-law phases and even the Kosterlitz-Thouless transition. In particular, the correlation length we extract from the numerical simulations shows a divergence that matches well with the hallmark exponential form of the Kosterlitz-Thouless transition $\xi\sim \exp(A/\sqrt{p-p_c})$. 
On the other hand the two replica model does not capture the observed critical behavior associated with the direct transition between the two area-law phases.

\subsection{Organization of the paper}
The rest of the paper is organized as follows.
In Section~\ref{sec:framework}, we explain the basic formalism and introduce a broad class of circuits that allows a mapping of the replicated dynamics to effective quantum ground state problems.
We discuss the basic structure and symmetry of the model and lay out the dictionary for translating between the quantities measured in the physical circuit and the corresponding operators in the effective ground state problem.
In Section~\ref{sec:Z2model}, we apply the framework to classify the phases of a random circuit model with $\mathbb{Z}_2$ symmetry.
We then demonstrate numerically that some of the new phases are found in the phase diagram of a concrete stabilizer circuit model.
In Section~\ref{sec:fermion}, we consider Gaussian fermionic circuits as outlined above. Section~\ref{sec:discussion} presents a broader discussion and summary of the results. In particular, we remark on connections between the quantum circuit ensemble discussed in this paper and the theory of spin glasses pioneered by Edwards and Anderson~\cite{edwards1975theory}.
Close analogies exist both at the level of emergent symmetries and in aspects of information theory. We shall also comment on fundamental differences between the problems associated with the quantum and dynamical nature of the circuit.

\section{Framework}
\label{sec:framework}
In this section, we introduce a framework for mapping the time evolution of an ensemble of quantum trajectories to an effective ground state problem. We identify the symmetry of this problem as an extension of the physical circuit symmetry by intrinsic dynamical symmetries.

\subsection{States and operators in duplicated Hilbert spaces}
We consider the dynamics of quantum systems undergoing unitary evolution interspersed by projective measurements.
The outcome of each projective measurement is  probabilistic, determined by the usual Born rules. This leads to stochastic dynamics of the unnormalized density matrix of the system, which for a particular sequence of measurement outcomes is given by
\begin{align}
    \tilde{\rho}_{m} (t) = U_k \dots U_2  P_{m_1} U_1 \rho_0 U_1^\dagger P_{m_1} U_2^\dagger \dots U_k^\dagger. 
\end{align}
Here $\rho_0$ is the initial state, $U_j$ are the set of unitary evolution, and $P_{m_j}$ are projection operators associated with measurement outcome $m_j$. 
Given $U_j$, the sequence of measurement outcomes $m = (m_1, m_2, \cdots )$ defines the trajectory of the wave function, which occurs with the probability $p_{m} = \tr \tilde{\rho}_m$.
The set of normalized quantum states $\rho_m \equiv \tilde{\rho}_m /p_m$ and their assigned probabilities $p_m$ form the measurement ensemble of the circuit. We are interested in certain steady-state properties of this ensemble. %

Recent works have shown that the steady states of quantum circuits with unitary gates and measurements can exhibit measurement-induced phase transitions, characterized by non-analyticities in various information theoretic quantities such as entanglement entropy, the global purity, and the Fisher information~\cite{li2018quantum,skinner2018measurement,li2019measurement,choi2019quantum,gullans2019dynamical,bao2020theory,jian2020measurement}.
These phase transitions have no signature in the averaged density matrix over quantum trajectories $\rho = \sum_m p_m \rho_m=\sum_m \tilde{\rho}_m$. 
In fact, it can be shown that $\rho$ approaches the maximally mixed state in the models considered in the literature~\cite{li2018quantum,skinner2018measurement,bao2020theory}. 

To characterize phase transitions in the measurement ensemble, one needs to consider the statistics of the ensemble, which is encoded in the time evolution of multiple copies of the trajectory density matrix, that is $\tilde{\rho}_m^{\otimes n}$.
In what follows, it will be convenient to view these objects as state vectors in a duplicated Hilbert space $\mathcal{H}^{(n)} = (\mathcal{H}\otimes\mathcal{H}^*)^{\otimes n}$. Thus, for example, the replicated un-normalized trajectory density matrix is denoted by
\begin{align}
\kket{\tilde{\rho}_m^{(n)}} \equiv \tilde{\rho}_m^{\otimes n}.
\end{align}

This state vector is evolved linearly by unitary and projection operators
\begin{align}
\label{eqn:replicated_operators}
\mathcal{U}^{(n)}_i \equiv (U_i\otimes U_i^*)^{\otimes n},\;\;\;
\mathcal{M}^{(n)}_{m_i} \equiv (P_{m_i} \otimes P_{m_i} )^{\otimes n},
\end{align}
defined through their action on the replicated density matrix 
\begin{align}
\mathcal{U}^{(n)}_i \kket{\tilde{\rho}_m^{(n)}} &\equiv \left(U_i \tilde{\rho}_m  U_i^\dagger\right)^{\otimes n}, \\
\mathcal{M}^{(n)}_{m_i} \kket{\tilde{\rho}_m^{(n)}} &\equiv \left(P_{m_i} \tilde{\rho}_m  P_{m_i}\right)^{\otimes n}.
\end{align}
If the state in the duplicated Hilbert space is not  factorizable, we use the linearity of the operators to define the actions of unitary gates and projections.

We will show that pertinent properties of the circuit ensemble are encoded in ensemble states defined as a sum over the replicated trajectory-states
\begin{align}
    \kket{\tilde{\rho}^{(n)}} 
    \equiv \sum_m \kket{\tilde{\rho}_m^{(n)}}.
\end{align}
This un-normalized ensemble state is very convenient because it undergoes a linear time evolution
\bea
\kket{\tilde{\rho}^{(n)}(t)} &=& \mathcal{V}^{(n)}(t)\kket{\tilde{\rho}^{(n)}_0}\\ &\equiv& \sum_m \mathcal{U}^{(n)}_t \cdots \mathcal{M}^{(n)}_{m_2}\mathcal{U}^{(n)}_2 \mathcal{M}^{(n)}_{m_1} \mathcal{U}^{(n)}_1\kket{\tilde{\rho}^{(n)}_0}.\nonumber
\eea
Later in this section, we present a class of models, in which this time evolution is exactly mapped to an imaginary time evolution generated by an effective quantum Hamiltonian, so that
\begin{align}
\kket{\tilde{\rho}^{(n)}(t)} = e^{-t H^{(n)}_{\text{eff}}}\kket{\rho^{(n)}_0}.
\end{align}
In these models, the properties of the ensemble state $\kket{\tilde{\rho}^{(n)}(t)}$ in late times are faithfully encoded in the ground state of $H^{(n)}_{\text{eff}}$. 

A key step in mapping of moments in the trajectory ensemble of the circuit to properties of the effective ground state  is to translate the trace operation in the replicated Hilbert space to the state-vector formalism. For example, we will need the trace $(\tr\tilde{\rho}_m)^n$ in order to normalize the trajectory states. In the state-vector notation, these traces can be expressed as an inner product with a reference state
\begin{align}
    \bbra{\mathcal{I}} = \mathds{1}^{\otimes{n}}= \sum_{\{\tau_\ell\}} \bbra{\tau_1\tau_1,\tau_2\tau_2,\ldots,\tau_n\tau_n},
    \label{eq:refI}
\end{align}
where each pair of $\tau_\ell$ labels the ket and bra state of copy $\ell$. Now, with the inner product between states in the replicated Hilbert space defined as $\bbrakket{\chi}{\sigma} \equiv \tr( \chi^\dagger \sigma)$ we can write the trace $(\tr\tilde{\rho}_m)^n = \tr( \tilde{\rho}_m^{\otimes n} ) = \bbrakket{\mathcal{I}}{\tilde{\rho}_m^{(n)}}$. 

The simplest quantities that involve higher moments of the density matrix are the average $k$-th moments of an observable $\hat{O}$ over the measurement ensemble and circuit realizations 
\be
O_k=\overline{\sum_m p_m \left[\frac{\text{tr} (\hat{O}\tilde{\rho}_m)}{\text{tr}\tilde{\rho}_m}\right]^k}.
\label{eq:O_n}
\ee
For example, when
dealing with circuits with a global $\mathbb{Z}_2$ symmetry (see Section~\ref{sec:Z2model}) it is natural to consider the fluctuations of a local order parameter $\hat{O}=X_i$, which is odd under the symmetry. 
In this case the object $O_2= \sum_m p_m \langle X_i\rangle_m^2$ is an Edwards-Anderson type order parameter that can detect the broken symmetry in individual trajectories~\cite{edwards1975theory}. 
More precisely, we'll be interested in the operator  $\hat{O}=X_i X_j$ that gives rise to the EA correlations $O_2= \sum_m p_m \langle X_i X_j\rangle_m^2$. Another example of an operator of interest is the $\mathbb{Z}_2$ parity on subsystem $A$, $\hat{O}=\prod_{j\in A} Z_j$. In this case, $O_2 = \sum_m p_m\langle\prod_{j\in A}Z_j\rangle_m^2$ measures the parity variance on $A$. 

The quantities given in Eq.~\eqref{eq:O_n} can be obtained from a replica limit.
To formally eliminate the denominator in Eq.~\eqref{eq:O_n}, we introduce a replica
index $n$ to write $p_m^{1-k}=\lim_{n\to 1}p_m^{n-k}$. Substituting into \eqref{eq:O_n}, we can express the $k$-th moment as a limit $n\to 1$ of the sequence of replica quantities
\begin{align}
O_k^{(n)} &= \frac{\text{tr}\left(\mathcal{O}^{(k)}
 \tilde{\rho}^{\otimes n}\right)}{\text{tr}\tilde{\rho}^{\otimes n}} =\frac{\bbra{\mathcal{I}}\mathcal{O}^{(k)}\kket{\tilde{\rho}^{(n)}}}{ \bbrakket{\mathcal{I}}{\tilde{\rho}^{(n)}} },\label{eq:Onq}
\end{align}
where we defined
\begin{align}
   {\mathcal{O}}_k^{(n)} \equiv \left[\bigotimes_{i = 1}^k (\hat{O} \otimes \mathds{1})\right] \; \otimes \; (\mathds{1} \otimes \mathds{1})^{\otimes n -k}.
\end{align}
The denominator in Eq.~\eqref{eq:Onq} is added to ensure that the quantities $O_k^{(n)}$ correspond to normalized boundary correlations in the effective model. It is exactly equal to unity in the replica limit $n\to 1$.

Subsystem purities can also be expressed in the same framework. The average $k$-th purity of a subsystem $A$ is given by 
\be
\mu_{k,A}
=\overline{\sum_m p_m {\text{tr}(\tilde{\rho}_{A,m}^k)\over  (\text{tr}\tilde{\rho}_{m})^k}}.
\ee
We can express it in the form \eqref{eq:Onq} by choosing the operator 
$\mathcal{O}^{(k)}$ 
 to be the subsystem cyclic permutation operator (of kets) 
\begin{align}
\label{eqn:cyclic_perm}
    \mathcal{C}_{\ell,A}^{(k)} &= \sum_{\{\alpha_i\}}  \bigotimes_{i=1}^k \left( \ket{\alpha_{i+1}}\bra{\alpha_{i}} \otimes \mathds{1} \right),
\end{align}
where $\alpha_{k+1} \equiv \alpha_1$ is assumed and the subscript $\ell$ implies acting on the left of the density matrices (i.e. permutation of kets). $\ket{\alpha_i}$ runs over all basis states of subsystem $A$, while $\mathcal{C}_{\ell,A}^{(k)}$ acts as the identity outside of subsystem $A$. That is, we have
\be
\mu_{k,A}^{(n)}= \frac{\bbra{\mathcal{I}}\mathcal{C}_{\ell,A}^{(k)}\kket{\tilde{\rho}^{(n)}}}{ \bbrakket{\mathcal{I}}{\tilde{\rho}^{(n)}} },
\ee
and $\mu_{k,A} = \lim_{n\to 1} \mu_{k,A}^{(n)}$.

In much of this paper we study the behavior of second moments in a two replica model (i.e. $k=2, n=2$). It is therefore worth writing the moments explicitly in this special case
\be
O_2^{(2)}=\frac{\bbra{\mathcal{I}}\mathcal{O}^{(2)}\kket{\tilde{\rho}^{(2)}}}{ \bbrakket{\mathcal{I}}{\tilde{\rho}^{(2)}} }=\frac{\sum_m p_m^2 \langle \hat{O}\rangle_m^{\,\, 2}}{\sum_m p_m^2},
\ee
and similarly
\be
\mu_{2,A}^{(2)}=\frac{\sum_m p_m^2 \tr(\rho_{A,m}^2)}{\sum_m p_m^2}.
\ee
We see that in the two replica model the trajectories are weighted by a distorted probability distribution $p_m^{(2)}=p_m^2/(\sum_{m'} p_{m'}^2)$.

Finally, for completeness, we discuss the computation of the von Neumann entanglement entropy within the state-vector formalism. For this purpose it is useful to treat the classical measurement device $M$ as a part of the extended system, as we have shown in Ref.~\cite{bao2020theory}. In this picture, the randomness of the measurement outcomes are encoded in the correlation between the system and measurement device. 

The average von Neumann entanglement entropy of subsystem $A$ over the possible measurement outcomes is
\begin{align}
S_A = \sum_m p_m S_{A,m} = \sum_m - p_m \tr ( \rho_{A, m} \log \rho_{A, m} ).
\end{align}
In the extended system,
we can express this average as the conditional entropy
\be
S_A = S(A|M) \equiv S_{MA}-S_M,
\ee
where the measurement device is characterized by a diagonal density matrix $\rho_M=\delta_{mm'}p_m$. In this formulation, we can further include an average over the random unitary gates and express the average conditional von Neumann entropy $\overline{S_A}$ as a limit of ``conditional R\'enyi entropies''~\cite{bao2020theory},
\begin{align}
\label{eqn:n_th_entropy}
S^{(n)}_A &= \frac{1}{1-n}\log\left(\,\overline{\tr\rho_{MA}^n}\,\right)-
\frac{1}{1-n}\log\left(\,\overline{\tr\rho_{M}^n}\,\right)
\nonumber\\
&=\frac{1}{1-n} \log \left(
\frac{\overline{\sum_m p_m^n \tr ( \rho_{A,m}^n )} }{\overline{\sum_m p_m^n}}
\right).
\end{align}
The properly averaged von Neumann entropy is restored in the replica limit: $\overline{S_A} = \lim_{n\rightarrow 1} S_A^{(n)}$. The conditional R\'enyi entropies have a simple expression within our formalism as the logarithm of a boundary matrix element. As in the case of purity, the relevant operator is the permutation $\mathcal{C}_{\ell,A}^{(n)}$ of forward propagating trajectories (kets) within subsystem $A$:
\begin{align}
    S_A^{(n)} = \frac{1}{1-n} \log \frac{\langle\langle \mathcal{I}|\mathcal{C}_{\ell,A}^{(n)}|\tilde{\rho}^{(n)}\rangle\rangle}{\langle\langle \mathcal{I}|\tilde{\rho}^{(n)}\rangle\rangle}.
    \label{eq:SA}
\end{align}
Note that Eq.~\eqref{eqn:n_th_entropy} implies that each of the conditional R\'enyi entropies involves weighting of the subsystem purities by the outcome probabilities to the $n$-th power. Thus, they are not identical to the R\'enyi entropies averaged over the measurement ensemble. 
However, as noted above, the von Neumann entropy obtained in the replica limit, is correctly weighted by the outcome probabilities.
In this paper, we mainly focus on $n = 2$ model, the conditional R\'enyi entropy and purity are related by $e^{-S^{(2)}_A} = \mu_{2,A}^{(2)}$.

\subsection{Enlarged dynamical symmetry}\label{sec:symmetry}
The dynamics $\mathcal{V}(t)$ of the quantum state in the duplicated Hilbert space exhibits an enlarged symmetry, which is an extension of the physical symmetry of the circuit by the dynamical symmetries inherent to the time evolution of identical copies of the density matrix.
In general, the dynamical symmetry does not commute with the physical circuit symmetry, and the combination of the two produces the enlarged symmetry of $\mathcal{V}(t)$. We shall see that this enlarged symmetry dictates the possible steady states of the circuit.

We say that the circuit has a physical symmetry $G$ if all the unitary gates $U_i$ and the projectors $P_{m_i}$ implementing the measurements commute with the symmetry generators $g\in G$. In this paper, for simplicity, we assume the symmetry group $G$ has a unitary representation and do not consider the anti-unitary circuit symmetries.
Note that the circuit symmetry is important for characterizing the steady state, however, unlike in purely unitary dynamics it does not give rise to a conserved quantity enforced at the level of individual trajectories. Given an initial state with indefinite quantum numbers associated with the symmetry, projective measurements may lead to the collapse of wave functions, such that quantum amplitudes or probabilities in each symmetry sector changes.
Once the system has a definite set of quantum numbers, then the measurements cannot change them. 

In the dynamics of $n$ replica copies, the physical circuit symmetry $G$ is duplicated to each of the forward and backward evolving branches, leading to a symmetry group $G^{\otimes n} \times G^{\otimes n}$.
We call the first and second $G^{\otimes n}$ the \emph{left} and \emph{right} circuit symmetry.
In addition, the time evolution $\mathcal{V}(t)$ of $n$ copies of the density matrix is invariant to permutation among the different copies of the identical circuit elements ($U_i$ and $P_{m_i}$) operating on the density matrix from the left and, independently, to permutation of the circuits elements acting from the right. 
The operators representing the left and right permutation symmetry are given respectively by 
\begin{align}
    \mathcal{C}_{\ell,\xi} &= \sum_{\{\alpha_a\} } \bigotimes_a \left( \ket{\alpha_{\xi(a)}}\bra{\alpha_a} \otimes \mathds{1}\right),\nonumber\\
    \mathcal{C}_{r,\xi} &= \sum_{\{\alpha_a\} } \bigotimes_a \left( \mathds{1} \otimes \ket{\alpha_{a}}\bra{\alpha_{\xi(a)}} \right),
\end{align}
where $\xi \in \mathcal{S}_n$ is a member of the permutation group of $n$ elements.
Here, we omit the superindex for $n$ replica copies for the simplicity of notation.
The left cyclic permutation $\mathcal{C}_\ell$, which is a special case of $\mathcal{C}_{\ell,\xi}$, was already introduced in Eq.~\eqref{eqn:cyclic_perm}.
The left and right permutation symmetries are independent from one another since $[\mathcal{C}_{\ell,\xi}, \mathcal{C}_{r,\eta}] = 0 $ for any $\xi, \eta \in \mathcal{S}_n$.
These permutation symmetries transform the replicated $G^{\otimes n}$ circuit symmetries into each other giving rise to $G^{\otimes n}\rtimes \mathcal{S}_n$ invariance, independently for the forward and backward branches.
The group extension $G^{\otimes n}\rtimes \mathcal{S}_n$ is also known as the wreath product $G \wr \mathcal{S}_n$.

Finally, there is one more dynamical symmetry associated with hermiticity of the density matrix, which the time evolution induced by $\mathcal{V}(t)$ must preserve. The operator representing this symmetry is the hermitian conjugation \begin{align}
    \mathbb{H}: c_{\alpha \beta \gamma \delta} \ket{\alpha}\bra{\beta} \otimes \ket{\gamma}\bra{\delta} 
    \mapsto
    c_{\alpha\beta \gamma \delta}^* \ket{\beta}\bra{\alpha} \otimes \ket{\delta}\bra{\gamma},
\end{align}
which is anti-linear (anti-unitary) and of order $2$, such that $\mathbb{H}^2 = \mathds{1}$.
We denote this symmetry by $\mathbb{Z}_2^\mathbb{H}$.
The physical circuit symmetries combine with the left and right permutation symmetries and hermiticity to give the complete dynamical symmetry
\begin{align}
\label{eqn:full_symmetry}
    \mathcal{G}^{(n)} = \left[ \left( G^{\otimes n} \rtimes \mathcal{S}_n\right) \times \left( G^{\otimes n} \rtimes \mathcal{S}_n\right) \right] \rtimes \mathbb{Z}_2^{\mathbb{H}}. 
\end{align}
Equivalently, $\mathcal{G}^{(n)}$ can be written in terms of the wreath product $\mathcal{G}^{(n)} = G \wr \mathcal{S}_n \wr \mathbb{Z}_2^\mathbb{H}$.

We make a few remarks.
While Eq.~\eqref{eqn:full_symmetry} is generically valid, the symmetry of $\mathcal{V}(t)$ can be even larger in the presence of additional constraints.
For example, in Section~\ref{sec:fermion}, we consider a fermionic circuit that preserves the Gaussianity of the fermionic wave function, which leads to a continuous symmetry.
In addition, while $\mathcal{G}^{(n)}$ itself is very large, the physically relevant symmetry can be a smaller subgroup of $\mathcal{G}^{(n)}$, depending on the details of the model. 
For instance, if $\mathcal{V}$ is averaged over a specific random circuit ensemble, the averaging process can constrain the state to be in faithful representations only of a subgroup of $\mathcal{G}^{(n)}$ thus reducing the available options for symmetry breaking. 

We also note that our analysis only concerns the symmetry of $\mathcal{V}(t)$ and does not address additional physical constraints such as the complete positivity of quantum channels.

\subsection{Effective quantum Hamiltonian}\label{sec:Heff}
\begin{figure}
    \centering
    \includegraphics[width=0.48\textwidth]{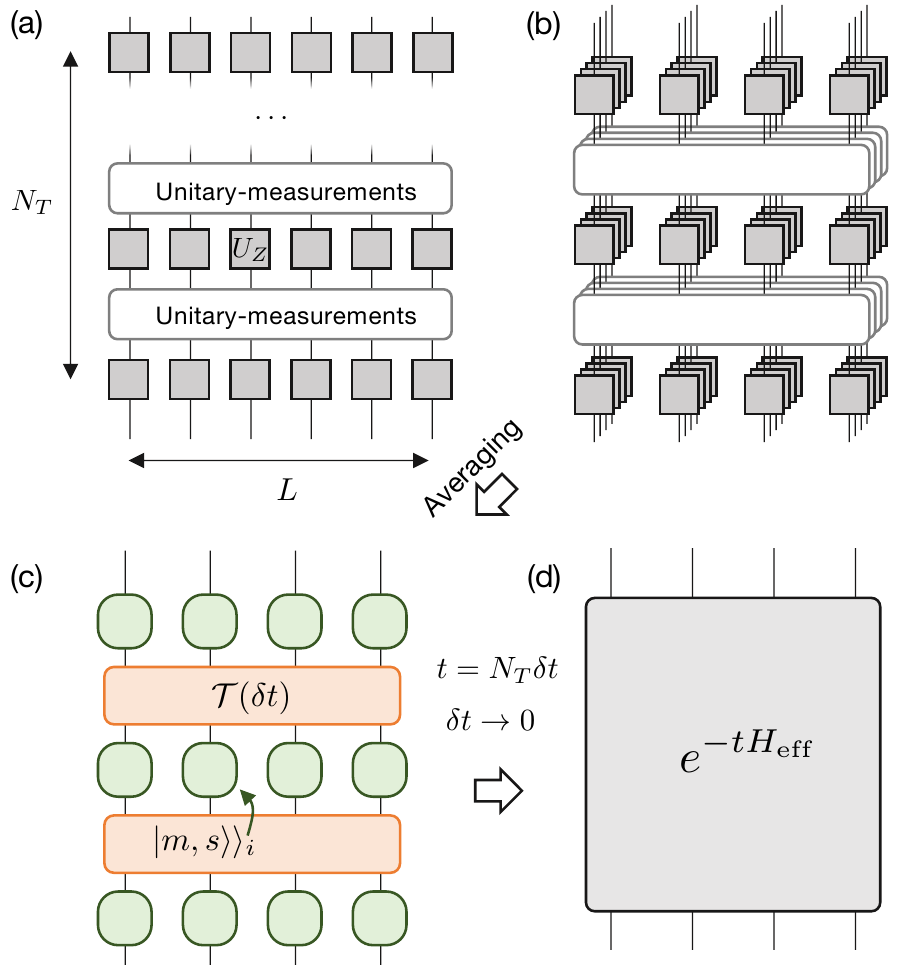}
    \caption{
    \label{fig:qubit_model}
    Mapping the dynamics of random quantum circuits to the imaginary time evolution with an effective Hamiltonian.
    (a) A quantum circuit with layers of random on-site $Z$-rotations $U_Z = e^{-\ri\theta_i Z_i}$.
    The inter-layer elements include measurements and unitary gates satisfying conditions needed for a hermitian effective Hamiltonian (see text).
    (b) The effective time evolution of a doubled density matrix is effected by four copies of the quantum circuit, two corresponding to forward propagation and two backward.
    (c,d) Averaging over the random $Z$ rotations projects on the reduced site Hilbert space $\kket{m,s}_{i}$, which can be interpreted as quantum spin states. The states at successive time steps are connected by the transfer matrix $\mathcal{T}(\delta t)$, which is generated by an effective Hamiltonian $H_{\text{eff}}$ in the limit $\delta t\to 0$.
    }
\end{figure}

In this section, we introduce an exact mapping of the dynamics $\mathcal{V}(t)$ in a broad class of qubit circuits to imaginary time evolution generated by an effective Hamiltonian $H_{\text{eff}}$, which inherits the enlarged dynamical symmetry.
Thus, the long-time steady states of the quantum circuits can be characterized by ground states of $H_{\text{eff}}$.

The structure of the quantum circuits we consider is shown in Fig.~\ref{fig:qubit_model}(a). Each ``time step" $\delta t$ is subdivided into layers.
The first layer in each time step is made of single-site random unitary phase rotations (i.e. rotation about the $Z$-axis). This is not a necessary step in our scheme, but we shall see that averaging over the random phases simplifies the effective model by projecting onto a reduced Hilbert space. The next layer consists of near identity unitary gates, such as $U_{ij}=\exp\left(-\ri\theta^{\alpha\beta}_{ij}\tau^\alpha_i\tau^\beta_j\right)$, where $\tau^\alpha_j$ with $\alpha=0,\ldots, 3$ represent the identity and Pauli operators on the qubit.  
The random couplings $\theta^{\alpha \beta}_{ij}$ are drawn from a symmetric Gaussian distribution with variance $\text{var}(\theta_{ij}^{\alpha\beta})\equiv J_{\alpha\beta} \delta t$.
This layer also contains projective measurements of Pauli string operators $M_\alpha$, applied with probability $p_\alpha = \Gamma_\alpha \delta t$. 

The replicated density matrix $\kket{\tilde{\rho}^{(n)}}$ is evolved by the operator $\mathcal{V}(t)$, consisting of the replicated circuit elements. We obtain a transfer matrix by averaging the evolution by one time step over the ``circuit ensemble'', namely the distribution of unitary gates and the probabilities $p_\alpha$ of applying measurement operators. The latter are also replicated on the $n$ forward and $n$ backward branches [e.g. the case of $n = 2$ is depicted in Fig.~\ref{fig:qubit_model}(b)]. 

For an infinitesimal time step, $\delta t \to 0$, the transfer matrix takes the form $\mathcal{T}=\overline{\mathcal{V}(\delta t)}=\exp({-\delta t H_{\text{eff}} })$. Hermiticity of $\mathcal{T}$ (and thus also of $H_{\text{eff}}$) is ensured if the distribution of the coupling constants in the unitary gates is symmetric about zero.
In the long-time evolution, the steady state $\kket{\tilde{\rho}^{(n)}} = \lim_{t\to\infty} e^{-tH\eff}\kket{\rho_0}$ is given by the ground state of $H\eff$.

For the circuits considered in this section, the effective Hamiltonian $H\eff$ has a set of mutually commuting local integrals of motion.
These are the local parity operators $\mathcal{X}_j = \prod_{a = 1}^n (X_{a}X_{\bar{a}})_j$, $\mathcal{Y}_j = \prod_{a = 1}^n (Y_{a}Y_{\bar{a}})_j$, and $\mathcal{Z}_j = \prod_{a = 1}^n (Z_{a}Z_{\bar{a}})_j$ (see appendix \ref{app:local_symmetry}). 
The subscript $a$ and $\bar{a}$ denote the Pauli operators acting on the $a$-th forward and backward branch, respectively.
The eigenstates of $H_{\text{eff}}$ are labeled by the eigenvalues of the  $\mathcal{X}_j$, $\mathcal{Y}_j$, and $\mathcal{Z}_j$ on all sites.

We will only need to characterize the ground state of $H_{\text{eff}}$ in the even sector of the local parities in order to compute the physical diagnostics in Eq.~\eqref{eq:O_n} and~\eqref{eqn:n_th_entropy}. 
The reason is that the reference state $\bbra{\mathcal{I}}$
has definite local parities on all sites, $\mathcal{X}_j = \mathcal{Z}_j = +1$ and $\mathcal{Y}_j = (-1)^n$.
Meanwhile, all the duplicated operators associated with diagnostics considered in this paper, e.g. $\mathcal{O}_2^{(2)} = (X\otimes \mathds{1})^{\otimes 2}$, preserve the local parities.
Therefore, only the ground state in the same local parity sector as the reference state $\kket{\mathcal{I}}$ has a nonvanishing contribution to the matrix element in Eq.~\eqref{eq:Onq} and~\eqref{eq:SA}.

We now turn to construct the effective Hamiltonian explicitly for the case of $n=2$ replicas.  %
Generalizing to higher replicas $n \geq 3$ is straightforward.
Consider first the single-site phase rotation at the entry to each time step. Averaged over the random phase, the two copy evolution through this layer takes the form
\begin{align}
\label{eqn:on-site_avg}
&\overline{e^{-i\theta_j Z_j}\otimes e^{i\theta_j Z_j}\otimes e^{-i\theta_j Z_j}\otimes e^{i\theta_j Z_j}}\nonumber\\ 
    &={1\over 2\pi}\int_{0}^{2\pi} d \theta_j e^{-\ri \theta_{j} \left(Z_{j1}- Z_{j \overline{1}}+Z_{j2}- Z_{j\overline{2}}\right)}\nonumber\\
    &= \sum_{k=1}^6 \kket{\varphi_k} \bbra{\varphi_k}.
\end{align}
The averaging over the identical rotation angle of the four branches  yields a delta function that implements a projection onto the six-dimensional subspace defined by vanishing of $Z_{j1}+Z_{j2}-Z_{j\overline{1}}-Z_{j\overline{2}}$. 
A convenient basis for this subspace, labeled by $\kket{m,s}$, is given explicitly by
\begin{align}
\kket{1,\pm} &=\frac{1}{\sqrt{2}}\left( \ket{1}\bra{1}\otimes \ket{0}\bra{0} \pm \ket{0}\bra{0}\otimes \ket{1}\bra{1}\right)\nonumber\\
\kket{0,\pm} &=\frac{1}{\sqrt{2}}\left( \ket{1}\bra{1}\otimes \ket{1}\bra{1} \pm \ket{0}\bra{0}\otimes \ket{0}\bra{0}\right)\nonumber\\
\kket{-1,\pm} &=\frac{1}{\sqrt{2}}\left( \ket{0}\bra{1}\otimes \ket{1}\bra{0} \pm \ket{1}\bra{0}\otimes \ket{0}\bra{1}\right).
\label{eqn:spin_basis}
\end{align}
The quantum numbers $s_i=\pm 1$ on site $i$ happen to be the eigenvalues of the local integrals of motion $\mathcal{X}_i$ (and of $\mathcal{Y}_i$). As discussed above, for $n = 2$, the reference state $\kket{\mathcal{I}}$ is in the even sector of local parities. 
Hence, we need to consider only the even sector $s_i=+1$. This leaves us with three states on every site $m=-1,0,1$, which can be represented by a spin-1 degree of freedom. 

The left permutation (swap) symmetry $\mathcal{S}_2$ has a simple action on the basis \eqref{eqn:spin_basis}
\begin{align}
\mathcal{C}_\ell: \kket{m,s} \leftrightarrow \kket{-m,s}.
\end{align}
Thus, spontaneous breaking of the $\mathcal{S}_2$ permutation symmetry will appear as onset of a spontaneous $z$-magnetization of the spin-1 degrees of freedom. 

We now turn to construct the transfer matrix operating between the states of the reduced Hilbert space
\begin{align}
    \bbra{\{m,s\}}\mathcal{T}\kket{\{m',s'\}}
    = 
     \bbra{\{m,s\}} 
     \prod_\nu \mathcal{M}_\nu
     \prod_\mu \mathcal{U}_\mu
     \kket{\{m',s'\}} .
     \label{eq:TransferMatrix}
\end{align}
Here, $\mathcal{U}_\mu$ and $\mathcal{M}_\nu$ represent the averaged unitary gates and measurements acting on the duplicated state. The averaging is over both the random couplings in the unitary gates and the probabilistic application of the measurements.
The indices $\mu, \nu$ run over all different unitary gates and measurements in the layers of the circuit that are part of a single time step $\delta t$.
All of these operators are designed to be close to unity. The deviation from unity after averaging is of order $\delta t$. 

The averaged measurement operators take the form
\be
\mathcal{M}_\nu=(1-\Gamma_\nu\delta t)\mathds{1}^{\otimes 4}+\Gamma_\nu\delta t\sum_{m_\nu = \pm}P_{m_\nu}^{\otimes 4},
\label{eq:Mavg}
\ee
where $\Gamma_\nu$ may be viewed as the rate at which a measurement of type $\nu$ is performed,
and $P_{\pm} = (1 \pm M_\nu)/2$ is the projection onto the $\pm$ eigenstate of the Pauli string operator $M_\nu$. 

The unitary gates operating on qubits are of the general form:   
$U_{j}=\exp(-\ri\theta_{j}^{\alpha\beta}\tau_j^\alpha\tau_{j+1}^\beta)$, where $\tau^\alpha_j$ denote local Pauli operators, including the identity. That is $\tau^0_i\equiv \mathds{1}_j$,  $\tau^1_j\equiv X_j$ etc. Averaging over the Gaussian distribution of $\theta_j^{\alpha\beta}$ with variance $J_{\alpha\beta}\, \delta t$ gives:
\begin{align}
\label{eq:Uavg}
\mathcal{U}_j&=\overline{U_j\otimes U_j^*\otimes U_j\otimes U_j^*}\\ 
    &= e^{- \frac{\delta t J_{\alpha\beta}}{2}\left(\tau_{j,1}^\alpha\tau_{j+1,1}^\beta
    -\tau_{j,\overline{1}}^\alpha\tau_{j+1,\overline{1}}^\beta
    +\tau_{j,2}^\alpha\tau_{j+1,2}^\beta
    -\tau_{j,\overline{2}}^\alpha \tau_{j+1,\overline{2}}^\beta\right)^2}.\nonumber
\end{align}

For an infinitesimal $\delta t$, the transfer matrix \eqref{eq:TransferMatrix} takes the form $\mathcal{T}=1-H_{\mathcal{M}}\delta t-H_{\mathcal{U}}\delta t +O(\delta t^2)$. The effective Hamiltonian $H\eff=H_{\mathcal{M}}+H_{\mathcal{U}}$ can be read from Eqs. \eqref{eq:Mavg} and \eqref{eq:Uavg} projected to the reduced Hilbert space of three states per site.

In Appendix~\ref{app:Hq_Z2} we exemplify a detailed construction of an effective Hamiltonian for the two replica dynamics of a simple circuit with $\mathbb{Z}_2$ symmetry. 
The result is a spin-1 model with $D_4 \times \mathbb{Z}_2^\mathbb{H}$ symmetry. 
In the next section, we discuss the possible steady-state phases of circuits with $\mathbb{Z}_2$ symmetry more generally. We present direct simulation of a concrete quantum circuit, which exhibits a subset of the possible steady-state phases.

\section{Qubit circuit with $\mathbb{Z}_2$ symmetry}
\label{sec:Z2model}

We now discuss how the phase structure of a quantum circuit with measurements is enriched by imposing a physical $\mathbb{Z}_2$ symmetry on the circuit elements.
To derive general results, we go beyond the circuit models discussed in Section~\ref{sec:Heff}. In particular, we do not enforce a projection to a reduced six-dimensional local Hilbert space.  

In Section~\ref{sec:z2_phases} below, we present the general symmetry analysis of the problem to identify the allowed phases and determine what are their sharp signatures in the physical probes of the circuit state. In Section~\ref{sec:z2_numerics}, we present the results of numerical simulations of a concrete circuit model with $\mathbb{Z}_2$ symmetry. The results demonstrate the establishment of at least two of the possible area-law phases and three of the distinct volume-law phases in the steady state phase diagram of this circuit model.

\subsection{Phases protected by enlarged symmetry}\label{sec:z2_phases}

The circuits we consider in this section operate on a one-dimensional array of qubits. 
Unitary gates in the circuit are generated by Pauli operators with random couplings drawn from symmetric distributions as considered in Section~\ref{sec:framework}. 
All circuit operations, unitary gates and measurements, commute with a global $\mathbb{Z}_2$ parity operator defined on a single chain
\begin{align}
    \hat{\pi} = \prod_{j = 1}^L Z_j.
    \label{eq:Z2parity}
\end{align}
To make the model more general, we allow additional chains of qubits, which are not transformed by the above parity operator. Thus, operations applied to qubits on the additional chains are not restricted by symmetry.

We first focus on the phases that arise in the simplest case of two replica copies (i.e. $n=2$). 
A model of two copies gives a natural framework to address fluctuations that are second moments of observables over the trajectory ensemble. 
This is not an exact approach because without extrapolating to the replica limit $n \to 1$ the weighting of trajectories is distorted. 
To validate key aspects of the phase structure predicted by the two-replica model we use direct simulations of a quantum circuit model. In particular  we confirm the existence of new phases enabled by the enlarged symmetry.

According to the analysis presented in Section~\ref{sec:symmetry}, the enlarged symmetry of the two-replica model should be $\mathcal{G}^{(2)}=(D_4\times D_4)\rtimes\mathbb{Z}_2^{\mathbb{H}}$.
The left (right) $D_4$ group consist of the $\mathbb{Z}_2$ symmetries of the two copies on the forward (backward) branch, compounded with the $\mathcal{S}_2$ permutation symmetry between the two copies, i.e.  $D_4=(\mathbb{Z}_2\times\mathbb{Z}_2)\rtimes \mathcal{S}_2$.
The $\mathbb{Z}_2$ symmetries of a single copy and branch are generated by product operators which we denote as $\prod_j(ZIII)_j$ (copy 1 forward branch), $\prod_j(IZII)_j$ (copy 1 backward),   $\prod_j(IIZI)_j$ (copy 2 forward) and $\prod_j(IIIZ)_j$ (copy 2 backward).

The effective symmetry that determines the phase structure can be reduced and simplified by accounting for additional constraints. 
Specifically, in our case, the effective Hamiltonian conserves a set of local parities $\mathcal{X}_j = (XXXX)_j$, $\mathcal{Y}_j=(YYYY)_j$ and $\mathcal{Z}_j=(ZZZZ)_j$ (see Appendix~\ref{app:local_symmetry}).
This leads to reduction of the effective symmetry $G^{(2)}$ as follows. First, the fact that the operators $\mathcal{X}_j$ have non-vanishing expectation values on every site implies breaking of all four single-branch $\mathbb{Z}_2$ symmetries already at the outset because the $\mathcal{X}_j$ anti-commutes with these symmetries.  
Second, as noted before, we seek the ground state in the even sector of $\mathcal{X}_j$, $\mathcal{Y}_j$ and $\mathcal{Z}_j$ on every site. The exchange symmetry $\mathcal{S}_2^X$,  generated by the product of left and right permutation $\mathcal{C}_\ell\mathcal{C}_r$, acts trivially in this sector (see Appendix~\ref{app:symmetry}). Therefore it cannot be broken and we can quotient it out. Third, 
we can also quotient out the symmetry $\prod_j\mathcal{Z}_j$ because it is a product of the local parities, which cannot be broken. 

The above considerations leave us with an effective global symmetry $\mathcal{G}\eff^{(2)} = D_4 \times \mathbb{Z}_2^{\mathbb{H}}$, where $D_4 = (\mathbb{Z}_2^{\Pi_L} \times \mathbb{Z}_2^{\Pi_1}) \rtimes \mathcal{S}_2$.
Here, $\mathbb{Z}_2^{\Pi_L}$ and $\mathbb{Z}_2^{\Pi_1}$ are generated by $\hat{\Pi}_L = \prod_j (ZIZI)_j$ and $\hat{\Pi}_1 = \prod_j (ZZII)_j$, respectively. %
In Appendix~\ref{app:Hq_Z2}, we explicitly derive the effective Hamiltonian for a concrete model, demonstrating the $D_4 \times \mathbb{Z}_2^\mathbb{H}$ symmetry.

The Hermitian conjugate $\mathbb{H}$ in the even parity sector is given by the complex conjugate $\mathcal{K}$ as shown in Appendix~\ref{app:symmetry} and commutes with the rest symmetry generators in $\mathcal{G}^{(2)}_{\text{eff}}$. Hence, $\mathbb{H}$ can be broken independently. 
Moreover, the reference state $\bbra{\mathcal{I}}$ and $\bbra{\mathcal{I}}\mathcal{O}_2$ associated with the physical diagnostics are symmetric under Hermitian conjugate, making them blind to its breaking.
Therefore, in this paper, we do not distinguish phases by the hermiticity breaking and consider the phases allowed by the $D_4$ symmetry.

We are now in a position to classify the possible steady states for the case of $n = 2$ through their correspondence to ground states of an effective Hamiltonian with $D_4$ symmetry. This includes broken symmetry states, characterized by the residual subgroups of $D_4$, as well as SPT phases protected by the $D_4$ effective symmetry. 
The results of this analysis are summarized in Table~\ref{tab:phases}. We shall also discuss the physical diagnostics in quantum circuits that can distinguish different phases. In Appendix~\ref{app:higher_replicas} we comment on the extension to higher replicas.
\begin{table*}[t!]
\centering

\begin{tabular}{|c|l|l|c|c|c|}
\hline
\hline
Entropy & Name & Residual symmetry & $\Pi^{(2)}(A)$ & $\chi_{EA}^{(2)}(i,j)$ & $\Phi^{(2)}$ \\
\hline
\multirow{3}{*}{Area law} & \cellcolor{gray!25} Symmetric phase (trivial/SPT) &\cellcolor{gray!25} $(\mathbb{Z}_2^{\Pi_L} \times \mathbb{Z}_2^{\Pi_1}) \rtimes \mathcal{S}_2$ & $\text{const.}$ & $\to 0$ & 0 \\ \cline{2-6}
& Coexistence phase (trivial/SPT) & $\mathbb{Z}_2^{\Pi_{L}} \times \mathcal{S}_2$ & $\text{const.}$ & $\text{const.}$ & 0 \\ \cline{2-6}
& \cellcolor{gray!25} Broken symmetry phase & \cellcolor{gray!25} $\mathcal{S}_2 $ & $\to 0$ & $\text{const.}$ & $\text{const.}$\\ \cline{2-6}
& Composite phase& $\mathbb{Z}_4^{\mathcal{C}_\ell\Pi_1}$ & $\text{const.}$ & $\text{const.}$ & 0 \\
\hline
\multirow{6}{*}{Volume law} & \cellcolor{gray!25} Symmetric phase & \cellcolor{gray!25} $\mathbb{Z}_2^{\Pi_L} \times \mathbb{Z}_2^{\Pi_1} $ & $\text{const.}$ & $\to 0$ & 0 \\ \cline{2-6}
& \cellcolor{gray!25} Featureless phase & \cellcolor{gray!25} $\mathbb{Z}_2^{\Pi_1}$ & $\to 0$ & $\to 0$ & $\text{const.}$ \\ \cline{2-6}
& Coexistence phase I & $\mathbb{Z}_2^{\Pi_1\Pi_L}$ & $\text{const.}$ & $\text{const.}$ & 0 \\ \cline{2-6}
& Coexistence phase II & $\mathbb{Z}_2^{\Pi_L} $ & $\text{const.}$ & $\text{const.}$ & 0 \\ \cline{2-6}
& \cellcolor{gray!25} Broken symmetry phase & \cellcolor{gray!25} $\emptyset$ & $\to 0$ & $\text{const.}$ & $\text{const.}$ \\ \cline{2-6}
& Composite phase & $\mathbb{Z}_2^{\mathcal{C}_\ell\Pi_L}$ & $\to 0$ & $\text{const.}$ & 0 \\
\hline
\hline
\end{tabular}

\caption{Possible phases of a qubit circuit with $\mathbb{Z}_2$ symmetry and their diagnostics. Four area-law and six volume-law phases are characterized by distinct residual symmetries given in the second column. The generators of the respective $\mathbb{Z}_2$ subgroups are specified in  superscript. $\hat{\Pi}_{L} = \prod_j (ZIZI)_j$ is the parity in the two forward branches. $\hat{\Pi}_{1} = \prod_j (ZZII)_j$ is the parity in the first copy. For each phase we give the long distance limit of the parity variance $\Pi^{(2)}(A)$ and EA correlation $\chi_{EA}^{(2)}(i,j)$ as well as the value of the Fisher information ``order parameter'' $\Phi^{(2)}=1-\frac{1}{2} F^{(2)}$ in the bulk. The phases we observe numerically in the model studied in Section~\ref{sec:z2_numerics} are shaded gray.}
\label{tab:phases}
\end{table*}

\subsubsection{``Conventional'' area-law phases}\label{sec:z2_area_phases}
We start from a working definition of the area-law regime as ground states of the effective Hamiltonian %
that preserve the $\mathcal{S}_2$ permutation symmetry generated by $\mathcal{C}_\ell$.\footnote{The left and right permutation $\mathcal{C}_\ell$ and $\mathcal{C}_r$ are identified because the exchange operation $\mathcal{C}_\ell\mathcal{C}_r$ acts as an identity matrix in the symmetric sector of the local symmetries.}
Indeed in these states domain walls of the $\mathcal{S}_2$ symmetry are condensed.  Therefore the sub-system swap operator that inserts such domain walls on either side of the sub-system is non-decaying implying non-decaying purity or $O(1)$ entanglement entropy. However, we shall see that the working definition needs to be modified slightly to account for one possible area-law state with broken $\mathcal{S}_2$ symmetry.

Under the working definition the different area-law states are characterized by distinct residual subgroups of $D_4$ that contain $\mathcal{S}_2$. 
Let us start from the most symmetric state, which retains the full $D_4$ symmetry. We first ask what are the ``charges'' (order parameters) that can condense and are invariant under the $\mathcal{S}_2$ permutation symmetry.

The first option is to condense the local charge $\mathcal{Q}_{1,j}\equiv (XXII)_j$ together with $(IXXI)_j$, which is related to $\mathcal{Q}_{1,j}$ by the $\mathcal{S}_2$ symmetry. If these two charges condense, then the charges 
$(IIXX)_j$ and $(XIIX)_j$ must also condense because of the definite local parities $\mathcal{X}_j$ in the ground state.
Condensing these charges breaks the $\hat{\Pi}_L$ and $\hat{\Pi}_1$ symmetry, leaving a residual global symmetry $\mathcal{S}_2$ in addition to the local symmetries $\mathcal{X}_j,\mathcal{Y}_j,\mathcal{Z}_j$.
We call this phase the broken symmetry area-law phase.

The second route for symmetry breaking, starting from the symmetric phase, is to condense the charges $(XI XI)_j$ and $(I XI X)_j$. This condensate breaks the symmetry $\hat{\Pi}_1$, but leaves intact the symmetry $\hat{\Pi}_L$, which commutes with the order parameter. We shall refer to this phase as the ``coexistence phase'' for reasons that will become clear below. From the coexistence phase, the system can further break the residual symmetry by condensing $\mathcal{Q}_{1,j}$ and its symmetry related charges defined above. This leads to the broken symmetry area-law phase discussed above.
We note that charges containing an odd number of Pauli-$X$ matrices, such as $XIII$ and $XXXI$, cannot condense because they anti-commute with the local parities $\mathcal{Y}_j$ and $\mathcal{Z}_j$, which cannot be broken. 

Let us discuss the physical interpretation of the area-law phases outlined above and how they can be distinguished by diagnostics of the circuit. We consider properties of the output state, such as the subsystem entanglement entropy, the Edwards-Anderson correlation function $\chi_{EA} \equiv \sum_m p_m \langle X_i X_j\rangle_m^2$ and the subsystem parity variance $\Pi(A) \equiv \sum_m p_m \langle\hat{\pi}(A)\rangle_m^2= \sum_m p_m \langle\prod_{j\in A} Z_j\rangle_m^2$. As discussed in Section~\ref{sec:framework}, these diagnostics map to different normalized matrix elements between the ground state and a reference state. %

Our measure of the purity and entanglement entropy is obtained from the overlap 
\be
e^{-S^{(2)}(A)}=\frac{\bbra{\mathcal{I}}\prod_{j\in A}{\mathcal{C}}_{\ell,j}\kket{\psi_{gs}}}{\bbra{\mI}\psi_{gs}\rangle\rangle}. 
\label{eq:expS2}
\ee
Here the operator acting on $\kket{\psi_{gs}}$ creates a pair of domain walls of the left permutation (swap) symmetry at the two edges of the region $A$. If the ground state is symmetric under swap, then these domain walls are condensed, implying 
$\prod_{j\in A}{\mathcal{C}}_{\ell,j} \kket{\psi_{gs}}\approx \lambda \kket{\psi_{gs}}$ for a long region $A$, leading to a length independent (area-law) entanglement entropy. 

The parity variance in region $A$ is similarly obtained by the matrix element of a nonlocal operator that creates domain walls of the $\mathbb{Z}_2$ symmetry generated by $\hat{\Pi}_L$, $\Pi^{(2)}(A)= \bbra{\mI}\prod_{j\in A} (ZIZI)_j\kket{\psi_{gs}}/\bbrakket{\mI}{\psi_{gs}}$. Finally, the Edwards-Anderson order parameter in the circuit maps to the the following matrix element in the effective model
$
\Psi_{EA}^{(2)}(i)=\bbra{\mI}(XIXI)_i\kket{\psi_{gs}}/\bbrakket{\mI}{\psi_{gs}}
$. Strictly speaking, we need to consider the long-range EA correlation $\chi^{(2)}_{EA}(i,j)=\bbra{\mI}(XIXI)_i(XIXI)_j\kket{\psi_{gs}}/\bbrakket{\mI}{\psi_{gs}}$ to detect the spontaneous symmetry breaking because the initial state time evolution and reference state are all symmetric with respect to $\hat{\Pi}_1$. To simplify the explanations below, however, we will assume the presence of an infinitesimal symmetry breaking field and thus refer to establishment of an EA order parameter.

To understand the behavior of these objects in the different phases, we need to determine, besides the symmetry of the ground state, also that of the reference state $\bbra{\mathcal{I}}$ in Eq.~\eqref{eq:refI}. 
It is easy to check that $\bbra{\mathcal{I}}$ is invariant under the $\mathbb{Z}_2^{\Pi_1}$ symmetry generated by $\hat{\Pi}_1 = \prod_j (ZZII)_j$, the $\mathcal{S}_2^X$ exchange symmetry, and the $\mathcal{Z}_2^\mathbb{H}$ hermiticity symmetry, while breaking all other global symmetries.
In particular, the reference state $\bbra{\mathcal{I}}$ is a condensate of the charges $(XXII)_j$, $(IIXX)_j$, $(YYII)_j$ and $(IIYY)_j$. For example, we have $\bbra{\mathcal{I}}(XXII)_j = \bbra{\mathcal{I}}$.

Consider first the signatures of the symmetric phase, which retains the full $D_4$ symmetry. Invariance with respect to $\hat{\Pi}_L$ means that the domain wall associated with this symmetry is condensed in the ground state, $\prod_{j\in A} (ZIZI)_j\kket{\psi_{gs}}\approx c \kket{\psi_{gs}}$. Thus, the subsystem parity variance $\Pi^{(2)}(A)$ remains constant independent of the size of $A$. The EA order parameter, on the other hand, vanishes because both the ground state and the reference state are even under $\hat{\Pi}_1$, whereas $XIXI$ is an odd operator with respect to this symmetry (anti-commutes with $ZZII$). 

Next, consider the broken symmetry phase, obtained by condensing $\mathcal{Q}_{1,j}=(XXII)_j$ and its symmetry related charges. 
When calculating the EA order parameter $\Psi^{(2)}_{EA}(i)$ we can replace $(XIXI)_i$ by $(IXXI)_i = (XXII)_i(XIXI)_i$ because the reference state is an eigenstate of $(XXII)_i$ with eigenvalue $1$. 
Hence, we expect a nonzero value of $\Psi^{(2)}_{EA}(i)$ owing to the condensation of $(IXXI)_i$ in the ground state $\kket{\psi_{gs}}$.
At the same time, the parity variance $\Pi^{(2)}(A)$ is expected to decay exponentially.  
The subsystem parity $\prod_{j\in A}(ZIZI)_j$ acts nontrivially on the broken symmetry ground state $\kket{\psi_{gs}}$, creating a domain of a symmetry related ground state within region $A$. Since $\bbra{\mathcal{I}}$ is a uniform broken symmetry state, its normalized overlap with the transformed region is exponentially decaying, $\Pi^{(2)}(A)\sim e^{-L_A /\xi}$ for $\xi\ll L_A\ll L$.\footnote{The broken symmetry state is strictly speaking a cat-like superposition of the two distinct broken symmetry states, leading to $\Pi^{(2)}(A)\sim e^{-L_A /\xi}+e^{-(L-L_A) /\xi}$ under periodic boundary condition.}

The symmetric and broken symmetry phases outlined above are the same as those discussed in Ref.~\cite{sang2020measurement} and observed there in simulations of a concrete model.
The physical intuition for the establishment of these phases 
is clear. In the broken symmetry phase, dominant nearest neighbor $XX$ measurements collapse the individual trajectories to a near product state with qubits polarized in random directions along $X$. As in a glassy phase, such random $X$ orientations are detected by averaging the square of individual trajectory expectation values $\Psi_{EA}=\sum_m p_m\langle X_i\rangle_m^2$. At the same time, the parity of a long region $A$ has a vanishing expectation value $\langle\hat{\pi}(A)\rangle_m=\langle \prod_{j\in A} Z_j\rangle_m\to 0$ in every trajectory. 
In the symmetric phase, dominant single-qubit $Z$ measurements collapse individual trajectories to near product states with definite, yet random parity on each qubit. This is detected by taking the variance over trajectories of the parity in region $A$: $\sum_m p_m \langle\hat{\pi}(A)\rangle_m^2 \to \text{const}$. 
At the same time, the EA order parameter vanishes in the symmetric phase because $\langle X_i\rangle_m=0$ in every trajectory.   

Besides these two phases, we have alluded to the possible existence of another, more peculiar phase, in which the symmetry is partially broken by condensing just the charges $XIXI$ and $IXIX$. Such a condensate implies the nonvanishing EA order parameter
$
\Psi_{EA}^{(2)}(i) \to \text{const}.
$
At the same time, the parity symmetry generated by $\hat{\Pi}_L$ remains intact, implying
nonvanishing parity variance on a long subsystem $\Pi^{(2)}(A)\to \text{const}$.

The coexistance of a nonvanishing EA order parameter with a non-decaying parity variance, allowed within the effective model, may seem paradoxical at first sight when considering the wave functions of individual trajectories. 
The EA order parameter must originate from trajectories $m$ that exhibit nonvanishing $\langle X_i\rangle_m\ne 0$. The parity expectation value $\langle \hat{\pi}(A) \rangle_m$ must vanish in these symmetry breaking trajectories.  Nonetheless, because we have an ensemble of trajectories, the EA order parameter and non-vanishing parity variance can coexist in separate trajectories. In theory, such a phase is established if symmetry breaking and symmetric trajectories both appear with nonvanishing probability in the ensemble. 
It is left as an open question if there are physical obstructions towards realizing such a coexistence phase.

Having considered area-law states which preserve the $\mathcal{S}_2$ symmetry, we now show that this is not a necessary condition for getting a 
phase with area-law entanglement. Specifically, consider breaking the $D_4$ symmetry while preserving a $\mathbb{Z}_4$ subgroup generated by the composite operator $\hat{\Pi}_1\mathcal{C}_\ell$. 
This state, which we term th ``composite area-law phase'', can be obtained from the symmetric phase by condensing the operator $(YZXI)_j$, symmetrized over the residual $\mathbb{Z}_4$ subgroup.
The key to calculate the entanglement entropy in this state is to note that the reference state is an eigenstate of $\hat{\Pi}_{1,j}=(ZZII)_j$ with eigenvalue $1$. This allows to make the replacement $\bbra{\mathcal{I}}\mathcal{C}_{\ell,j}=\bbra{\mathcal{I}}\hat{\Pi}_{1,j}\mathcal{C}_{\ell,j}$ when calculating the subsystem purity to obtain  
\begin{align}
    e^{-S^{(2)}(A)} = \frac{\bbra{\mathcal{I}} \prod_{j\in A}\left(\hat{\Pi}_{1,j}\mathcal{C}_{\ell,j}\right)\kket{\psi_{gs}}}{\bbrakket{\mathcal{I}}{\psi_{gs}}} \to \text{const}.
\end{align}
The subsystem purity tends to a constant for large $L_A$ because the operator appearing in the matrix element creates domain walls of the unbroken $\mathbb{Z}_4$ symmetry at the two boundaries of $A$.

Besides area-law entanglement, the ``composite phase" is also characterized by non-decaying EA correlations and subsystem parity variance. 
To see the former we use the fact that $\bbra{\mI}$ is an eigenstate of $(ZZII)_j$ to replace the operator $(XIXI)_j$ with $\ri (YZXI)_j=(ZZII)_j(XIXI)_j$ in the EA correlation
\be
\chi_{EA}^{(2)}(i,j) = \frac{\bbra{\mathcal{I}} \ri(YZXI)_i \ri(YZXI)_j \kket{\psi_{gs}}}{\bbrakket{\mathcal{I}}{\psi_{gs}}}\to \text{const}.
\ee
This is non-decaying because the operator $(YZXI)_j$ overlaps with the order parameter of the state as discussed above. 
The parity variance is non-decaying because the symmetry $\hat{\Pi}_L = (\hat{\Pi}_1\mathcal{C}_\ell)^2$ is preserved.

We note that the residual $\mathbb{Z}_4$ symmetry enforces an exact relation between different EA correlations, which distinguishes the composite area-law phase from the coexistence phase.
To see this we consider a distinct EA correlation $\chi'_{EA}(i,j) \equiv \sum_m p_m \langle X_i Y_j \rangle_m^2$.
The corresponding quantity in the two-replica model 
can be written as $\chi'^{(2)}_{EA}(i,j) = \bbra{\mathcal{I}} \ri(YZXI)_i (-\ri)(XZYI)_j \kket{\psi_{gs}}/\bbrakket{\mathcal{I}}{\psi_{gs}}.$
Here, we replace $(YIYI)_j$ with $(-\ri)(XZYI)_j = (ZZII)_j(YIYI)_j$.
In the composite phase, we have an exact relation $\chi^{(2)}_{EA} = \chi'^{(2)}_{EA}$ because charges $\ri(YZXI)$ and $(-\ri)(XZYI)$ related by the conjugation of $\hat{\Pi}_1\mathcal{C}_\ell$ are condensed in the ground state with the same amplitude.
However, in the coexistence phase, $\chi'^{(2)}_{EA}(i,j)$ and $\chi^{(2)}_{EA}(i,j)$ are both nonvanishing but generally of different values.
Further reducing the $\mathbb{Z}_4^{\mathcal{C}_\ell\Pi_1}$ symmetry will give rise to volume-law phases discussed in the subsection~\ref{sec:z2_volume_phases}. 

\subsubsection{SPT area-law phases}\label{sec:z2_spt_phases}

In this section, we point to the possibility of establishing {\it symmetry protected topological} (SPT) area-law  phases protected by the effective $D_4$ symmetry. 
To identify and characterize such phases, we employ the decorated domain wall picture of Ref.~\cite{chen2014symmetry}.
Starting from the broken symmetry area-law phase, symmetry can be restored by condensing the domain walls of the symmetry $\hat{\Pi}_L$. For the ensuing phase to be distinct from the trivial symmetric phase the domain walls must condense only while bound to charges of an independent symmetry. In this case the only option is the $\mathcal{S}_2$ permutation symmetry. Note that condensing the $\mathcal{S}_2$ charges bound to a domain wall, does not lead to breaking of the $\mathcal{S}_2$ symmetry in this case because the condensed object is nonlocal. Indeed, a dual description of the same phase is a condensate of the $\mathcal{S}_2$ domain walls bound to the charges of the symmetry $\hat{\Pi}_L$. The state established in this way is a fully symmetric SPT phase, protected by the $\mathbb{Z}_2^{\Pi_L}\times \mathcal{S}_2$ symmetry. 
Starting from this SPT phase we can obtain another one by condensing the charges $XIXI$ and $IXIX$, which commute with the protecting symmetry. This gives the ``coexistence" SPT phase.

We emphasize that the enlarged dynamical symmetry is essential for establishing these topological phases as the $\mathbb{Z}_2$ symmetry of the physical circuit does not, on its own, allow an SPT phase in a one-dimensional system. This is different from the SPT state obtained in a measurement-only model in Ref.~\cite{lavasani2021measurement}, which is protected by the physical
$\mathbb{Z}_2\times \mathbb{Z}_2$ symmetry of the measurement operators. 

Let us confirm that these are indeed area-law phases, by considering the boundary overlap of the operator that creates a pair of $\mathcal{S}_2$ domain walls, $\mathcal{D}_{i_\ell}$ and $\mathcal{D}_{i_r}$, at the ends of the region $A$ \eqref{eq:expS2}, i.e. $\bbra{\mI}\mathcal{D}_{i_\ell} \mathcal{D}_{i_r}\kket{\psi_{gs}}$, where $\mathcal{D}_{i} = \prod_{j\leq i}\mathcal{C}_{\ell,j}$. Unlike in the trivial phase, the $\mathcal{S}_2$ domain walls are not individually condensed in the effective ground state. So, one might be tempted to conclude that the overlap decays exponentially, implying a volume-law state. However, the area law is saved by the fact that the reference state $\bbra{\mathcal{I}}$ breaks the $\hat{\Pi}_L$ symmetry, i.e. it is a condensate of the respective charge $\mathcal{Q}_{1,j}$. This allows us to extract such charges from the condensate and write 
\begin{align}
e^{-S^{(2)}(A)}&=\frac{\bbra{\mI}\mathcal{D}_{i_\ell}\mathcal{D}_{i_r}\kket{\psi_{gs}}}{\langle\langle\mI\kket{\psi_{gs}}}\\
&=\frac{\bbra{\mI}(\mathcal{Q}_{1,i_\ell}\mathcal{D}_{i_\ell}) (\mathcal{Q}_{1,i_r}\mathcal{D}_{i_r})\kket{\psi_{gs}}}{\langle\langle\mI\kket{\psi_{gs}}}\to \text{const}.\nonumber
\end{align}
In the last line, we used the fact that the $\mathcal{S}_2$ domain wall coupled to the $\hat{\Pi}_L$ charge (i.e. $\mathcal{Q}_{1,j}$) is condensed in the ground state. 

Finally we note that these SPT phases cannot be distinguished from their trivial counterparts using the bulk probes discussed above. 
We can show that the subsystem parity variance $\Pi_A^{(2)}$ is non-decaying using the dual argument to that used for the purity. 
$\Pi_A^{(2)}$ is given as a boundary matrix element of the $\hat{\Pi}_L$ string operator $\prod_{j\in A}\hat{\Pi}_{L,j}$, which creates a pair of domain walls on the two ends of subsystem $A$. While these domain walls are not individually condensed, a bound state of a $\hat{\Pi}_L$ domain wall and an $\mathcal{S}_2$ charge is condensed. Since $\bbra{\mI}$ breaks the $\mathcal{S}_2$ symmetry, when calculating the matrix element we can always extract $\mathcal{S}_2$ charges $\bbra{\mI}$ and attach to the domain walls. This implies non-decaying subsystem parity. 
Note that this argument implies that the parity string is identical to the string order parameter due to the broken symmetry in the reference state. Thus string order parameters are also ruled out as diagnostics to distinguish the SPT and trivial phases. 

Nor do the EA correlations able to distinguish the SPT phases from the trivial ones. The EA correlations decay exponentially in the symmetric SPT phase, as in the trivial state, because both the ground state and the reference state $\bbra{\mI}$ are symmetric under $\hat{\Pi}_1$, while the EA order parameter is odd under this symmetry. In the ``coexistence'' SPT phase the order parameter  $(XIXI)_j$ is condensed leading to nondecaying EA correlations, as in the trivial coexistence phase.

We conclude that the SPT phases cannot be distinguished by bulk probes. In Section~\ref{sec:Fisher} we show that a different type of diagnostic, when applied to the edge of the system, allows to distinguish between the SPT and trivial phases.

\subsubsection{Volume-law phases}\label{sec:z2_volume_phases}

Based on the criterion we developed above for establishing an area-law phase, we expect that a volume-law phase is established if the $\mathcal{S}_2$ permutation symmetry is broken along with all composite symmetries $g_\mathcal{I}\,\mathcal{C}_\ell$, where $g_\mathcal{I}$ is a symmetry present in the reference state $\bbra{\mathcal{I}}$. 
One example of such composite symmetry is $\hat{\Pi}_1\mathcal{C}_\ell$.

To identify the volume-law phases, we start from the fully symmetric area-law phase with the effective $D_4$ symmetry and determine the possible charges that can be condensed to reduce the symmetry in a way that $\mathcal{S}_2$ and required composite symmetries are broken.

The most straightforward option to break symmetry is to condense the $\mathcal{S}_2$ charges that are neutral under the parity symmetries $\hat{\Pi}_1$ and $\hat{\Pi}_L$.
This gives the symmetric volume-law phase with a residual symmetry $\mathbb{Z}_2^{\Pi_L} \times \mathbb{Z}_2^{\Pi_1}$. Due to the presence of the symmetries $\hat{\Pi}_L$ and $\hat{\Pi}_1$, the EA order and parity variances behave exactly as they do in the symmetric area-law phase.

One way to further break symmetry is to condense the pair of charges $(XXII)_j$ and $(IIXX)_j$, which are related by the local conserved parities $\mathcal{X}_j$.
A second way is to condense the charges $(XIIX)_j$ and $(IXXI)_j$, similarly related by $\mathcal{X}_j$. Recall that in the area-law phase the first pair of charges was related to the second pair by the $\mathcal{S}_2$ permutation symmetry, which is broken in the volume-law phase. Thus, the phase transition, which in the area-law phase involved simultaneous condensation of all four charges, generically splits into two transitions. Which intermediate phase is established between the symmetric and the broken symmetry volume-law states depends on which pair of charges condenses first.

Yet another way to break symmetry, starting from the symmetric volume-law phase, is to condense the charges $XIXI$ and $IXIX$ to establish a third distinct intermediate phase. These three routes of breaking successive symmetries starting from $\mathbb{Z}_2^{\Pi_L} \times \mathbb{Z}_2^{\Pi_1}$ are summarized as follows
\begin{align}
    1)\, &\mathbb{Z}_2^{\Pi_L} \times \mathbb{Z}_2^{\Pi_1} & &\rightarrow ~\mathbb{Z}_2^{\Pi_1} & &\rightarrow ~\emptyset,\nonumber\\
    2)\, &\mathbb{Z}_2^{\Pi_L} \times \mathbb{Z}_2^{\Pi_1} & &\rightarrow ~\mathbb{Z}_2^{\Pi_1\Pi_L} & &\rightarrow ~ \emptyset,\nonumber\\
        3)\, &\mathbb{Z}_2^{\Pi_L} \times \mathbb{Z}_2^{\Pi_1} & &\rightarrow ~\mathbb{Z}_2^{\Pi_L} & &\rightarrow ~\emptyset.\label{eq:vol_routes}
\end{align}
We note that these phases are not necessarily realized in a given model, and one may be able to make physical arguments why certain phases can be hard or even impossible to realize in the quantum circuit.

Route 1 is realized if, starting from the symmetric volume-law phase, $XXII$ and $IIXX$ are condensed first. Because this condensate breaks the $\hat{\Pi}_L$ and $\hat{\Pi}_1\hat{\Pi}_L$ symmetry, the subsystem parity variance decays exponentially in this phase. Furthermore, we cannot create the EA order parameter $XIXI$ out of any pair of condensed charges. Hence, the EA order also vanishes. We therefore term this phase the featureless volume-law phase. It is worth pointing
out that this state has exactly the same symmetry breaking pattern as the maximally mixed state $\kket{\mI}=\mathds{1}^{\otimes 2}$. Hence, we can view this state as being smoothly connected to a conventional thermal state.
From this phase, the system can go into the broken symmetry volume-law phase by condensing the charges  $XIIX$ and $IXXI$.

Route 2 is realized if the charges $XIIX$ and $IXXI$ are the first to condense starting from the symmetric volume-law phase, leaving the symmetry generated by $\hat{\Pi}_1\hat{\Pi}_L$ intact. 
The ensuing intermediate phase exhibits both long-range EA order and nondecaying parity variance.
The former is seen by writing the EA order parameter as the product $XIXI=(XXII)(IXXI)$. The first factor is condensed in $\bbra{\mI}$, whereas the second factor is condensed in $\kket{\psi_{gs}}$. The latter is seen by factoring $\hat{\Pi}_L(A)=\hat{\Pi}_1(A)(\hat{\Pi}_1(A)\hat{\Pi}_L(A))$.
We call this phase the volume-law coexistence phase I.
The broken symmetry phase is reached from here by further condensing the charges $XXII$ and $IIXX$

Finally, route 3 is realized, starting from the symmetric volume-law phase, by condensing the charges $XIXI$ and $IXIX$. As in the area-law case, this again gives rise to coexistence of long-range EA order and nondecaying parity variance. We thus term this phase the volume-law coexistence phase II.

There is yet one more volume-law phase, obtained from the symmetric {\em area-law} phase by breaking the $\mathcal{S}_2$ symmetry while retaining a composite $\mathcal{Z}_2$ symmetry generated by $\mathcal{C}_\ell\hat{\Pi}_L$.
This is achieved by condensing the order parameter $ZYXI$ symmetrized over the residual symmetry $\mathcal{C}_\ell\hat{\Pi}_L$ and the local symmetries. 
This ``composite" volume-law phase, as we term it, exhibits an exponentially decaying parity variance because the ground state breaks the $\hat{\Pi}_L$ symmetry along with all the composite symmetries $g_\mathcal{I}\,\hat{\Pi}_L$.
At the same time, there is a long-range EA order, which can be seen by factoring $XIXI=(YYII)\ri(ZYXI)$. The first factor is condensed in $\bbra{\mathcal{I}}$. The second is a component of the order parameter of this phase. 
Further reducing the residual symmetry leads to the broken symmetry volume-law phase.

\subsection{Fisher Information as a probe of circuit states}\label{sec:Fisher}

So far we have considered equal-time properties of the output state, which translate to boundary matrix elements in the effective model. An alternative diagnostic approach, which is more natural to implement experimentally, focuses on the probability distribution of measurement outcomes obtained along the time evolution. In Ref.~\cite{bao2020theory}, we showed that the sensitivity of the measurement outcome distribution to changes in the initial state is a direct probe of the emergence of an encoding state. This sensitivity to initial conditions, is measured by a quantity known as the Fisher information in the distribution. The Fisher information is depressed below its maximal attainable value upon entering the volume-law phase, indicating that information about the perturbation to the initial state remains partially hidden and thus protected from the measurements. Within the effective model of random unitary circuits, the behavior of the Fisher information is a direct probe of spontaneous breaking of the $\mathcal{S}_n$ permutation symmetries established in the volume-law phases~\cite{bao2020theory}. 

The Fisher information probe can be adapted to characterize the symmetry enriched phases of the circuit. Specifically, we consider the sensitivity of the measurement outcome distribution to a symmetry breaking perturbation of the initial state. We will show that the corresponding Fisher information allows to distinguish between different  broken symmetry states and even identify SPT phases.

Let us start by giving a precise definition of the probing scheme. In the first, initialize the system in a pure state with definite parity $\ket{\psi_0} = \bigotimes_{j = 1}^L \ket{0}$. The next step is to apply a weak parity changing perturbation at site $i$, $U_{\theta,i} = e^{-\ri \theta X_i}$. Then operate the hybrid circuit on this state.
We denote by $p_{\theta,m}$ the distribution of measurement outcomes resulting from the initial state perturbed by $U_\theta$, dropping the site index $i$ for simplicity.
The Fisher information, which quantifies the sensitivity of the measurement outcome distribution to the perturbation $\theta$ is defined as
\begin{align}
    F = \frac{\rd^2}{\rd \theta^2}D_{KL}(p_{0,m}||p_{\theta,m})\Big|_{\theta = 0}.
\end{align}
$D_{KL}$, the Kullback-Leibler (KL) divergence, measures the distinguishability of two probability distributions $D_{KL}(p_{0,m}||p_{\theta,m}) = \sum_m p_{0, m} \log(p_{0,m}/p_{\theta,m})$.
The first order derivative of $D_{KL}$ is zero as $D_{KL}$ is non-negative and vanishes at $\theta = 0$.

In close analogy with the von-Neumann entropy we analyze KL divergence by defining a replica sequence
\begin{align}
    D^{(n)} \equiv \frac{1}{1-n} \log \left(\frac{\sum_m p_{0,m} p_{\theta,m}^{n-1}}{\sum_m p_{0,m}^n} \right),
\end{align}
where $D_{KL}$ is recovered in the replica limit $n \to 1$, i.e. $D_{KL} = \lim_{n \to 1} D^{(n)}$.
The $n = 2$ replica quantity $D^{(2)}$ can be formulated in our framework as
\begin{align}
    D^{(2)} &= \lim_{t \to \infty} -\log\left(\frac{\bbra{\mathcal{I}}e^{-t H_{\text{eff}}} \mathcal{U}_{\theta,X} \kket{\psi_0}}{\bbra{\mathcal{I}}e^{-t H_{\text{eff}}} \kket{\psi_0}}\right) \nonumber\\
    &= -\log\left(\frac{\bbra{\psi^\mathcal{I}_{gs}} \mathcal{U}_{\theta,X} \kket{\psi_0}}{\bbrakket{\psi^{\mathcal{I}}_{gs}}{\psi_0}}\right),
\end{align}
where $\kket{\psi_0} = \bigotimes_{j = 1}^L \kket{0000}$ represents the initial state $\ket{\psi_0}$ in the duplicated Hilbert space $\mathcal{H}^{(2)}$, and $\mathcal{U}_{\theta,X}=\mathds{1} \otimes \mathds{1} \otimes U_{\theta,i} \otimes U_{\theta,i}^*$ is the unitary operation in $\mathcal{H}^{(2)}$ generated by the single-qubit rotation $U_{\theta,i}$ acting on one copy of the density matrix.
The ground state $\bbra{\psi^{\mathcal{I}}_{gs}} = \lim_{t\to\infty} \bbra{\mathcal{I}}e^{-tH_{\text{eff}}}$ is obtained from long imaginary time evolution of the reference state. 
We distinguish this state by the superscript $\mathcal{I}$ from the ground state obtained by evolving the initial state $\kket{\psi_0}$.
There is a subtle difference between these two states in the broken symmetry phase. Because the reference state $\bbra{\mathcal{I}}$ has broken symmetries, the ground state evolved from it can break symmetries explicitly. By contrast, the state $\kket{\psi_{gs}}$ in the broken symmetry phase is a macroscopic superposition of the two broken-symmetry states (i.e. a cat state). 

From the KL divergence, we define the $n$-th Fisher information as $F^{(n)} \equiv \rd^2_\theta D^{(n)}\rvert_{\theta = 0}$.
Thus, the second Fisher information $F^{(2)}$ is given by a boundary matrix element 
\begin{align}
F^{(2)} &\equiv \rd_\theta^2 D^{(2)} = 2\frac{\bbra{\psi^\mathcal{I}_{gs}} 1-\mathcal{O}_{X,i} \kket{\psi_0}}{\bbrakket{\psi^\mathcal{I}_{gs}}{\psi_0}}\nonumber\\
&=2\left( 1- \frac{\bbra{\psi^\mathcal{I}_{gs}} \widetilde{\mathcal{O}}_{X,i} \kket{\psi_0}}{\bbrakket{\psi^\mathcal{I}_{gs}}{\psi_0}}\right).
\label{eq:F2}
\end{align}
Here $\mathcal{O}_{X,i} = [(IIXX)_i+(XXII)_i]/2$ and in the last equality we  
claim that $\mathcal{O}_{X,i}$ can be replaced with the operator 
$\widetilde{\mathcal{O}}_{X,i} = [(IIXX)_i + (XXII)_i + (XIIX)_i + (IXXI)_i]/2$ without changing the Fisher information. This is because $XIIX$ and $IXXI$ are odd under the symmetry $\hat{\Pi}_1$, whereas  $\bbra{\mathcal{I}}$, $\kket{\psi_0}$ and $\exp(-t H_{\text{eff}})$ are even. Therefore, the matrix element of these components vanishes.\footnote{Note that even if the ground state spontaneously breaks the $\hat{\Pi}_1$ symmetry, it manifests in $\bbra{\psi^\mathcal{I}_{gs}}$ through emergence of a cat state because the reference state is symmetric under $\hat{\Pi}_1$.} 

For convenience, we define the ``order parameter''
\be
\Phi^{(2)}\equiv 1- {1\over 2} F^{(2)}=\frac{\bbra{\psi^\mathcal{I}_{gs}} \widetilde{\mathcal{O}}_{X,i} \kket{\psi_0}}{\bbrakket{\psi^\mathcal{I}_{gs}}{\psi_0}}.\label{eq:order_parameter}
\ee
Because $\widetilde{\mathcal{O}}_{X,i}$ is symmetric under both left and right swap, while it is odd under $\hat{\Pi}_L$, the order parameter can detect breaking of circuit symmetries that anti-commute with $XXII$ and $IIXX$, 
while it is insensitive to breaking of the $\mathcal{S}_2$ permutation symmetry. 

It is important to note that this behavior is special to circuits with physical $\mathbb{Z}_2$ symmetry. In particular, the replacement of $\mathcal{O}_{X,i}$ with $\widetilde{\mathcal{O}}_{X,i}$ can be made only if $H_{\text{eff}}$ inherits the Ising symmetry $\hat{\Pi}_1$ from the circuit. To define a probe that differentiates between area-law and volume-law phases as in Ref.~\cite{bao2020theory}, we would need to consider the Fisher information, which measures the sensitivity of the measurement outcomes to a different perturbation, which respects the circuit symmetries. Note also that the matrix element $\Phi^{(2)}$ is necessarily positive in the broken symmetry state because $\bbra{\psi^\mathcal{I}_{gs}}$ breaks the symmetry in a definite way inherited from $\bbra{\mI}$, which is a positive weight superposition of symmetric and anti-symmetric states. 

The behavior of $\Phi^{(2)}$ in the different phases is summarized in Table~\ref{tab:phases}. From the information theoretical perspective, $\Phi^{(2)}>0$ (i.e. Fisher information below its maximal value) means that the information about the perturbation to the initial state remains partially hidden from the measurement results.

The Fisher information can furthermore allow  detection of the SPT area-law phases by directly probing edge states. 
Recall that the reference state $\bbra{\mI}$ provides symmetry breaking initial conditions to the imaginary time evolution $\bbra{\mathcal{I}}e^{-tH_{\text{eff}}}\to \bbra{\psi^\mI_{gs}}$. In the SPT phase, the broken symmetry can propagate to infinite time by coupling to the edge zero mode. This implies the order parameter has a spatial dependence, and $\Phi^{(2)}_i>0$ for $i$ near the edge, but decaying exponentially into the bulk of the qubit chain. It is also possible to probe the broken permutation symmetry at the edge of the SPT phase by measuring the Fisher information associated with a perturbation that respects the physical circuit symmetry.

Crucially, this detection scheme for the edge modes of the SPT phase is efficient and scalable in the sense that the number of repetitions of the experiment needed to gather enough data does not grow with the system size. 
Since we are probing a gapped state with a finite correlation length $\xi$ in space and time, the natural expectation is that it is enough to produce the Fisher information from a partial measurement record including only a finite space-time bubble of radius $r\gg\xi$ away from the perturbation. Indeed, within the effective theory, discarding measurement results creates a bulk field that breaks the $\mathcal{S}_2$ symmetry in the discarded region outside of the bubble~\cite{bao2020theory}. The effect of these symmetry breaking conditions on the perturbed site in the bubble decays exponentially with $r/\xi$, leading to a correspondingly small effect on the Fisher information. 
Thus, the required number of repetitions is expected to scale as $e^\xi$, allowing an efficient detection.

\subsection{Numerical demonstration of transitions between volume-law phases}\label{sec:z2_numerics}
\begin{figure}
    \centering
    \includegraphics[width=0.45\textwidth]{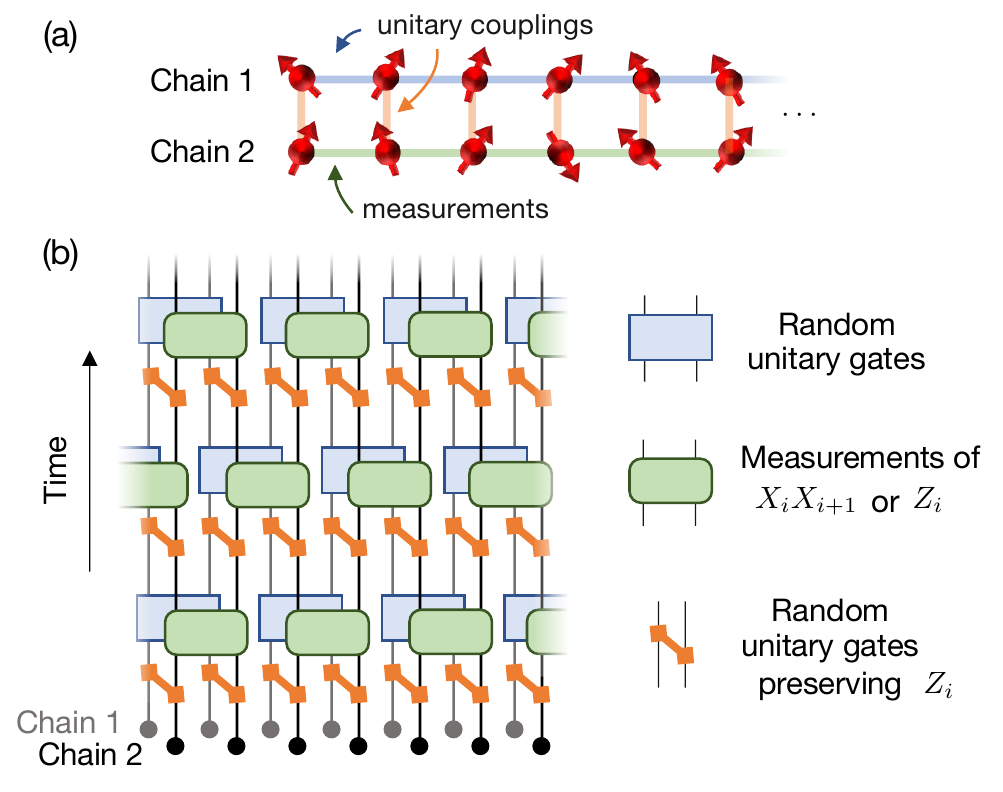}
    \caption{
    \label{fig:volume_law_model}
    (a) Two chain model investigated numerically. Internal dynamics of chain-1 is generated by purely unitary two-qubit gates, while that of chain-2 by pure projective measurements (of $Z_i$ or $X_i X_{i+1}$). The
    two chains are coupled by  random unitary gates applied with probability $q$.
    All gates and measurements commute with the  $\mathbb{Z}_2$-parity symmetry of chain 2.
    (b)
    Circuit diagram for the model.
    Every time step consists of three types of operations: 
    two-qubit unitary gates within chain-1 (blue boxes) are randomly drawn from the Clifford group; 
    measurements in $X_iX_{i+1}$- or $Z_i$-basis are performed on chain-2 (green boxes with rounded corners) with the probabilities $r$ and $1-r$, respectively; 
    and the inter-chain gates (orange bonds) are randomly drawn from the Clifford elements preserving $Z_i$ on the chain-2 and applied to the circuit with the probability $q$. }
\end{figure}

Having classified the possible phases of quantum circuits enriched by the physical $\mathbb{Z}_2$ circuit symmetry, we now demonstrate that parts of this phase structure is realized in a concrete circuit model illustrated in Fig.~\ref{fig:volume_law_model}. 
We simulate the circuit directly without relying on replica tricks. The results therefore give further confidence that the effect of the enlarged dynamical symmetry persists to the physical replica limit $n\to 1$.

\begin{figure*}
    \centering
    \includegraphics[width=0.85\textwidth]{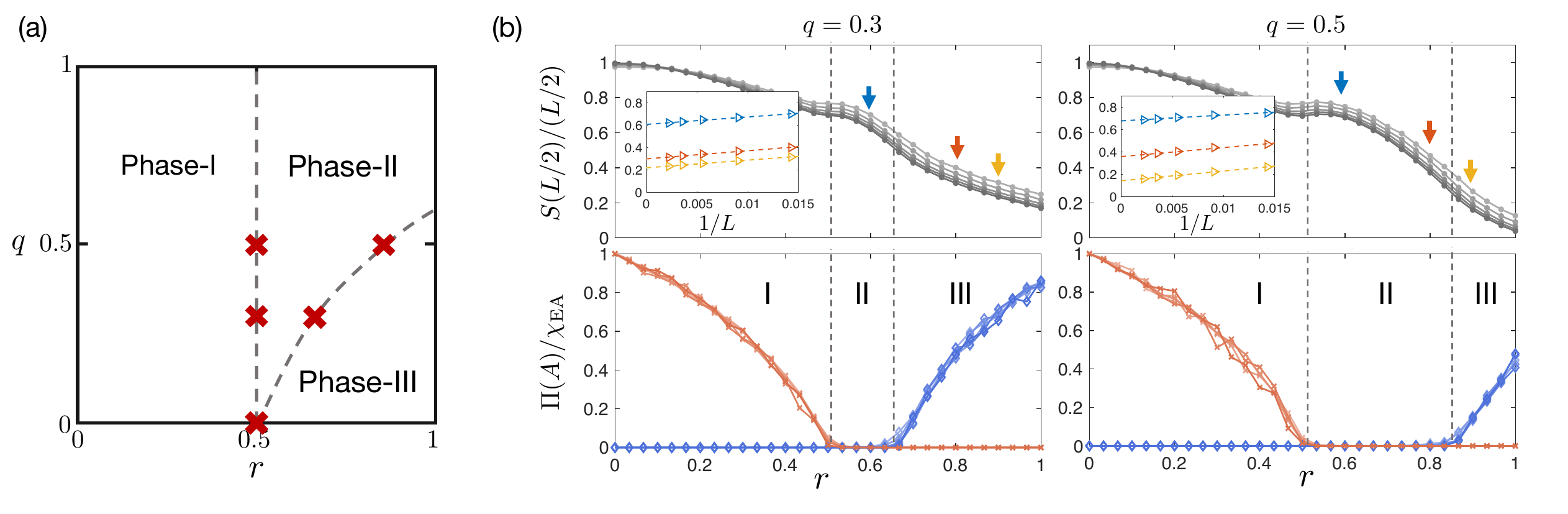}
    \caption{
    \label{fig:volume_law}
    (a) Phase diagram of the two chain model. $q$ is the inter-chain coupling. $r$ is the probability of measuring $X_iX_{i+1}$ on chain 2 (versus $1-r$ of measuring $Z_i$). The red crosses indicate phase boundaries extracted from the numerics. All three phases exhibit volume-law entanglement.
    (b) Numerical simulation results. Top panels: Half-chain entanglement entropy normalized by volume. Insets: system size dependence of the entanglement entropy density at points $r=0.6,0.8,0.9$ (indicated by arrows) indicates convergence to a non-vanishing value as $L\to\infty$. Bottom: Behavior of the Edwards-Anderson correlation function $\chi_{EA}(L/2)$ (blue diamonds) and parity variance $\Pi(A=L/2)$ (red crosses)
    as a function of $r$ for two values of $q$.  The nearly overlapping lines correspond to different system sizes $L=70, 110, 180, 280, 450$.
    }
\end{figure*}

Our model consists of two coupled $L$-qubit chains in a ladder geometry, undergoing a hybrid dynamics that consists of unitary gates and projective measurements.
We restrict the set of unitary gates to the elements of the two-qubit Clifford group, which allows efficient simulation of large systems~\cite{aaronson2004improved,chandran2015semiclassical}. 
This restriction should not affect any qualitative features of our analysis as Clifford gates are statistically indistinguishable from Haar random unitary gates up to third moments, i.e. Clifford group forms unitary 3-design~\cite{webb2015clifford}.  

The circuit dynamics can be summarised as follows. First, all unitary gates and measurements respect the global $\mathbb{Z}_2$ symmetry generated by $\hat{\pi} = \prod_j Z_j$, where $Z_j$ are  the Pauli-$Z$ operators acting on the qubit at site $j$ in chain-2. Thus, the qubits of chain-1 are neutral with respect to the $\mathbb{Z}_2$ symmetry.
The time evolution internal to chain-1 is generated purely by unitary gates, which are unrestricted by symmetry. 
The time evolution internal to chain-2 is generated purely by measurements: with probability $r$ a pair of qubits at sites $j$ and $j+1$ are measured in the $X_j X_{j+1}$-basis or, with probability $1-r$ the qubit at site $j$ is measured in the $Z_j$-basis. These measurements are denoted by the green boxes in the figure.
Finally, a coupling between the two chains is implemented by two-qubit unitary gates on the rungs of the ladder, applied with probability $q$ per gate at each time step.   These gates, represented by orange rungs in Fig.~\ref{fig:volume_law_model}, are drawn uniformly from one of the 384 elements in the Clifford group that commute with $Z_j$ in the chain-2.
Thus, the parameter $q$ controls the unitary coupling strength between two chains, and $r$ tunes the ratio between different kinds of measurements on chain-2. The system is initialized in the state $\ket{\psi_0} = \ket{0}^{\otimes 2L}$, and we are interested in its steady-state properties. 

We can easily understand the steady-state behavior of our model in the decoupled chains limit $q=0$.
In this case, chain-1 evolves as a random unitary circuit, in which the entanglement entropy of any subsystem grows linearly until it saturates to its maximum~\cite{nahum2017quantum,von2018operator}.
In contrast, chain-2 undergoes measurement-only dynamics.
Numerical simulations and analytical arguments in Ref.~\cite{sang2020measurement} have shown that this leads to two area-law phases separated by a phase transition at $r_c = 1/2$. For $r<r_c$, chain-2 is in the symmetric area-law phase characterized by nondecaying parity variance and exponentially decaying EA correlations. Chain-2 with $r>r_c$ realizes the broken symmetry area-law phase with long-range EA correlations and exponentially decaying parity variance.

When the two decoupled chains are considered as a single system, they trivially realize a volume-law entanglement scaling (due to chain-1), which undergoes $\mathbb{Z}_2$ symmetry breaking phase transition (due to chain-2) at $r=r_c$.
Within our formalism these two volume-law phases are represented by gapped ground states of an effective Hamiltonian. Thus, they are expected to be robust to (at least) weak coupling between the chains.\footnote{By the same token, the phases are also robust to any other small perturbation to the circuit, which respects the symmetry.} 
Furthermore, we can use the symmetry analysis of Section~\ref{sec:z2_phases} to predict which phases would be realized. 

Since chain-1 has broken $\mathcal{S}_2$ symmetry, its leading effect at weak coupling $q\ll 1$ is to exert an effective $\mathcal{S}_2$-symmetry breaking field on chain-2, which does not affect the properites associated with the parity symmetry.
Thus, from the symmetry perspective the state established for $r<r_c$ at an infinitesimal coupling is identical to the symmetric volume-law phase  (row 5 in Table~\ref{tab:phases}). Similarly, the state established for $r>r_c$ at an infinitesimal coupling is identical to the broken symmetry volume-law phase (row 9 in Table~\ref{tab:phases}). 

Finally, coupling to the symmetry breaking field produced by chain-1 is relevant at the critical point separating the two area-law phases of chain-2. As discussed in subsection~\ref{sec:z2_area_phases}, the critical point at $q=0$ is characterized by condensation of the charges $\mathcal{Q}_1=XXII$ and $\mathcal{Q}_2=IIXX$ together with $\mathcal{Q}_1'=IXXI$ and $\mathcal{Q}_2'=XIIX$. Because the two pairs of charges are connected by the $\mathcal{S}_2$ symmetries, they must condense together when these symmetries are present. However, once we turn on the field breaking the $\mathcal{S}_2$ symmetries, the critical point will generically split into two critical points, one involving condensation $\mathcal{Q}_1$ and $\mathcal{Q}_2$ and the other involving condensation of $\mathcal{Q}_1'$ and $\mathcal{Q}_2'$. %
This gives rise to an intermediate phase in which only one of the two pairs of charges is condensed (the featureless phase in row 6 or the coexistence phase I in row 7 of Table~\ref{tab:phases}). Which intermediate phase is realized  depends on the charges that acquire a lower energy, which in turn may depend on the details of the model.

The persistence of the two distinct volume-law phases and the emergence of an intermediate phase between them in the coupled-chain model are nontrivial predictions of the effective theory, which we now turn to test numerically. To this end, we simulate the time evolution of systems with length $L\le 450$. We evaluate $\Pi(A)$, $\chi_{EA}$, and the volume-normalized half-system entanglement entropy, $S(L/2)/(L/2)$ averaged over 
$\sim 500$ random circuit realizations. 
To focus on the long-range and late-time behaviors, we use $i_\ell \approx L/3$ and $i_r \approx 2L/3$ for evaluating $\chi_{EA}(i_\ell,i_r)$ and $\Pi(A)$ (as the left and right edges of $A$) and we evaluate all quantities at time $T \approx 3L$.\footnote{More precisely, we choose $T = 3L+1-a$ with $a=0$ or $a=1$ depending on whether $L$ or $L+2$ is a multiple of $4$, respectively. We always choose even $L$. This choice is introduced in order to avoid artifacts associated the gates acting across the half-chain boundaries at the very last time step. We did similar regularizations for the choice of $i$ and $j$ for $\Pi(A)$ and $\chi_{EA}$.}

The results of our numerical simulations are summarized in
Fig.~\ref{fig:volume_law}. For the decoupled chains ($q=0$), we find, as expected, a single phase transition tuned by the ratio $r$ of bond to site measurements. Upon increasing the inter-chain coupling to $q=0.3$ and $q=0.5$, we observe a splitting into two critical points and emergence of an intermediate phase. As seen in the bottom of panel (b), the intermediate phase is characterized by vanishing of both the the EA correlations and the parity variance, consistent with the featureless phase in row 6 of Table~\ref{tab:phases}. Physically, this may be interpreted as a result of the unitary inter-chain coupling countering the effect of the $X_iX_{i+1}$ measurements, favoring the state without long-range order. We see that for sufficiently large $q$ there is no broken symmetry state for any value of $r$. At the same time, the unitary coupling $q$ involving a single-site $Z_i$ operator on chain 2 commutes with the sub-system parity. Therefore introducing this coupling does not change the point at which non-decaying sub-system parity variance is established. 
It would be interesting to check if coupling the chains by various measurements that preserve the $\mathbb{Z}_2$ symmetry instead of unitary gates could favor a volume-law coexistence state (row  7 in Table~\ref{tab:phases}) as the intermediate phase. 

The top panels of Fig.~\ref{fig:volume_law}(b) confirm that all the observed phases are indeed characterized by volume-law scaling of the entanglement entropy. 
For large values of $r$, we observe a slow drift of the entropy-density $S(L/2)/(L/2)$ as a function of system sizes [inset of Fig.~\ref{fig:volume_law}(b)].
We attribute this drift to the additive contribution $S_0$ of chain-2 to $S(L/2)$.
Indeed, plotted as a function of $1/L$, the entanglement entropy density shows a clear linear dependence, $S(L/2)/(L/2) = 2S_0 / L + s$ with a finite entropy density $s>0$ in the limit $L\rightarrow \infty$ (the dotted lines in the inset).

The establishment of long-range order in a one-dimensional system with extensive entropy, may appear, at first sight, contradictory to theorems prohibiting such order at any non-vanishing temperature. But, this is not a thermal system.

As we mentioned above, an appropriate description of the volume-law phase at weak coupling ($q\ll 1$) is to replace chain-1 by an effective field that breaks the $\mathcal{S}_2$ permutation symmetries. In the physical circuit, this corresponds to coupling chain-2 to an effective infinite temperature bath in place of chain-1. We are then interested in the trajectories of the mixed state in chain-2,  connected by coupling $q$ to an infinite temperature bath and subject to local measurements. For each individual trajectory, the measurements act as couplings to  zero temperature bath, wherein the measurement projection collapse the local wave function into a definite quantum state with zero entropy. Because of this coupling to two very different baths, each trajectory is very far from being in thermal equilibrium and establishment of long-range order is not prohibited.

One of the nontrivial results of this section is that measurements can facilitate the establishment of quantum order in one dimension, even in states with volume-law entanglement entropy. We may further argue that this protection extends also to topological order and edge states.
To this end, we note that our model can be also interpreted as the system of Majorana fermions on chain-2
via Jordan-Wigner transformation. On-site $Z$ measurements map to measurement of the Majorana parity $\ri \gamma_{2i-1}\gamma_{2i}$ on odd bonds, whereas the $XX$ measurements map to measurements of parity on even bonds $\ri \gamma_{2i}\gamma_{2i+1}$. Chain-1 is coupled to chain-2 with unitary coupling $q$ to the Majorana bond parities. The broken symmetry volume-law phase of the qubit system (phase III in Fig.~\ref{fig:volume_law}) translates to a topological volume-law state with free Majorana edge modes. The symmetric volume-law phase (phase I in Fig.~\ref{fig:volume_law}) translates to a volume-law phase which is distinct from a thermal state by having subsystem fermion-parity variance that is nondecaying with subsystem size. Finally, phase II in Fig.~\ref{fig:volume_law} is smoothly connected to the conventional thermal phase of the Majorana system, where there is no long-range correlation and the fermion parity of a subsystem fluctuates.

\section{Gaussian fermionic circuits}\label{sec:fermion}

We turn to investigate hybrid quantum circuits that operate on fermionic degrees of freedom rather than on qubits. Specifically we consider quadratic gates and measurements that preserve the Gaussianity of the fermionic wave function.
Such circuits cannot sustain a volume-law state for any finite rate of measurements~\cite{cao2018entanglement,fidkowski2020dynamical}. However, recent numerical work indicated that a critical phase with entanglement that scales as $\log(L)$ is established at low measurement rates~\cite{alberton2020trajectory,sang2020entanglement}.

Here, we provide a simple description of this critical phase using the framework developed in Section~\ref{sec:framework} to map the dynamics to an effective ground state problem. While we describe the general structure of the theory for any number of replicas $n$, we carry out a more detailed study only of the case of $n=2$. This theory predicts a measurement-induced Kosterlitz-Thouless (KT) transition from the critical phase into two area-law phases, which are distinguished by the sign of the vortex fugacity. These are identified as a trivial and a topological area-law phase.
We map pertinent properties of the circuit dynamics to universal boundary operators in this theory, thus facilitating a detailed comparison to exact numerical simulations of the circuit. The numerical results, obtained for systems of size $L\le 160$, are consistent with a KT transition.

Before proceeding we note that the two area-law phases, a topological and a trivial phase 
have been previously discussed in the context of measurement only models~\cite{nahum2020entanglement,lang2020entanglement}. %
Furthermore, Sang et al.~\cite{sang2020entanglement} argued that adding a specific set of gates to the measurement only model, allows a mapping to the completely packed loop model with crossings~\cite{nahum2013loop}, which also hosts a critical state known as the Goldstone phase. We comment on the connections of our results to these loop model predictions at the end of the section.

\subsection{Model}\label{sec:fermi_model}

\begin{figure}
    \centering
    \includegraphics[width=0.48\textwidth]{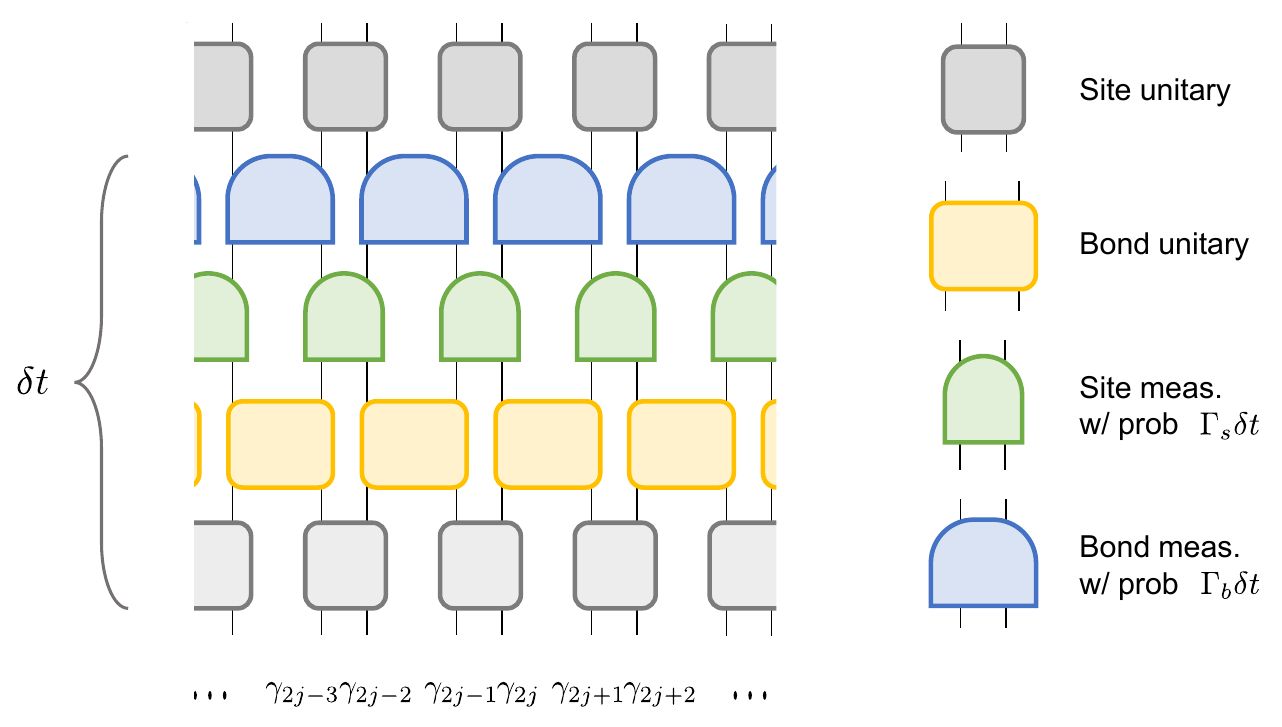}
    \caption{Gaussian fermionic circuit. Each time step $\delta t$ consists of sequential application of one layer of site unitary gates, bond unitary gates, site measurements, and bond measurements. The quadratic site and bond parity measurements are applied with probabilities $\Gamma_{s}\delta t$ and  $\Gamma_b\delta t$ respectively.}
    \label{fig:fermionic_ckt}
\end{figure}

The general structure of the circuits we consider here is shown in Fig.~\ref{fig:fermionic_ckt}. Circuit elements are operating on a one-dimensional chain of Majorana modes $\gamma_{2j-1}$ and $\gamma_{2j}$. We view such pairs of ``Majorana sites" $(2j-1, 2j)$ as making up a single ``physical site" $j$, hosting a single complex fermion $f_j = (\gamma_{2j-1} + \ri \gamma_{2j})/2$ on site $j$. For simplicity, we take the initial state to be the vacuum of all fermionic modes, i.e. $f_j|\text{vac}\rangle_f = 0~~\forall j$.

The unitary elements of the circuit are generic Gaussian gates operating on the nearest neighbor Majorana pairs
\begin{align}
    U_{s,j} = e^{\theta_{s,j}\gamma_{2j-1}\gamma_{2j}}, \quad U_{b,j} = e^{\theta_{b,j}\gamma_{2j}\gamma_{2j+1}}.
\end{align}
We call $U_{s,j}$ the site unitary and $U_{b,j}$ the bond unitary as they operate on physical sites and bonds, respectively. These gates are depicted as gray and yellow boxes in Figure~\ref{fig:fermionic_ckt}.

Similar to the qubit circuits discussed in the previous section, drawing the couplings $\theta_{s,j}$ and $\theta_{b,j}$ from symmetric distributions ensures a mapping to imaginary time evolution with an effective Hermitian Hamiltonian. 
Here, we shall draw the bond coupling $\theta_{b,j}$ from a Gaussian distribution $N(0, \sigma^2_J)$ with zero mean and variance $\sigma^2_J$, while the site coupling 
$\theta_{s,j}$ is taken from a uniform distribution over $[0, 2\pi)$. %
We note that this specific choice of the distributions is made to simplify the exact mapping and the resulting effective Hamiltonian, and it is not crucial for our general discussion in this section.

After each layer of unitary gates in the circuit, measurements are made on a certain fraction of the fermion parity operators defined on sites and bonds:
\begin{align}
    \hat{\pi}_{s,j} = -\ri \gamma_{2j-1}\gamma_{2j}, \quad \hat{\pi}_{b,j} = -\ri\gamma_{2j}\gamma_{2j+1}.
\end{align}
Measurements of the site parity are performed with a probability $p_s$ at each time step and those of bond parity with probability $p_b$.
Upon the measurement of  $\hat{\pi}_{a,j}$ with $a \in \{s,b\}$, the quantum state evolves according to the Born rules: projection onto a definite parity eigenstate
$P_\pm |\Psi\rangle/\sqrt{\langle \Psi | P_\pm |\Psi\rangle}$ with probability $\langle \Psi | P_\pm |\Psi\rangle$, where $P_\pm=(1\pm \hat{\pi}_{a,j})/2$.

We investigate the quantum state at the output of this circuit after a long time evolution of $N_T\gg 1$  steps. As before, we are interested in quantities that are nonlinear in the system density matrix, such as the entanglement entropy of a subsystem $A$ or the variance of fermion parities taken over the ensemble of trajectories:
\be
\Pi_{a,A}\equiv \overline{\sum_{m} p_{m} \Big\langle \prod_{j\in A}\hat{\pi}_{a,j}\Big\rangle^2_{m}}.
\label{eq:parity_corr}
\ee
The sum is over different trajectories (runs of the circuit) characterized by a specific set of measurement outcomes $m$ and the angled brackets represent a quantum expectation value in the output state of that trajectory.

\subsection{Enlarged Symmetry}\label{sec:fermi_symm}
The entanglement entropy and other quantities that depend on high moments of observables can be derived from the average replicated density matrix $\kket{\tilde{\rho}^{(n)}}$.
Thus, it is important to identify the symmetry of the dynamics of $n$ replicas, which will allow us to classify distinct steady state phases. Due to the special structure of Gaussian fermionic circuits, this symmetry is different from the dynamical symmetry identified for the qubit circuits in Section~\ref{sec:symmetry}.

To facilitate the analysis, we introduce second quantized operators acting in the duplicated Fock space.
The Majorana operators $\gamma_j$ acting on the wave-function of the physical circuit are extended to $\gamma_{j,\alpha\sigma}$, where $\alpha \in \{1,2, \dots, n\}$ indicates the replica on which this Majorana operates and $\sigma \in \{\uparrow, \downarrow\}$ indicates forward and backward branch of the evolution, respectively.  Note that the mapping from tensor product operators, such as $\gamma_j\otimes\mathds{1}\otimes\mathds{1}\otimes\mathds{1}$ or $\mathds{1}\otimes\gamma_j\otimes\mathds{1}\otimes\mathds{1}$ to the second quantized operators in the duplicated Hilbert space requires extra phase factors to ensure anti-commutation relations between all fermionic operators.
The precise mapping is given in Appendix~\ref{app:correspondence}.

The evolution of the replicated density matrix in a single time step of the fermionic circuit in Fig.~\ref{fig:fermionic_ckt} is given by 
\begin{align}
    \kket{ \tilde{\rho}^{(n)}(t+\delta t)} = \mathcal{M}^{(n)}_b \mathcal{M}^{(n)}_s \mathcal{U}_b^{(n)} \mathcal{U}_s^{(n)} \kket{ \tilde{\rho}^{(n)}(t)}.
\end{align}
As discussed in Section~\ref{sec:framework},  $\mathcal{M}^{(n)}_{s/b}$ is a linear operation on the unnormalized quantum state $\kket{\tilde{\rho}^{(n)}}$ generated by a layer of site/bond measurements performed with probabilities $p_{s/b}$.

We first focus on the the dynamical symmetry in the case of pure unitary evolution, i.e. when $p_{s} = p_b =0$. 
The generators of the unitary time steps take the form of decoupled quadratic Majorana Hamiltonians in the duplicated Fock space:
\begin{align}
    h_s^{(n)} &= \sum_{\alpha = 1}^n \sum_{j = 1}^L \ri \theta_{s,j}  \left(\gamma_{2j-1, \alpha \uparrow} \gamma_{2j, \alpha \uparrow} - \gamma_{2j-1, \alpha \downarrow} \gamma_{2j, \alpha \downarrow}\right), \nonumber\\
    h_b^{(n)} &= \sum_{\alpha = 1}^n \sum_{j = 1}^L \ri \theta_{b,j} \left(\gamma_{2j, \alpha \uparrow} \gamma_{2j+1, \alpha \uparrow} - \gamma_{2j, \alpha \downarrow} \gamma_{2j+1, \alpha \downarrow}\right).
    \label{eq:hamiltonians}
\end{align}
Using the Majorana operators of the $2n$ chains, it is natural to construct a local $so(2n)$ algebra on every Majorana site $\ell$, generated by 
$\Gamma^{\alpha\sigma,\beta\sigma'}_{\ell} =\frac{\ri}{4}[\gamma_{\ell,\alpha\sigma},\gamma_{\ell,\beta\sigma'}]
$.
From these local objects, we construct a global $so(2n)$ algebra with elements
\be 
\Gamma^{\alpha\sigma,\beta\sigma'}=\sum_{j = 1}^L \left({\sigma\sigma'}\Gamma^{\alpha\sigma,\beta\sigma'}_{2j-1}+\Gamma^{\alpha\sigma,\beta\sigma'}_{2j}\right),
\label{eq:so2n}
\ee
which commute with the generators~\eqref{eq:hamiltonians} of the unitary dynamics. Note that here and below $\sigma = \uparrow$ ($\sigma = \downarrow$) is identified with $\sigma = +1$ ($\sigma = -1$).

The $\Gamma$ matrices generate proper rotations among the $2n$ species of Majorana fermions. In addition we have the single-branch fermion parity symmetries 
$\hat{\Pi}_{\alpha,\sigma} = \prod_{j = 1}^L \ri\gamma_{2j-1,\alpha,\sigma}\gamma_{2j,\alpha,\sigma}$, which constitute improper rotations, e.g. $\hat{\Pi}_{1,\ua}$ that maps $\gamma_{j,1,\uparrow} \mapsto -\gamma_{j,1,\uparrow}$.
Thus, altogether the purely unitary dynamics on $n$ copies have an $O(2n)$ symmetry.

In addition, the Hamiltonian admits a fermionic time-reversal symmetry $\mathbb{T}$, under which $\mathbb{T}\gamma_{\alpha,\ua}\mathbb{T}^{-1} = \gamma_{\alpha,\da}$, $\mathbb{T}\gamma_{\alpha,\da}\mathbb{T}^{-1} = -\gamma_{\alpha,\ua}$, and $\mathbb{T}\,\ri\,\mathbb{T}^{-1} = -\ri$. We conclude that the unitary dynamics exhibits the symmetry $O(2n)\rtimes \mathbb{Z}_2^{\mathbb{T}}$. 
Note that the effective time-reversal symmetry is not exactly the Hermiticity symmetry, which appears as a nonlocal transformation in the fermionic duplicated Hilbert space (see Appendix~\ref{app:correspondence}).

Adding measurements reduces the symmetry. Only the subset of the generators~\eqref{eq:so2n} having $\sigma=\sigma'$ commute with the measurement operators. These generators, supported only on either the forward or the backward branches, form a $so(n)\oplus so(n)$ algebra and give rise to a $O(n)\times O(n)$ symmetry. 
Together with fermionic time reversal, the symmetry of the $n$-copy unitary evolution with measurements is $[O(n)\times O(n)]\rtimes \mathbb{Z}_2^{\mathbb{T}}$. 

It is worth pointing out that the dynamical symmetry of the pure unitary evolution is the same as the static symmetry in the case of free fermions in the symplectic class AII~\cite{Ryu2007,Fu2012}. Moreover, the addition of measurements here breaks the $O(2n)$ symmetry in the same way as adding a non-Hermitian quadratic coupling does in the free fermion problem. There is, however, an important difference in taking the replica limit, which in our case is $n\to 1$, whereas it is $n\to 0$ for the disordered fermion ground state problem.   

We note on passing that a measurement-only model (i.e. no unitary gates) also has an enlarged dynamical symmetry $O(2n) \rtimes \mathbb{Z}_2^{\mathbb{T}}$ like the purely unitary model. In this case, the $O(2n)$ symmetry generators are given as simple zero momentum sums of the local generators. That is, the first term in Eq.~\eqref{eq:so2n} does not have an additional sign $\sigma \sigma'$.

Similar to the case of the qubit circuits considered in Section~\ref{sec:Z2model}, the effective global symmetry is reduced when we account for the presence of local integrals of motion. Specifically, the fermion gates and measurements commute with all the local operators  $\mathcal{R}_\ell = \prod_{\alpha = 1}^n \ri \gamma_{\ell,\alpha\uparrow}\gamma_{\ell,\alpha\downarrow}$ that measure the total parity of fermions from all species at Majorana site $\ell$ (see Appendix~\ref{app:local_symmetry}). 
Thus $\kket{\tilde{\rho}^{(n)}}$ evolves to a ground state of $H_{\text{eff}}$ with definite eigenvalues of $\mathcal{R}_\ell$. Such a state necessarily breaks the single branch parity symmetries that anti-commute with the $\mathcal{R}_\ell$.

Having removed the improper rotations generated by the  single-branch parities, we are left with the effective symmetry $SO(2n) \rtimes \mathbb{Z}_2^{\mathbb{T}}$ for the purely unitary circuit. With measurements, the effective symmetry is reduced to $\mathcal{G}^{(n)}_{f,\text{eff}} \equiv (SO(n) \times SO(n)) \rtimes (\mathbb{Z}_2 \times \mathbb{Z}_2^{\mathbb{T}})$, where the first $\mathbb{Z}_2$ is generated by fermion parity in both the forward and backward branch of the first copy, i.e. $\hat{\Pi}_1 = \hat{\Pi}_{1\ua}\hat{\Pi}_{1\da}$. Finally, we note that the global symmetry $\prod_{\alpha = 1}^n\hat{\Pi}_\alpha=\prod_\ell \mathcal{R}_\ell$, which is part of $\mathcal{G}^{(n)}_{f,\text{eff}}$, cannot be broken because it is a product of local integrals of motion.
We now elaborate on how this scheme applies to the case of two replicas. The dynamical $O(2) \times O(2)$ symmetry in presence of measurements, implies two conserved $U(1)$ charges, which we can write as the total occupation numbers of complex fermion modes
\be
N_\sigma= \sum_{j = 1}^L N_{\sigma,j} = \sum_{j = 1}^L c^\dagger_{2j-1,\sigma}c_{2j-1,\sigma}+c^\dagger_{2j,\sigma}c_{2j,\sigma},
\label{eq:charge}
\ee
The complex fermions are not local to a single copy of the system and are rather defined as superpositions of Majorana modes on the two copies:
\begin{align}
    c_{2j-1,\uparrow} = \frac{\gamma_{2j-1,1\uparrow} + \ri\gamma_{2j-1,2\uparrow}}{2}, c_{2j,\uparrow} = \frac{\gamma_{2j,2\uparrow}-\ri\gamma_{2j,1\uparrow}}{2}, \nonumber \\
    c_{2j-1,\downarrow} = \frac{\gamma_{2j-1,1\downarrow}-\ri\gamma_{2j-1,2\downarrow}}{2}, c_{2j,\downarrow} = \frac{\gamma_{2j,2\downarrow} + \ri \gamma_{2j,1\downarrow}}{2}.\label{eq:c_fermion}
\end{align}
The $U(1)$ symmetries are supplemented by the single branch $\mathbb{Z}_2$ parity symmetries to give the two copies of $O(2) = U(1) \rtimes \mathbb{Z}_2$.
These parity transformations act on the two conserved fermion species independently as particle-hole transformations, namely $c_{j,\sigma} \mapsto (-1)^j c_{j,\sigma}^\dagger$. %
Together with the time-reversal we have the full symmetry $\mathcal{G}^{(2)}_{f} \equiv [O(2)\times O(2)]\rtimes \mathbb{Z}_2^{\mathbb{T}}$.

For purely unitary dynamics, the symmetry is enlarged to  $O(4) \rtimes \mathbb{Z}_2^{\mathbb{T}} = (SO(4) \rtimes \mathbb{Z}_2)\rtimes \mathbb{Z}_2^{\mathbb{T}}$. The $SO(4)$ symmetry has a useful and intuitive representation in terms of two $SU(2)$ symmetries, which stems from the isomorphism $SO(4)\cong [SU(2)\times SU(2)]/\mathbb{Z}_2$. 
One of the $SU(2)$ symmetries is associated with rotations between the forward and backward branches, labeled as up and down spins. We call this the $\Sigma$ spin symmetry. The other $SU(2)$ symmetry is a charge ``$\eta$-symmetry''~\cite{yang1990so}. The respective generators of the spin and charge symmetries can be written explicitly as
\begin{subequations}\label{eq:su2_generators}
\begin{align}
\Sigma_j^+ &\equiv c_{2j-1,\uparrow}^\dagger c_{2j-1,\downarrow} + c_{2j,\uparrow}^\dagger c_{2j,\downarrow},\\
\Sigma_j^- &\equiv \left(\Sigma_j^+\right)^\dagger,\\
\Sigma_j^z &\equiv \frac{1}{2} \sum_\sigma \sigma \left( c_{2j-1,\sigma}^\dagger c_{2j-1,\sigma} + c_{2j,\sigma}^\dagger c_{2j,\sigma}\right),\\
\eta_j^+ &\equiv - c_{2j-1,\uparrow}^\dagger c_{2j-1,\downarrow}^\dagger + c_{2j,\uparrow}^\dagger c_{2j,\downarrow}^\dagger,\\
\eta_j^- &\equiv \left(\eta_j^+\right)^\dagger,\\
\eta_j^z &\equiv \frac{1}{2} \sum_\sigma \left( c_{2j-1,\sigma}^\dagger c_{2j-1,\sigma} + c_{2j,\sigma}^\dagger c_{2j,\sigma}-1\right).
\end{align}
\end{subequations}
Here, the ladder operators $\Sigma^{\pm}_j = \Sigma^x_j \pm \ri \Sigma^y_j$ and $\eta^{\pm}_j = \eta^x_j \pm \ri \eta^y_j$.
The $SU(2)\times SU(2)$ is quotient by $\mathbb{Z}_2$ because the total $\eta$ and $\Sigma$ spins must add to an integer spin representation to give the required $SO(4)$ symmetry. Finally, the $SO(4)$ symmetry is supplemented by the time-reversal symmetry and the single branch parity symmetry to complete the full $O(4) \rtimes \mathbb{Z}_2^\mathbb{T}$ symmetry.

When acting on the spin and $\eta$-operators, the single-branch fermion parity transformation exchanges two sets of $SU(2)$ generators, i.e. $\Sigma_{j}^{\pm,z} \mapsto \eta_{j}^{\pm,z}$ and $\eta_{j}^{\pm,z} \mapsto \Sigma_{j}^{\pm,z}$,
and the time reversal flips the sign of spin operators, while keeping charge $\eta$-operators invariant, i.e. $\mathbb{T}\Sigma_j^{\pm,z}\mathbb{T}^{-1} = -\Sigma_j^{\pm,z}$ and $\mathbb{T}\eta_j^{\pm,z}\mathbb{T}^{-1} = \eta_j^{\pm,z}$.

As explained above, the ground state of the effective Hamiltonian must break the single-branch symmetries and thus exhibits a reduced effective  symmetry. 
For purely unitary dynamics, the full symmetry $O(4) \rtimes \mathbb{Z}_2^\mathbb{T}$ is thus reduced to $SO(4) \rtimes \mathbb{Z}_2^\mathbb{T}$.
With measurements, the reduced symmetry is $\mathcal{G}^{(2)}_{f,\text{eff}} \equiv (U(1) \times U(1)) \rtimes (\mathbb{Z}_2\times \mathbb{Z}_2^{\mathbb{T}})$.
Here the first $\mathbb{Z}_2$ is generated by the particle-hole transformation $\mathbb{P}$ for both conserved fermion species, which maps $\mathbb{P} \eta_{j}^{\pm,z}\mathbb{P}^{-1} \mapsto -\eta_{j}^{\mp,z}$ and $\mathbb{P} \Sigma_{j}^{\pm,z}\mathbb{P}^{-1} \mapsto -\Sigma_{j}^{\mp,z}$.

\subsection{Effective 1D quantum Hamiltonian}\label{sec:fermi_mapping}

We now carry out the program developed in Section~\ref{sec:framework} to map the dynamics of the averaged replicated density matrix  $\kket{\tilde{\rho}^{(2)}}$ to effective imaginary time evolution with a one-dimensional quantum Hamiltonian.
Almost everything proceeds exactly as detailed in Section~\ref{sec:framework} for qubit circuits. 
Here we elaborate on distinct features in our fermionic model related to the enlarged dynamical symmetry.

As a first step in the program, we consider averaging over probabilistic measurements and bond unitary gates. 
The site unitary gates, designed in this model to project onto a reduced Hilbert space, will be considered later. It is straightforward to integrate over the Gaussian distribution of the bond coupling $\theta_{b,j}$ to obtain the averaged bond unitary acting on the doubled density matrix
\begin{align}
    \mathcal{U}_{b,j} =
    \exp\left[-\frac{J_b\delta t}{2} \bigg(\sum_{\alpha,\sigma} \sigma\,\ri \gamma_{2j,\alpha\sigma}\gamma_{2j+1,\alpha\sigma} \bigg)^2\right],
    \label{eq:unitary_2}
\end{align}
where we have denoted the variance $\overline{\theta_{b,j}^2}\equiv J_b \delta t$.
The averaged measurement in the duplicated Hilbert space takes the form  
\begin{align}
    \mathcal{M}_{a,j} = (1-\Gamma_a\delta t) 
    + \Gamma_a\delta t \sum_{m = \pm} P_{a,j,m}^{\otimes 4}, \label{eq:meas_2}
\end{align}
where $a\in \{s,b\}$ indicates the site and bond measurements implemented by the projections $P_{s,j,\pm} = (1\pm \hat{\pi}_{s,j})/2$ and $P_{b,j,\pm} = (1\pm \hat{\pi}_{b,j})/2$ on measurement outcomes $m=\pm$. $\Gamma_a$ is the measurement rate so that the measurement probability in a time step is $p_a = \Gamma_a\delta t$. 
Crucially, the evolution operators 
$\mathcal{U}_{b,j}$ and  $\mathcal{M}_{a,j}$ are both Hermitian. Thus in the limit $\delta t\to 0$ they describe imaginary time evolution over an infinitesimal time step, generated by an effective quantum Hamiltonian  $H_\textrm{eff} = H_\mathcal{U} + H_{\mathcal{M}}$.

We could have also included site unitary gates 
\begin{align}
    \mathcal{U}_{s,j} = 
    \overline{\exp\left[\sum_{\alpha,\sigma} \theta_{s,j}\sigma\ri \gamma_{2j-1,\alpha\sigma}\gamma_{2j,\alpha\sigma} \right]}
\end{align}
on the same footing in this program, averaging over $\theta_{s,j}$ drawn from a Gaussian distribution of coupling constants.
This would result in a complete effective Hamiltonian written in terms of the Majorana operators on the four chains.
However, to obtain a simpler effective model without changing the essential structure and symmetries of the problem, we choose the site coupling constants $\theta_{s,j}$ to be uniformly distributed on $[0,2\pi)$.
In this case, exactly as in Section~\ref{sec:framework},  averaging over the site unitaries implements a projection on a six-dimensional local Hilbert space on physical sites (i.e. Majorana sites $2j-1,2j$). However, in the fermionic system, it is advantageous to use a different basis of the reduced Hilbert space, which makes use of the fermionic symmetries. 

We recall the above observation that unitary gates, including the single-site gates, commute with generators of the $SU(2)$ spin ($\Sigma$) and $SU(2)$ charge ($\eta$) symmetry. Therefore, the basis of the projected Hilbert space can be organized into multiplets of the $\Sigma$ and $\eta$ spins on each physical site $j$. Three of the basis states $\kket{m}_\eta$ with $m=\pm1,0$ form the $\eta$ triplet which transform as the $\Sigma$ singlet.
The other three states  $\kket{m}_\Sigma$ with $m=\pm1,0$ form the $\Sigma$ triplet and transform as the $\eta$ singlet.
These basis states can be written explicitly in terms of the second quantized conserved fermion operators defined in Eq.~\eqref{eq:c_fermion}:
\begin{subequations}
\begin{align}
    & \kket{-}_\eta = \kket{\text{vac}}, \\
    & \kket{\,0\,}_\eta = \frac{1}{\sqrt{2}}\left(c_{2j-1,\uparrow}^\dagger c_{2j-1,\downarrow}^\dagger - c_{2j,\uparrow}^\dagger c_{2j,\downarrow}^\dagger \right) \kket{\text{vac}}, \\
    & \kket{+}_\eta = c_{2j-1,\uparrow}^\dagger c_{2j,\uparrow}^\dagger c_{2j-1,\downarrow}^\dagger c_{2j,\downarrow}^\dagger \kket{\text{vac}}, \\
    &\kket{-}_\Sigma = c_{2j-1,\downarrow}^\dagger c_{2j,\downarrow}^\dagger \kket{\text{vac}},\\
    &\kket{\,0\,}_\Sigma = \frac{1}{\sqrt{2}}\left(c_{2j-1,\downarrow}^\dagger c_{2j,\uparrow}^\dagger + c_{2j-1,\uparrow}^\dagger c_{2j,\downarrow}^\dagger \right) \kket{\text{vac}}, \\
    &\kket{+}_\Sigma = c_{2j-1,\uparrow}^\dagger c_{2j,\uparrow}^\dagger \kket{\text{vac}},
\end{align}
\end{subequations}
where $\kket{\text{vac}}$ is defined by the annihilation $c_{j,\sigma} \kket{\text{vac}}=0$.
Intuitively we can view the system as hosting on each physical site a single spin-1 particle, which can be either an $\eta$ or a $\Sigma$ spin.
The occupation of $\eta$ or $\Sigma$ triplet at site $j$ corresponds to distinct eigenvalues of the local conserved quantities: $\mathcal{R}_{2j-1} = \mathcal{R}_{2j} = \pm 1$ for $\eta$ and $\Sigma$, respectively.

The effective Hamiltonian that generates the imaginary time evolution directly follows from Eqs.~\eqref{eq:unitary_2} and~\eqref{eq:meas_2} projected on the reduced six-dimensional local Hilbert space. %
After straightforward algebra, we find that $H_\text{eff}$ simplifies to 
\begin{align}
    \label{eq:Heff}
    H_\text{eff} = H_{\text{eff},\eta} + H_{\text{eff},\Sigma},
\end{align}
where each term in $H_{\text{eff},\eta}$ ($H_{\text{eff},\Sigma}$) acts nontrivially only for $\eta$-spins ($\Sigma$-spins) and otherwise annihilates the wave function. We can write this explicitly as
\begin{align}
    H_{\text{eff},\kappa} &= \sum_j \Delta \left( S_{\kappa,j}^z\right)^2\label{eq:Heff_k}\\
    &- \sum_j J_\perp \left( S_{\kappa,j}^x S_{\kappa,j+1}^x + S_{\kappa,j}^y S_{\kappa,j+1}^y \right) + J_z S_{\kappa,j}^z S_{\kappa,j+1}^z,\nonumber
\end{align}
where the spin-1 operators $\vec{S}_{\kappa,j} = (S_{\kappa,j}^x, S_{\kappa,j}^y,S_{\kappa,j}^z)$ annihilate the singlet state of the species $\kappa=\eta,\Sigma$.
The coupling constants in this Hamiltonian are determined from the parameters of the unitary gates and the measurements as follows:  $\Delta = \Gamma_s$, $J_\perp = 2J_b + \Gamma_b/4$, and $J_z =2J_b - \Gamma_b/4$. 

Let us make a few remarks on the effective Hamiltonian.
First, $H_\text{eff}$ conserves the total $z$ components as well as the parities associated with $S_{\kappa,j}^z \mapsto - S_{\kappa,j}^z$ transformation of $\eta$ and $\Sigma$ spins, separately.
Besides, $H_\text{eff}$ is the same in $\eta$ and $\Sigma$ sector and therefore invariant under the exchange of $\eta$ and $\Sigma$ spins.
Thus, $H_\text{eff}$ respects the symmetry $[(U(1) \rtimes \mathbb{Z}_2) \times (U(1) \rtimes \mathbb{Z}_2)]\rtimes \mathbb{Z}_2^{\mathbb{T}}$ as expected.
Second, the occupation of $\eta$ and $\Sigma$ triplet states is locally conserved at every site.
Note that the number of local integrals of motion is reduced from $2L$ to $L$ upon averaging over the unitary gates and projection unto the six-dimensional local Hilbert space $\mathcal{P}\mathcal{R}_{2j-1} \mathcal{P}= \mathcal{P}\mathcal{R}_{2j}\mathcal{P}$. Accordingly, the dynamics is partitioned into $2^L$ sectors corresponding to configurations of the $\eta$ and $\Sigma$ occupations. 

In our case, the physical circuit is initialized in the vacuum state of the fermions $f_j$.
When extended to the duplicated Hilbert space, this state can be written using the conserved fermions as
\begin{align}
    \kket{\rho_0}_j &= \frac{1}{2}\left(c_{2j-1,\uparrow}^\dagger + c_{2j,\uparrow}^\dagger\right)\left(c_{2j-1,\downarrow}^\dagger - c_{2j,\downarrow}^\dagger\right) \kket{\text{vac}}_{j}\nonumber\\
    &=\frac{1}{\sqrt{2}}\left(\kket{0}_\eta-\kket{0}_\Sigma\right).
\end{align}
We see that the initial state is already in the reduced Hilbert space and equally populates both $\eta$ and $\Sigma$ triplet states at every site.

In the long imaginary time evolution, the wave function is dominated by the global ground states of the Hamiltonian \eqref{eq:Heff} among all $2^L$ sectors. The ground states are found in the two uniform sectors, i.e. the $\eta$ sector with all sites occupied by $\eta$ spins ($\mathcal{R}_\ell = 1$) and the $\Sigma$ sector with all sites occupied by $\Sigma$ spins ($\mathcal{R}_\ell = -1$).
Thus, the $\lim_{t\rightarrow \infty} \kket{\rho{(t)}} \approx \kket{\psi_{gs}}_\eta + \kket{\psi_{gs}}_\Sigma$

The steady state phases of the circuit can be classified based on the reduced  effective symmetry within either one of the two relevant sectors. The symmetry is reduced because in each sector the two $U(1)$ charges are not independent. In the $\eta$ sector every site is a sigma singlet with $N_{\ua,j} = N_{\da,j}$, so only $N_\ua+N_\da$ is a non trivial charge. In the $\Sigma$ sector two fermions occupy each site, so $N_{\ua,j} + N_{\da,j} = 2$, so only $N_\ua-N_\da$ is a non trivial charge.
Hence, either sector only forms a faithful representation of $U(1) \rtimes (\mathbb{Z}_2 \times \mathbb{Z}_2^\mathbb{T})$, which is a subgroup of $\mathcal{G}_{f,\text{eff}}^{(2)}$.
One can verify $U(1) \rtimes (\mathbb{Z}_2 \times \mathbb{Z}_2^\mathbb{T})$ is exactly the symmetry of $H_{\text{eff}}$ in either sector. 
For vanishing measurement rates, owing to local conserved quantities, each sector only forms a faithful representation of either $SU(2)$ spin or charge symmetry, reducing the effective symmetry down to $O(3) = SO(3) \rtimes \mathbb{Z}_2^\mathbb{T}$.\footnote{The symmetry is $SO(3)$ rather than $SU(2)$ because the ground state occupies an integer representation.}
Indeed in this limit the effective Hamiltonian in each of the two sectors is the Heisenberg ferromagnet, which manifests the $O(3)$ symmetry.
Finally, we establish the correspondence between the ground state properties of $H_\textrm{eff}$ and the steady states of quantum circuits by mapping  various physical observables measured on the circuit to the matrix elements of different boundary operators in the effective spin model.
As explained in Section~\ref{sec:framework}, properties related to the second moments of the density matrix (conditional on the measurement device) generally take the form 
\begin{equation}
O^{(2)} \leftrightarrow{\bbra{\mathcal{I}}\mathcal{O}\kket{\rho^{(2)}}\over \bbrakket{\mathcal{I}}{\rho^{(2)}}},
\label{eq:O2}
\end{equation}
where $\bbra{\mathcal{I}}$ is the reference state that implements the (doubled) trace operation in the duplicated Hilbert space, i.e. $\bbrakket{\mathcal I}{\rho^{(2)}}=\tr \rho^{(2)}$. 

\begin{figure}[t]
    \centering
    \includegraphics[width=0.48\textwidth]{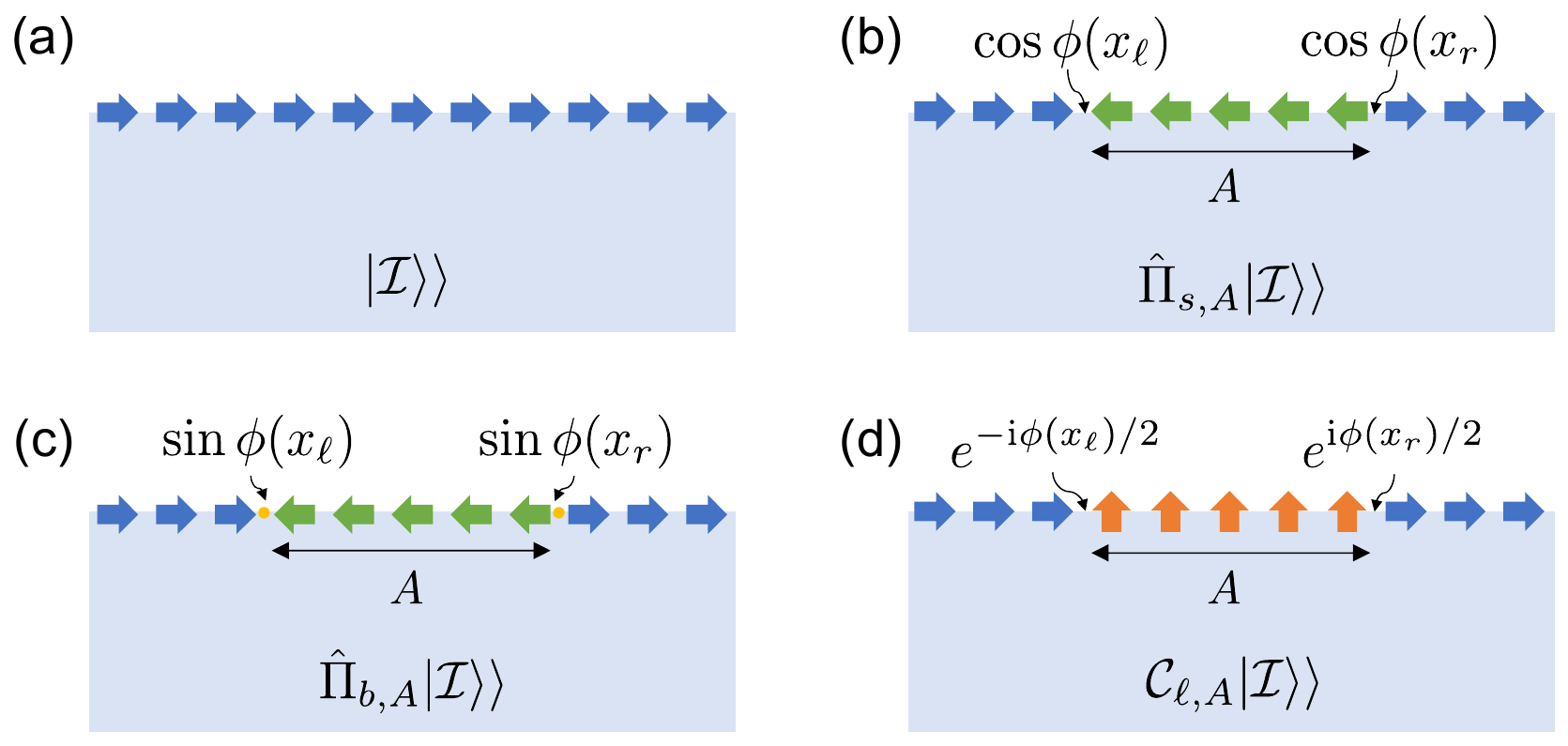}
    \caption{Boundary conditions imposed on the $U(1)$ phase by the boundary operators corresponding to different probes. (a) The reference state $\kket{\mathcal{I}}$ imposes a symmetry breaking state with $U(1)$ phase locked to zero (blue arrows). (b) The site parity variance $\hat{\Pi}_{s,A}$ rotates the phase by $\pi$-phase in region $A$, which creates a vortex  anti-vortex pair at the edges of the region. (c) The bond parity variance $\hat{\Pi}_{b, A}$ similarly affects a $\pi$ phase rotation, creating a vortex anti-vortex pair at the edges of $A$. However, in this case, the vortex creation operators are bound to a $\mathbb{Z}_2$ charge (of the symmetry $\phi\to -\phi$). (d) The swap operator $\mathcal{C}_{\ell,A}$ rotates the phase in $A$ by $\pi/2$ (orange arrows), creating a pair of half-vortices at edges of the region.
    }
    \label{fig:boundary_states}
\end{figure}

Within the effective description, the state $\bbra{\mathcal{I}}\mathcal{O}$ can be viewed as the boundary condition to the imaginary time evolution at the latest time. The action of $\mathcal{O}$ to the left transforms the boundary conditions set by the reference state $\bbra{\mathcal{I}}$ in the region on which $\mathcal{O}$ is supported. 
The kind of boundary condition imposed is dictated by the  symmetries of $\bbra{\mathcal{I}}$. %
We show in appendix~\ref{app:boundary} that the reference state breaks the $U(1)$ symmetry in both the $\eta$ and $\Sigma$ sector, exhibiting the long range order $\bbra{\mathcal{I}}\eta^y_i\eta^y_j\kket{\mathcal{I}}\to c$ and $\bbra{\mathcal{I}}\Sigma^y_i\Sigma^y_j\kket{\mathcal{I}}\to c$  as $|i-j|\to \infty$. Thus, $\bbra{\mathcal{I}}$ effectively
imposes a symmetry breaking boundary condition on the imaginary time evolution by $H_{\text{eff}}$ [see Fig.~\ref{fig:boundary_states}(a)].

The various operators we use to probe the circuit transform the broken symmetry boundary conditions in different ways. Consider first the site parity variance $\Pi^{(2)}_{s,A}$ on subsystem $A$, which is mapped to the boundary matrix element \eqref{eq:O2} of the operator 
\begin{align}
\hat{\Pi}_{s,A}&=\prod_{j\in A} (-\ri\gamma_{2j-1,1,\ua}\gamma_{2j,1,\ua})(-\ri\gamma_{2j-1,2,\ua}\gamma_{2j,2,\ua})\nonumber \\
&=\prod_{j\in A}e^{-\ri \pi (\eta^z_j+ \Sigma^z_j)}.
\end{align}
This operator rotates the $U(1)$ order parameters for both $\eta$ and $\Sigma$ spins by $\pi$ everywhere in subsystem $A$.
The ensuing boundary condition is illustrated schematically in Fig.~\ref{fig:boundary_states}(b).
We'll see that this is equivalent to inserting a pair of topological defects (vortices) at the edges of the region $A$.

The bond parity variance operator $\hat{\Pi}_{b,A}$ can be obtained by simply translating $\hat{\Pi}_{s,A}$ by half of a physical site (one Majorana site). 
Thus, the operator leads to the same $\pi$ rotation of the $U(1)$ orders of the boundary state within region $A$, but it is bound to a $\mathbb{Z}_2$ charge (of the symmetry $\phi \to -\phi$)
on each edge of the region [see Fig.~\ref{fig:boundary_states}(c)]. 
As we will see in the next section, this difference has nontrivial implications on the behavior of the bond parity variance in the different phases.

The entanglement entropy of a subsystem is associated with the matrix element of a swap operator $\mathcal{C}_{\ell,A}$ as discussed in Section~\ref{sec:framework}.
Because $\mathcal{C}_{\ell,A}$ is nonlocal when written in terms of fermionic operators, we replace it with a local operator $\mathcal{\widetilde C}_{\ell,A}$ which has exactly the same action as $\mathcal{C}_{\ell,A}$  on the reference state, namely $\mathcal{\widetilde C}_{\ell,A}\kket{\mathcal{I}} = \mathcal{C}_{\ell,A}\kket{\mathcal{I}}$ (see  Appendix~\ref{app:swap}).
The operator $\mathcal{\widetilde C}_{\ell,A}$ can be simply expressed in terms of the $f$-fermions defined on the forward branches and in terms of the $\eta$ and $\Sigma$ spins,
\begin{align}
    \widetilde{\mathcal{C}}_{\ell,A} =& \prod_{j \in A} f^\dagger_{j,1,\uparrow} f_{j, 2, \uparrow} + f_{j,1,\uparrow} f_{j,2,\uparrow}^\dagger + \frac{1}{2} \left(1 + \hat{\Pi}_{j,1,\ua} \hat{\Pi}_{j,2,\ua}\right) \nonumber \\
    =& \prod_{j\in A} e^{-\ri\frac{\pi}{2}(\Sigma_j^z + \eta_j^z)},
\end{align}
where $\hat{\Pi}_{j,\alpha,\ua} = 1 - 2f_{j,\alpha,\ua}^\dagger f_{j,\alpha,\ua}$ is the fermion parity in the forward branch of copy $\alpha$ at site $j$. 
Since $\mathcal{\widetilde C}_{\ell,A}$ only operates on the forward branches, it naturally breaks the time-reversal symmetry of spin-$1/2$ fermions.
Specifically, the unitary operator $\widetilde{\mathcal{C}}_{\ell,A}$ rotates both $U(1)$ order parameters in the region $A$ by an angle $\pi/2$.
We'll see below that this is equivalent to inserting a pair of half vortices at the edges of region $A$ leading to the boundary state illustrated in Fig.~\ref{fig:boundary_states}(d).

\subsection{Phases of the effective model}\label{sec:fermi_phases}

The effective Hamiltonian $H_\text{eff}$ consists of two identical and decoupled spin-1 Hamiltonians $H_{\text{eff},\kappa}$, each acts on the $\kappa=\eta,\Sigma$ sectors.
The ground states and phase transitions of $H_{\text{eff},\kappa}$ have been extensively investigated numerically~\cite{chen2003ground} and admit a simple long-wavelength description~\cite{schulz1986phase}. 

In the physically relevant parameter regimes, $\Gamma_b, \Gamma_s, J_b \geq 0$, the ground state phase diagram of $H_{\text{eff},\kappa}$ contains three phases, as illustrated schematically in Fig.~\ref{fig:fermi_introductory}(a).
For vanishing measurement rates, $H_{\text{eff},\kappa}$ is a $O(3)$ ferromagnet, which admits a fluctuation-free broken symmetry state.
This is a singular point in the phase diagram corresponding to the volume-law state established for purely unitary dynamics of free fermions. 
For any non-zero measurement rate, the $O(3)$ symmetry is broken down to $U(1)\rtimes (\mathbb{Z}_2 \times \mathbb{Z}_2^\mathbb{T})$ leading to a critical phase with algebraic long-range order. 
Increasing the measurement rates $\Gamma_s$ and $\Gamma_b$ beyond a critical threshold leads to two possible gapped phases, depending on the ratio between the two measurement processes.
The site measurements $\Gamma_s$ contribute to the $\Delta$-term (also known as single-ion anisotropy), leading to a trivial gapped phase when they are dominant.
The bond measurements $\Gamma_b$, on the other hand, leads to the Haldane gapped phase, which is an SPT phase of the spin-1 model.

The above phases and phase transitions can be captured within a long wavelength theory.
First, note that the imaginary time evolution with the effective Hamiltonian has a coarse-grained description in terms of the 2d XY model $S_{XY}={K\over 2}\int dx d\tau (\nabla\theta)^2$ with $\theta = \theta + 2\pi$. We can gain even more insight, however, by framing the long-wavelength theory of the XY model in the form of the one-dimensional sine-Gordon Hamiltonian
\be
H={1\over 2}\int dx \left[ K (\nabla{\hat\theta})^2 + {1\over K}(\nabla\hat{\phi})^2\right] - g\int dx \cos(2\hat{\phi}).
\label{eq:SG}
\ee
Here, $\hat{\theta}$ is related to the $U(1)$ phase in the $XY$ model (though it is not compact). $\nabla\hat{\phi}/\pi$ is the long wavelength fluctuation of the conserved charge conjugate to $\hat{\theta}$, so that $[\hat{\theta}(x),\hat{\phi}(x')]=\ri\pi\Theta(x-x')$. The coupling $g$ is related to the vortex fugacity implicit in the XY model. This model is invariant to a shift of $\hat{\phi}\to\hat{\phi}+\pi$, which corresponds to a translation by a lattice constant, and to $\hat{\phi}\to -\hat{\phi}$ related to the $\mathbb{Z}_2$ ``particle-hole" symmetry. 

In the critical phase of the spin-1 Hamiltonian, the renormalized Luttinger parameter $K>2$ and the coupling $g$ is irrelevant in the long wavelength limit. 
As the microscopic parameters are varied, the system eventually undergoes a Kosterlitz-Thouless transition into a gapped phase at a critical universal value $K_c=2$. The long wavelength theory distinguishes the two gapped phases through the sign of the coupling $g$. If $g>0$, the dual field $\hat{\phi}$ is locked to $\phi=0$ giving rise to a trivial gapped phase.
If $g<0$, then $\hat{\phi}$ is  locked to $\phi=\pi/2$, which corresponds to the SPT (Haldane) phase. Note that the two phases are related through a shift by half a lattice constant (equivalently, one Majorana site).

We now turn to discuss how the different phases of the effective model are probed by the boundary operators. As noted in the previous section, the action of these operators on the boundary state imposes the boundary conditions depicted in Fig.~\ref{fig:boundary_states}. Hence, they correspond to boundary condition changing operators in the long wavelength theory.
The logarithm of the boundary overlap~\eqref{eq:O2} is directly related to the excess free energy of the effective 2d XY model with the modified boundary conditions relative to the free energy with the uniform boundary conditions set by the reference state. 

Consider first the parity
string operators $\hat{\Pi}_{s,A}$ and $\hat{\Pi}_{b,A}$, which enter the calculation of the parity variances.
These operators affect a $\pi$-phase flip on region $A$, which creates a pair of vortices on the boundary of the 2d XY model.
In the long wave length limit, the vortex insertion is affected by the operators $e^{\pm\ri\hat{\phi}}$. 
To compute the site parity string $\hat{\Pi}_{s,A}$, we must choose the symmetric combination $\cos\hat{\phi}(x) $ at each end because the site parity string preserves the particle-hole symmetry on every site. Thus, in the long wavelength limit we can write $\hat{\Pi}_{s,A}\sim \cos\phi(x_\ell)\cos\phi(x_r)$, where $x_\ell, x_r$ are the two ends of the region $A$. The bond parity string is obtained from the site parity string through translation by half a physical site (one Majorana site), implemented by a shift $\hat{\phi}\to\hat{\phi}+\pi/2$ in the long wavelength theory. Thus, $\hat{\Pi}_{b,A}\sim \sin\phi(x_\ell)\sin\phi(x_r)$. This is consistent with the fact that, on each end of the region $A$, the bond parity string carries a residual parity operator of half a physical site, which is anti-symmetric with respect to the particle-hole symmetry.
The swap string, which enters the calculation of the entanglement entropy, affects a $\pi/2$ rotation of the $U(1)$ phase, which creates a pair of half-vortices on the boundary of the 2d XY model. In the long wavelength theory, this is achieved by the operators $\mathcal{\widetilde C}_{\ell,A} \sim e^{-\ri \phi(x_\ell)/2} e^{\ri\phi(x_r)/2}$.

Having established the long wavelength form of the operators associated with the different boundary observables, we can determine the behavior of the boundary matrix element in the different phases. 
In the critical phase, the distinction between the site and bond parity strings is not important. Both decay as a power law, dictated by the scaling dimensions of boundary vortex insertions $\Pi_{a,A}^{(2)} \sim |x_\ell-x_r|^{-K/4}$~\cite{giamarchi2003quantum}. 
The behavior of the entanglement entropy on the other hand is dictated by a pair of half-vortex insertions at the boundary, each having half the scaling dimension of a vortex insertion. Thus, we expect $e^{-S_A^{(2)}} \sim \abs{x_\ell - x_r}^{-K/16}$.
So, while the decay exponents vary continuously in the the long wavelength theory, the theory predicts a universal ratio of $4$ between the decay exponent of the parity variance and that of $\exp(-S^{(2)}_A)$ in the critical phase. Furthermore, the KT transition is expected to occur at a universal value of the stiffness $K_c=2$, which implies a critical exponent $\alpha^c_\Pi=1/2$ for the decay of the parity variance and $\alpha^c_S=1/8$ for $\exp(-S_A^{(2)})$.

In the low energy fixed points corresponding to gapped ground states, $\hat{\phi}(x)$ is nonfluctuating; it is locked to $\phi=0$ in the trivial phase and to $\phi=\pi/2$ in the SPT phase. Hence, we can immediately deduce the action of parity and swap strings on the ground states. 
The swap does not vanish in either phase, thus giving $\exp(-S^{(2)}_A)\to \text{const}$ at long distances in both the trivial and SPT state. 
The two phases are distinguished by the action of the parity strings. The site parity string, which involves $\cos\hat{\phi}$, acts as a constant on the trivial ground state and vanishes on the SPT ground state at the fixed point. 
Conversely, the bond parity string acts as a constant on the SPT ground state and vanishes on the trivial ground state. 
Therefore, in the trivial phase, we expect $\Pi_{s,A}^{(2)} \to \text{const}$ and $\Pi_{b,A}^{(2)} \to \exp(-|x_\ell-x_r|/\xi)$. In the SPT phase, this is inverted, $\Pi_{b,A}^{(2)} \to \text{const}$, whereas $\Pi_{s,A}^{(2)} \to \exp(-|x_\ell-x_r|/\xi)$.
In the following section, we examine the predictions of the long wave-length theory for the various phases by numerically simulating the quantum circuit.

\subsection{Numerical simulation in Gaussian fermionic circuits}\label{sec:fermi_numerics}

\begin{figure*}[t!]
    \includegraphics[width=0.98\textwidth]{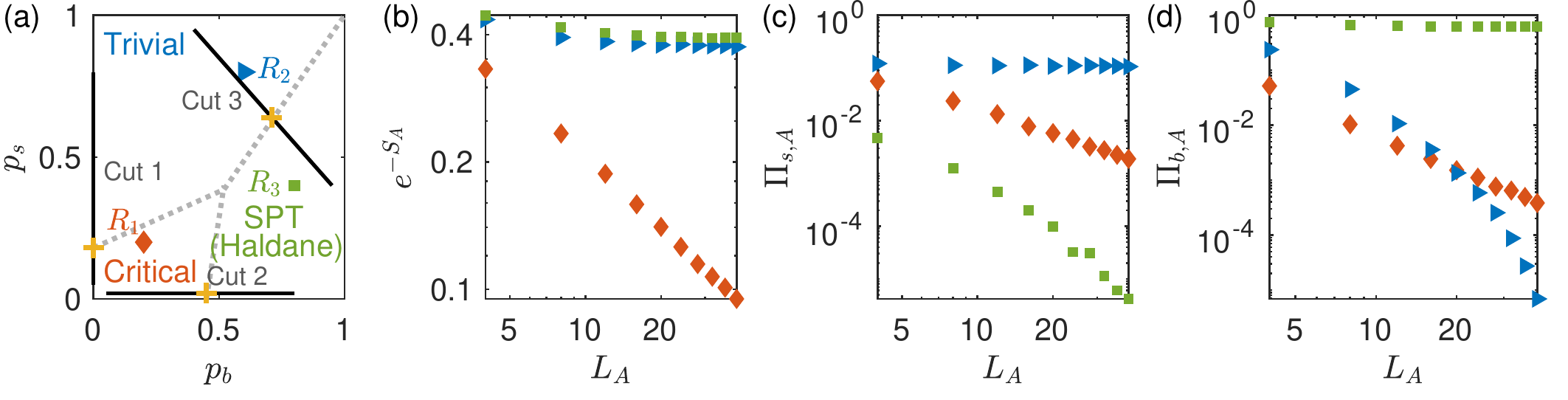}
    \caption{Entanglement entropy and subsystem parity variances in three phases of Gaussian fermionic circuits. (a) Schematic phase diagram in the space of bond measurement versus site measurement probabilities ($p_b$ v.s. $p_s$).
    Yellow markers represent numerically extracted critical points along marked cuts. (b-d) Results for representative points of the three phases $R_1:(0.2,0.2)$, $R_2:(0.6,0.8)$, $R_3:(0.8,0.4)$ shown on the phase diagram. (b) Entanglement entropy $e^{-S_{A}}$ shows power-law decay (logarithmic entanglement scaling) with subsystem size $L_A$ in the critical phase ($R_1$, red diamonds) and a constant (area law) in the trivial and SPT phases. (c) Subsystem site parity variance $\Pi_{s,A}$ shows a power-law decay with $L_A$ in the critical phase (red diamonds), faster than power-law in the SPT phase (green squares) and is nondecaying in the trivial phase (blue triangles). (d) Subsystem bond-parity variance $\Pi_{b, A}$ shows a power-law decay in the critical phase (red diamonds), exponential decay in the trivial phase and is nondecaying in the SPT phase. 
    The numerical results are obtained with system size $L = 160$ and averaged over $400$ random circuit realizations and measurement outcomes.}
    \label{fig:fermi_introductory}
\end{figure*}

The fermions circuit defined in Section~\ref{sec:fermi_model} can be simulated efficiently because it preserves the Gaussianity of the wave function in each quantum trajectory~\cite{terhal2002classical}. This allows us to test the predictions 
of the effective model by direct numerical
calculation of the circuit dynamics. 
Technically, it is enough to propagate the $O(N^2)$ two point functions $G_{ij} \equiv \langle \Psi | \ri \gamma_i \gamma_j |\Psi \rangle - \ri \delta_{ij}$, which fully determine the Gaussian wave function. 
Indeed, all quantities of interest, including the subsystem entanglement entropy $S_A$ and parity  variances $\Pi_{s,A}$ and $\Pi_{b, A}$, can be directly read out from $G_{ij}$. We perform the calculation with system sizes of up to $L=160$ sites using periodic boundary conditions to facilitate better finite size scaling. We operate the circuit to depth $3L$ to achieve a steady state.

We explore a two-dimensional phase space defined by the site and bond measurement probabilities as illustrated in Fig.~\ref{fig:fermi_introductory}(a). 
The behavior of the three observables we extract indicates the establishment of three phases. 
Panels (b-d) of the figure demonstrate this for three representative points in the phase diagram. At the point $(p_b,p_s)=(0.2,0.2)$ (red diamond), we observe the expected behavior of the critical phase with entanglement entropy scaling as $\log L_A$ and power-law decay of the parity variances. The point $(p_b,p_s)=(0.6, 0.8)$ (blue triangle) exhibits the behavior we expect in the trivial phase with entanglement entropy and the site parity variance saturating to a constant, while the bond parity variance decaying exponentially. Finally, the point $(p_b, p_s)=(0.8, 0.4)$ (green square) shows a constant entanglement entropy and bond parity variance as well as an exponential decaying site parity variance, as expected in the SPT phase. 

\begin{figure*}[t!]
    \includegraphics[width=0.98\textwidth]{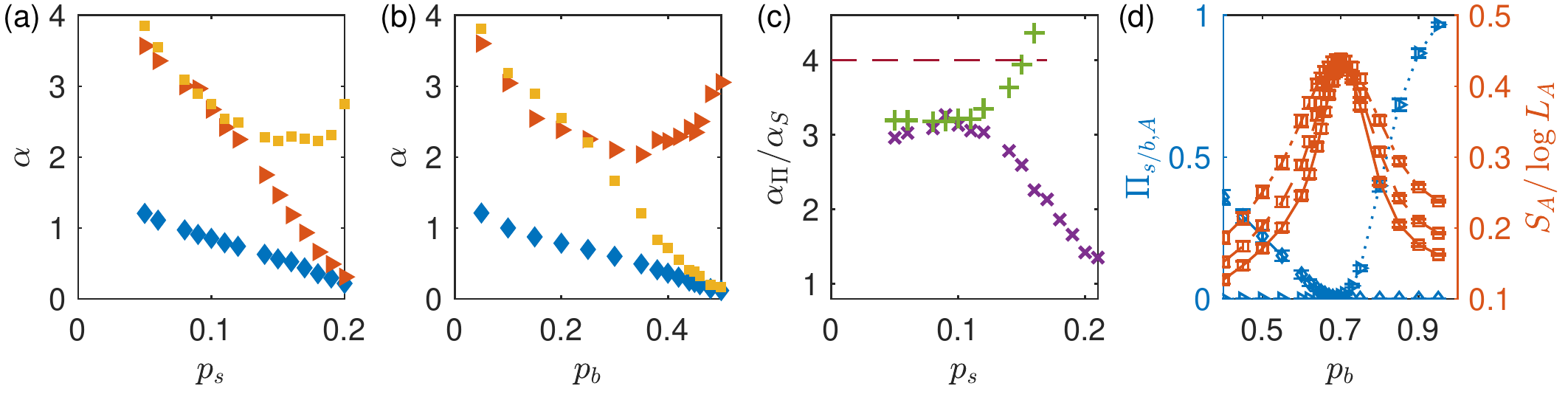}
    \caption{Tuning across the critical points on the three cuts shown in Fig. \ref{fig:fermi_introductory}(a): cut 1 ($p_b = 0$), 2 ($p_s = 0.02$), 3 ($p_b+p_s = 1.35$). (a, b) Fitted power-law decay exponent of $e^{-S_A}$ ($\alpha_S$, blue diamonds), site parity variance ($\alpha_{\Pi,s}$, red triangles), and bond parity variance ($\alpha_{\Pi,b}$, yellow squares) along cut 1 [panel (a)] and cut 2 [panel (b)]. The horizontal axis represents the tuning parameter along each cut. The exponents are extracted from simulations with the system size $L = 160$ averaged over $400$ random circuit realizations and measurement outcomes.
    (c) Ratio between exponents $\alpha_{\Pi,s} / \alpha_S$ and $\alpha_{\Pi,b}/\alpha_S$ along cut 1. Results for $\alpha_{\Pi,s} / \alpha_S$ (purple crosses) and $\alpha_{\Pi,b}/\alpha_S$ (green ``+''). The red dashed line represents the prediction of the two-replica model for the critical phase: $\alpha_{\Pi} / \alpha_S = 4$.
    (d) Parity variances (blue, left vertical axis) and normalized entanglement entropy (red, right vertical axis) along cut 3 from $(p_b, p_s) = (0.4, 0.95)$ to $(0.95, 0.4)$. $\Pi_{s,A}$ (blue dashed line) and $\Pi_{b,A}$ (blue dotted line) are for $L_A = L/2$. The plotted entanglement entropy is taken for subsystem size $L_A = L/2$ and normalized by $\log L_A$. The three red lines correspond to system sizes $L = 40, 80, 160$; the peak gets sharper with increasing system size and indicates logarithmic entanglement scaling only at the critical point.
    The numerical results are averaged over $400$ random circuit realizations and measurement results.}
    \label{fig:phase_transition}
\end{figure*}

Having confirmed these basic behaviors, we turn to investigate the critical phase and the transitions from this phase into the two gapped states. To this end, we vary the measurement probabilities $p_s$ and $p_b$ along the two cuts depicted as cut 1 and cut 2 in Fig.~\ref{fig:fermi_introductory}(a).
Along these cuts, we aim to fit the subsystem size dependence of $\Pi_{s,A}$, $\Pi_{b,A}$ and $\exp(-S_A)$ to a power law of $L_A$. More precisely, since we anticipate that the critical phase has a conformal field theory description, we can account for the finite system size and periodic boundary conditions by fitting the decay of these objects to a power law of the conformal coordinate $z_{A} = L\sin(\pi L_A/L)$.
Figure~\ref{fig:phase_transition}(b,c) shows the variation of the fitted exponents $\alpha_{\Pi,s}$, $\alpha_{\Pi,b}$, and $\alpha_{S}$ along the cuts. We see that
the exponents $\alpha_{\Pi,s}$ and $\alpha_{\Pi,b}$ coincide and decrease together on increasing the measurement rate toward the transition, as expected in the critical phase. 
The exponents then start to deviate, presumably near the critical point, in order to match the opposite behavior of the two parities in the gapped phases. For example, in the trivial phase, we expect $\alpha_{\Pi,s}\to 0$, which encodes $\Pi_{s,A}\to \text{const}$, and $\alpha_{\Pi,b}\to\infty$ in order to mimic $\Pi_{b,A}\sim \exp(-L_A/\xi)$. 

In the range of tuning parameter, where the two parity exponents coincide, we find that the ratio between these exponents and the one controlling the decay of $\exp(-S_A)$ is pinned to a constant value. 
The ratio, as seen in Fig.~\ref{fig:phase_transition}(c), is pinned to $\alpha_\Pi/\alpha_S\approx 3$, which deviates from the ratio of $4$ predicted by the low energy effective theory. This discrepancy could be related to the difference between the von Neumann entropy computed here and the second conditional R\'enyi entropy calculated in the effective model.

We now focus on the phase transition from the critical phase to the trivial phase along cut 1 in Fig.~\ref{fig:fermi_introductory}(a). To establish the universality class, we analyze the behavior of the correlation length and compare it to the hallmark exponential divergence associated with a KT transition. 

We extract the correlation length $\xi$ from the behavior of the entanglement entropy in the area-law phase near the transition since it gives much cleaner data than the parity variances. %
On distances small compared to the correlation length, i.e. $L_A\ll \xi$, we expect the critical behavior
\begin{align}
    S(z_A, p_s, L) = b(p_s)+\alpha_S(p_s) \log z_A, \label{eq:entropy_critical}
\end{align}
while for $L_A\gg \xi$ we should see a saturation to $b(p_s)+\alpha_S(p_s)\log(z_\xi)$. Here, $z_\xi = L \sin(\pi \xi/L)$ is the conformal coordinate for the correlation length $\xi$.

\begin{figure}
    \centering
    \includegraphics[width = 0.48\textwidth]{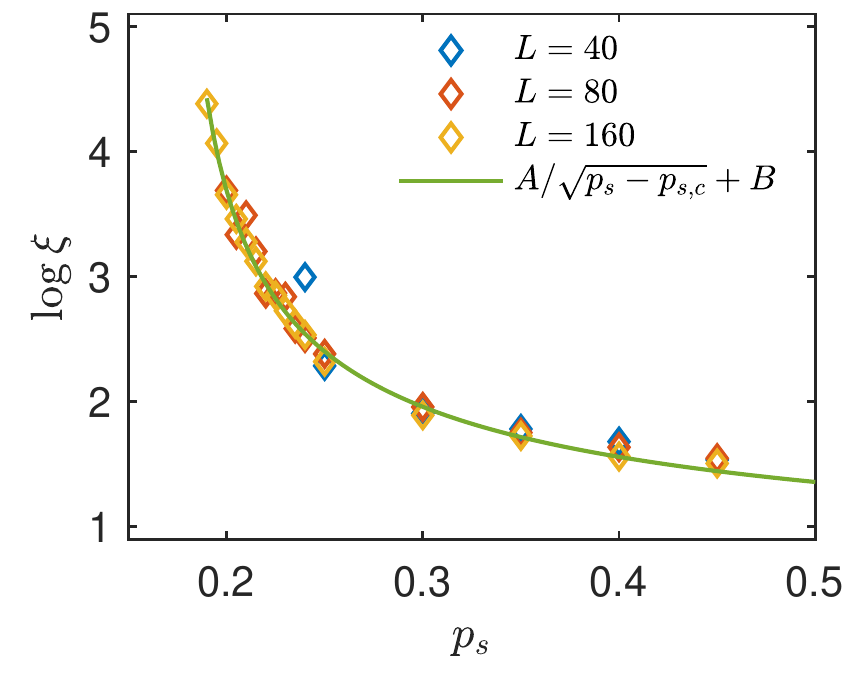}
    \caption{Correlation length $\xi$ in the trivial area-law phase along cut 1 indicated on Fig.~\ref{fig:fermi_introductory}(a). 
    The green line represents the fitted correlation length according to the scaling $\log \xi(p_s) = A/\sqrt{p_s - p_{s,c}} + B$ in the KT transition with $A = 0.58$, $p_{s,c} = 0.17$, and $B = 0.34$. 
    The results for $L = 40, 80, 160$ are denoted by blue, red, and yellow diamonds, respectively.
    The data used to extract $\xi(p_s)$ is presented in Appendix~\ref{app:n_corr_len}.
    }
    \label{fig:corr_len}
\end{figure}

To accurately obtain $\xi$, we first fit $b(p_s)$ and $\alpha_S(p_s)$ from the short distance behavior of $S(z_A,p_s,L)$.
We then use the interpolating function
\begin{align}
    S(z_A, p_s, L) = \frac{\alpha_S(p_s)}{\beta} \log\left( z_\xi^\beta \tanh\frac{z_A^\beta}{z_\xi^\beta} \right) + b(p_s) \label{eq:entropy_conj}
\end{align}
to obtain $\xi(p_s,L)$ for different values of $p_s$ and system sizes $L=40,80,160$. The parameter $\beta$ is added to improve the fit and details of the procedure are given in Appendix~\ref{app:n_corr_len}.
The fitted values of the correlation length are presented in Fig.~\ref{fig:corr_len} at all points for which $\xi\leq L/2$. We find that $\xi(p_s,L)$ fall close to the same curve for the different values of $L$. The figure shows a fit of the data for the largest length $L=160$ to the KT form of the correlation length $\xi(p_s) = \exp(A/\sqrt{p_s - p_{s,c}} + B)$.

The results are consistent with a KT transition within the range of system sizes reached and also provide an estimate of the critical point $p_{s,c} = 0.17$. We note that the finite size corrections to the critical stiffness $K_c$ in the KT transition exhibit a slow logarithmic decay with system size. Therefore, it is not surprising that the critical exponents governing the decay of the various boundary correlations are larger than the expected thermodynamic value.

Having analyzed the critical phase and the phase transition into the area-law phases, we turn to the phase transition between the two area-law states along cut 3 in Fig.~\ref{fig:fermi_introductory}(a).
The variation of the half system entanglement entropy normalized by $\log(L/2)$ is shown by the red curves in Fig.~\ref{fig:fermi_introductory}(d) for the three system sizes $L=40,80,160$. The normalized value is seen to be size independent at a single point and the curve appears to sharpen with increasing system size. This gives strong evidence for single critical point along the cut as expected from the $O(2)$ long-wavelength description \eqref{eq:SG}. Thus we do not find an indication of a generic critical phase intervening between the two area-law phases as suggested based on a relation to the completely packed loop model~\cite{sang2020entanglement}.

The details of the transition, however, also do not precisely match the predictions of the effective theory \eqref{eq:SG}. On the critical line separating the two area-law phases, the effective theory predicts a stiffness constant $K<2$, which is continuously decreasing with increasing measurement rate along this line. Accordingly the coefficient of the logarithmic half-system entropy is expected to be $\alpha_S< 1/8$, which is violated by the observed value of $\alpha_S\approx 0.43$.

The variation of the half-system site and bond parity variances across the transition behave broadly as expected, but again show a mismatch with the effective theory upon more detailed comparison. These variations are shown by the blue curves in the figure for a single system size of $L=160$. The two string ``order parameters'' are seen to switch roles between the two phases, as expected. 
However, the sharp suppression of both order parameters at the critical point is unexpected. The effective model implies 
a slow decay of the parity variance with an exponent $\alpha_\Pi<1/2$. This should produce a rather broad finite size coexistence region of the two ``order parameters''~\cite{berg2008rise}, which is not observed in the numerical results.

\subsection{Beyond the two-replica theory}
In our analysis of the Gaussian fermion circuits we focused on the two-replica theory without taking the replica limit $n\to 1$. Nonetheless, several key predictions of this model are supported by the numerical results obtained from direct simulation of fermion circuits. These includes the establishment of a critical phase characterized by logarithmic entanglement entropy with variable coefficient $\alpha_S$, the connection between $\alpha_S$ and the power-law decays of the site and bond parity variances, and the KT transition separating the critical phase from the area-law states. 
At the same time, the numerical results suggest that the phase transition between the two area-law states is not captured correctly within the two-replica model. 

To gain insights into the success and the limitations of this model, we consider an approximate scheme to evaluate $O_2$ in Eq.~\eqref{eq:O_n} in terms of a statistical mechanics model with quenched disorder.
The second moment $O_2$ is again given by
\be
O_2=\overline{\sum_m p_m(U)\left(\frac{\text{tr}\left(
{\hat{O}\,\tilde{\rho}}_m(U)\right)}{\text{tr}\left(\tilde{\rho}_m(U)\right)}\right)^2}.
\label{eq:O2U}
\ee
Here $U$ runs over the different realizations of unitary gates, and we explicitly note the dependence of the trajectory state $\tilde{\rho}_m(U)$ on $U$. 

Now, if $p_m(U) \equiv \text{tr}\left(\tilde{\rho}_m(U)\right)$ has a narrow distribution over $U$, we can approximate it by its average $p_m=\overline{p_m(U)}$ and perform the average over unitaries in Eq.~\eqref{eq:O2U} before averaging over measurement outcomes to obtain
\be
O_2\approx \sum_m p_m\frac{\bbra{\mI}\mathcal{O}^{(2)}\kket{\rho^{(2)}_m}}{\langle\langle\mI\kket{\rho^{(2)}_m}}.\label{eq:O2approx}
\ee
In this expression the measurements enter as quenched disorder. Typical trajectories $m$ consist of imaginary time evolution with an effective Hamiltonian generated by the averaged unitary dynamics, interrupted by isolated measurements located in random space-time positions. 
The isolated measurements reduce the $O(3)$ symmetry of the Heisenberg ferromagnet down to $O(2)$. We note that a similar scheme, which avoids the replica limit at the expense of introducing disorder, was considered in Ref.~\cite{zhou2019emergent} for purely unitary circuits.

Such a model with quenched disorder has a true symmetry $O(2)$.
In the case of low measurement rates, the stiffness is large enough such that the only relevant effect of the disorder is the reduction of the symmetry from $O(3)$ to $O(2)$; the space-time randomness is expected to be irrelevant in the ensuing $O(2)$ critical phase. In contrast the disorder does become relevant beyond the KT transition. Thus, taking a proper replica limit may be important for describing the transition between the two area-law states, which explains the failure of the two-replica theory to describe this critical behavior correctly.

It remains an interesting challenge to evaluate the second moment $O_2$ in the proper replica limit $n\to 1$.
One perturbative approach to the problem suggested in Ref.~\cite{Fu2012} in the context of disordered fermion models, is through an expansion of the $O(n)$ nonlinear sigma model in $\epsilon=2-n$. 
Alternatively, Sang et al.~\cite{sang2020entanglement} proposed to utilize a mapping of certain Majorana fermion circuits to the completely packed loop model with crossings~\cite{nahum2013loop}. Although the more generic model we considered does not map exactly to the loop model, it is possible that the two models are in the same universality class.
The numerical results at the available system sizes do not appear consistent with this possibility. These results indicate a direct transition between the two area-law states, while in the loop model the two area-law states are generically separated by a finite region of the critical phase.

\section{Discussion}\label{sec:discussion}

The quantum circuits we have considered in this paper generate the time evolution of many-body wave-functions or density matrices through a series of unitary gates and local measurements. Such a time evolution gives rise to a special kind of ensemble, whose members are the quantum states corresponding to all possible histories of measurement outcomes. 
Our goal has been to characterize the phases that can emerge as steady states in this new kind of ensemble.
A key observation of this work is that, unlike in thermal equilibrium or in quantum ground states, the symmetry that governs the classification of phases in quantum circuits is not simply the symmetry of the physical interactions imposed by the circuit elements. Rather, many statistical properties of the ensemble in steady states are dictated by an enlarged symmetry group, which combines the physical circuit symmetry with intrinsic dynamical symmetries of the problem. 

The enlarged symmetry emerges because distinctive properties of the circuit steady state can be seen only in the fluctuations of physical observables over the different measurement histories, while a simple average over all possible measurement outcomes looks trivial. 
In particular, the variance of physical observables, which is the minimal nontrivial fluctuation, is encoded in the time evolution of two identical copies of the density matrix. A symmetry to permute quantum states between the two copies naturally arises and combines with the physical circuit symmetries.  

This symmetry enlargement has an illuminating analogy in Anderson's pioneering work on spin glasses (see also Ref.~\cite{nahum2020measurement}). Edwards and Anderson famously pointed out that spin glass order is characterized by the random frozen magnetic moment, which can be diagnosed by 
the variance of the spin moment over the ensemble of random samples $q_{EA}=\overline{\langle s_i\rangle^2}$~\cite{edwards1975theory}.
In close analogy with the diagnostics, we considered for the quantum circuits, the EA order parameter can be expressed as a standard average (correlation function) in an effective probability distribution constructed from two identical copies of the system. This doubling naturally gives rise to a permutation symmetry between replicas, which is reminiscent of the permutation symmetries in our system. 

The analogy with the spin glass also extends to information theoretic aspect of the problem. %
We have shown that broken symmetry phases of the circuit are states that encode and protect information on symmetry breaking in the initial state. Similarly, the onset of an EA order parameter in a spin glass signals the emergence of an error correcting code, allowing to encode information in random broken symmetry spin configurations~\cite{sourlas1989spin,nishimori2001statistical}. 

There is however a fundamental difference between the quantum measurement ensemble and the spin glass problem. Due to the quantum nature of the circuit, fluctuations in the measurement ensemble are obtained from the time evolution of doubled density operators rather than from (static) probability distributions. 
This leads to a richer symmetry, since the forward (ket) and backward (bra) evolution of the density operators are associated with independent permutation symmetries. 
Spontaneous breaking of these inherently quantum symmetries signals a capacity to encode quantum information~\cite{choi2019quantum,gullans2019dynamical,bao2020theory,fan2020self,ippoliti2020entanglement}.

To facilitate the classification of phases, we formulated a mapping between the steady states of the circuit ensemble and the ground states of an effective Hamiltonian, which inherits the enlarged dynamical symmetry. This leads to establishment of phases that would not have been possible in presence of the physical circuit symmetry alone. Here is a summary of the most striking consequences. (i) A topological area-law phase can form in a one-dimensional model with only $\mathbb{Z}_2$ physical circuit symmetry, owing to protection from an enlarged symmetry $\mathbb{Z}_2\times \mathcal{S}_2$ that combines the circuit symmetry and the dynamical permutation symmetry between trajectories. We argued that this state admits an efficient  experimental detection scheme using the Fisher information associated with perturbations applied at the edge.
Demonstrating the existence of this state in a numerical simulation of a concrete model, or alternatively showing a fundamental obstruction to realize it, remains a challenge for future work. (ii) Several distinct volume-law phases, characterized by different broken symmetry patterns, arise as ground states of a circuit with $\mathbb{Z}_2$ symmetry. We demonstrate the existence of three of these volume-law entangled states in numerical simulations of a one-dimensional model.
An interesting conclusion from this result is that the quantum measurements allow to stabilize orders, which could not have existed in a finite entropy equilibrium state. We argue, as a corollary, that measurements can even facilitate the establishment of topological order and protected edge modes in a state with volume-law entanglement. 
(iii) A Gaussian fermionic circuit with only $\mathbb{Z}_2$ fermion parity conservation exhibits a critical phase facilitated by a $U(1)$ dynamical symmetry of two copies of the density matrix. We predict a Kosterlitz-Thouless transition to area-law states above a critical rate of measurement. 
We show numerical evidence supporting this transition.

It is worth noting that the predictions of the effective model rely on analysis of quantum dynamics with two replicas without taking the formal replica limit $n\to 1$.  
The rather detailed agreement with exact numerical results may suggest that two replicas play a special role in this problem because the minimal diagnostics of the ensemble are fundamentally second moments. It is worth exploring alternative schemes to average over the ensemble, which avoid introducing auxiliary replicas, possibly at the expense of retaining space-time disorder~\cite{zhou2019emergent}.
We also remark that our classification of phases in this paper was based on symmetry alone. It would be important to understand if additional constraints on the dynamics, such as complete positivity of quantum channels, could inhibit the establishment of certain states. 

Arguably the most pressing challenge for the theory right now is to connect with the rapidly advancing experimental work on quantum circuits. An important first step would be to extend the theoretical framework to more realistic models, which include, at a minimum, decoherence and possibly other dissipative processes in addition to unitary gates and measurements. Such an extension would also be interesting from the theoretical perspective because decoherence changes the symmetry of the problem.
Any weak coupling to the environment explicitly breaks the independent permutation symmetries among replicated quantum states propagating forward or backward, and it only respects simultaneous permutations of forward and backward states. 
We have noted that the separation of the two permutation symmetries is a signature of quantumness that is absent in classical settings, such as the replica theory of a spin glass.
Nonetheless, the universal behavior and encoding power may not immediately collapse to that of a classical circuit.
There is another quantum symmetry associated with the exchange between forward and backward propagating states, namely the hermiticity of the density matrix, which is neither removed nor trivialized immediately by the addition of decoherence.
A nontrivial hermiticity symmetry, in principle distinguishes the quantum dynamics of density matrices from classical probability distributions even in the presence of decoherence. 
It would be interesting to explore under what conditions the hermiticity symmetry can be spontaneously broken and to understand the implications of such order on the capacity for encoding quantum information.

We close by recalling a dialogue that had not occurred in Paris in the 1920s. Hopefully it sums up our message clearly:

{\scriptsize FITZGERALD:} Quantum circuits are different from spin glasses.

{\scriptsize HEMINGWAY:} Yes, they have more symmetry.

\begin{acknowledgments}
We thank David Huse, Michael Gullans and Zack Weinstein for useful discussions.
We acknowledge support from the NSF QLCI program through grant number OMA-2016245. SC acknowledges support from the Miller Institute for Basic Research in Science. EA is supported in part by the Mike Gyorgy Chair in Physics at UC Berkeley.
\end{acknowledgments}
\bibliography{refs}

\begin{thebibliography}{50}%
\makeatletter
\providecommand \@ifxundefined [1]{%
 \@ifx{#1\undefined}
}%
\providecommand \@ifnum [1]{%
 \ifnum #1\expandafter \@firstoftwo
 \else \expandafter \@secondoftwo
 \fi
}%
\providecommand \@ifx [1]{%
 \ifx #1\expandafter \@firstoftwo
 \else \expandafter \@secondoftwo
 \fi
}%
\providecommand \natexlab [1]{#1}%
\providecommand \enquote  [1]{``#1''}%
\providecommand \bibnamefont  [1]{#1}%
\providecommand \bibfnamefont [1]{#1}%
\providecommand \citenamefont [1]{#1}%
\providecommand \href@noop [0]{\@secondoftwo}%
\providecommand \href [0]{\begingroup \@sanitize@url \@href}%
\providecommand \@href[1]{\@@startlink{#1}\@@href}%
\providecommand \@@href[1]{\endgroup#1\@@endlink}%
\providecommand \@sanitize@url [0]{\catcode `\\12\catcode `\$12\catcode
  `\&12\catcode `\#12\catcode `\^12\catcode `\_12\catcode `\%12\relax}%
\providecommand \@@startlink[1]{}%
\providecommand \@@endlink[0]{}%
\providecommand \url  [0]{\begingroup\@sanitize@url \@url }%
\providecommand \@url [1]{\endgroup\@href {#1}{\urlprefix }}%
\providecommand \urlprefix  [0]{URL }%
\providecommand \Eprint [0]{\href }%
\providecommand \doibase [0]{http://dx.doi.org/}%
\providecommand \selectlanguage [0]{\@gobble}%
\providecommand \bibinfo  [0]{\@secondoftwo}%
\providecommand \bibfield  [0]{\@secondoftwo}%
\providecommand \translation [1]{[#1]}%
\providecommand \BibitemOpen [0]{}%
\providecommand \bibitemStop [0]{}%
\providecommand \bibitemNoStop [0]{.\EOS\space}%
\providecommand \EOS [0]{\spacefactor3000\relax}%
\providecommand \BibitemShut  [1]{\csname bibitem#1\endcsname}%
\let\auto@bib@innerbib\@empty
\bibitem [{\citenamefont {Landsman}\ \emph {et~al.}(2019)\citenamefont
  {Landsman}, \citenamefont {Figgatt}, \citenamefont {Schuster}, \citenamefont
  {Linke}, \citenamefont {Yoshida}, \citenamefont {Yao},\ and\ \citenamefont
  {Monroe}}]{landsman2019verified}%
  \BibitemOpen
  \bibfield  {author} {\bibinfo {author} {\bibfnamefont {K.~A.}\ \bibnamefont
  {Landsman}}, \bibinfo {author} {\bibfnamefont {C.}~\bibnamefont {Figgatt}},
  \bibinfo {author} {\bibfnamefont {T.}~\bibnamefont {Schuster}}, \bibinfo
  {author} {\bibfnamefont {N.~M.}\ \bibnamefont {Linke}}, \bibinfo {author}
  {\bibfnamefont {B.}~\bibnamefont {Yoshida}}, \bibinfo {author} {\bibfnamefont
  {N.~Y.}\ \bibnamefont {Yao}}, \ and\ \bibinfo {author} {\bibfnamefont
  {C.}~\bibnamefont {Monroe}},\ }\href@noop {} {\bibfield  {journal} {\bibinfo
  {journal} {Nature}\ }\textbf {\bibinfo {volume} {567}},\ \bibinfo {pages}
  {61} (\bibinfo {year} {2019})}\BibitemShut {NoStop}%
\bibitem [{\citenamefont {Arute}\ \emph {et~al.}(2019)\citenamefont {Arute},
  \citenamefont {Arya}, \citenamefont {Babbush}, \citenamefont {Bacon},
  \citenamefont {Bardin}, \citenamefont {Barends}, \citenamefont {Biswas},
  \citenamefont {Boixo}, \citenamefont {Brandao}, \citenamefont {Buell} \emph
  {et~al.}}]{arute2019quantum}%
  \BibitemOpen
  \bibfield  {author} {\bibinfo {author} {\bibfnamefont {F.}~\bibnamefont
  {Arute}}, \bibinfo {author} {\bibfnamefont {K.}~\bibnamefont {Arya}},
  \bibinfo {author} {\bibfnamefont {R.}~\bibnamefont {Babbush}}, \bibinfo
  {author} {\bibfnamefont {D.}~\bibnamefont {Bacon}}, \bibinfo {author}
  {\bibfnamefont {J.~C.}\ \bibnamefont {Bardin}}, \bibinfo {author}
  {\bibfnamefont {R.}~\bibnamefont {Barends}}, \bibinfo {author} {\bibfnamefont
  {R.}~\bibnamefont {Biswas}}, \bibinfo {author} {\bibfnamefont
  {S.}~\bibnamefont {Boixo}}, \bibinfo {author} {\bibfnamefont {F.~G.}\
  \bibnamefont {Brandao}}, \bibinfo {author} {\bibfnamefont {D.~A.}\
  \bibnamefont {Buell}},  \emph {et~al.},\ }\href@noop {} {\bibfield  {journal}
  {\bibinfo  {journal} {Nature}\ }\textbf {\bibinfo {volume} {574}},\ \bibinfo
  {pages} {505} (\bibinfo {year} {2019})}\BibitemShut {NoStop}%
\bibitem [{\citenamefont {Zhong}\ \emph {et~al.}(2020)\citenamefont {Zhong},
  \citenamefont {Wang}, \citenamefont {Deng}, \citenamefont {Chen},
  \citenamefont {Peng}, \citenamefont {Luo}, \citenamefont {Qin}, \citenamefont
  {Wu}, \citenamefont {Ding}, \citenamefont {Hu} \emph
  {et~al.}}]{zhong2020quantum}%
  \BibitemOpen
  \bibfield  {author} {\bibinfo {author} {\bibfnamefont {H.-S.}\ \bibnamefont
  {Zhong}}, \bibinfo {author} {\bibfnamefont {H.}~\bibnamefont {Wang}},
  \bibinfo {author} {\bibfnamefont {Y.-H.}\ \bibnamefont {Deng}}, \bibinfo
  {author} {\bibfnamefont {M.-C.}\ \bibnamefont {Chen}}, \bibinfo {author}
  {\bibfnamefont {L.-C.}\ \bibnamefont {Peng}}, \bibinfo {author}
  {\bibfnamefont {Y.-H.}\ \bibnamefont {Luo}}, \bibinfo {author} {\bibfnamefont
  {J.}~\bibnamefont {Qin}}, \bibinfo {author} {\bibfnamefont {D.}~\bibnamefont
  {Wu}}, \bibinfo {author} {\bibfnamefont {X.}~\bibnamefont {Ding}}, \bibinfo
  {author} {\bibfnamefont {Y.}~\bibnamefont {Hu}},  \emph {et~al.},\
  }\href@noop {} {\bibfield  {journal} {\bibinfo  {journal} {Science}\ }\textbf
  {\bibinfo {volume} {370}},\ \bibinfo {pages} {1460} (\bibinfo {year}
  {2020})}\BibitemShut {NoStop}%
\bibitem [{\citenamefont {Skinner}\ \emph {et~al.}(2018)\citenamefont
  {Skinner}, \citenamefont {Ruhman},\ and\ \citenamefont
  {Nahum}}]{skinner2018measurement}%
  \BibitemOpen
  \bibfield  {author} {\bibinfo {author} {\bibfnamefont {B.}~\bibnamefont
  {Skinner}}, \bibinfo {author} {\bibfnamefont {J.}~\bibnamefont {Ruhman}}, \
  and\ \bibinfo {author} {\bibfnamefont {A.}~\bibnamefont {Nahum}},\
  }\href@noop {} {\bibfield  {journal} {\bibinfo  {journal} {arXiv preprint
  arXiv:1808.05953}\ } (\bibinfo {year} {2018})}\BibitemShut {NoStop}%
\bibitem [{\citenamefont {Li}\ \emph {et~al.}(2018)\citenamefont {Li},
  \citenamefont {Chen},\ and\ \citenamefont {Fisher}}]{li2018quantum}%
  \BibitemOpen
  \bibfield  {author} {\bibinfo {author} {\bibfnamefont {Y.}~\bibnamefont
  {Li}}, \bibinfo {author} {\bibfnamefont {X.}~\bibnamefont {Chen}}, \ and\
  \bibinfo {author} {\bibfnamefont {M.~P.}\ \bibnamefont {Fisher}},\
  }\href@noop {} {\bibfield  {journal} {\bibinfo  {journal} {Physical Review
  B}\ }\textbf {\bibinfo {volume} {98}},\ \bibinfo {pages} {205136} (\bibinfo
  {year} {2018})}\BibitemShut {NoStop}%
\bibitem [{\citenamefont {Chan}\ \emph {et~al.}(2019)\citenamefont {Chan},
  \citenamefont {Nandkishore}, \citenamefont {Pretko},\ and\ \citenamefont
  {Smith}}]{chan2018weak}%
  \BibitemOpen
  \bibfield  {author} {\bibinfo {author} {\bibfnamefont {A.}~\bibnamefont
  {Chan}}, \bibinfo {author} {\bibfnamefont {R.~M.}\ \bibnamefont
  {Nandkishore}}, \bibinfo {author} {\bibfnamefont {M.}~\bibnamefont {Pretko}},
  \ and\ \bibinfo {author} {\bibfnamefont {G.}~\bibnamefont {Smith}},\ }\href
  {\doibase 10.1103/PhysRevB.99.224307} {\bibfield  {journal} {\bibinfo
  {journal} {Phys. Rev. B}\ }\textbf {\bibinfo {volume} {99}},\ \bibinfo
  {pages} {224307} (\bibinfo {year} {2019})}\BibitemShut {NoStop}%
\bibitem [{\citenamefont {Choi}\ \emph {et~al.}(2019)\citenamefont {Choi},
  \citenamefont {Bao}, \citenamefont {Qi},\ and\ \citenamefont
  {Altman}}]{choi2019quantum}%
  \BibitemOpen
  \bibfield  {author} {\bibinfo {author} {\bibfnamefont {S.}~\bibnamefont
  {Choi}}, \bibinfo {author} {\bibfnamefont {Y.}~\bibnamefont {Bao}}, \bibinfo
  {author} {\bibfnamefont {X.-L.}\ \bibnamefont {Qi}}, \ and\ \bibinfo {author}
  {\bibfnamefont {E.}~\bibnamefont {Altman}},\ }\href@noop {} {\bibfield
  {journal} {\bibinfo  {journal} {arXiv preprint arXiv:1903.05124}\ } (\bibinfo
  {year} {2019})}\BibitemShut {NoStop}%
\bibitem [{\citenamefont {Gullans}\ and\ \citenamefont
  {Huse}(2019)}]{gullans2019dynamical}%
  \BibitemOpen
  \bibfield  {author} {\bibinfo {author} {\bibfnamefont {M.~J.}\ \bibnamefont
  {Gullans}}\ and\ \bibinfo {author} {\bibfnamefont {D.~A.}\ \bibnamefont
  {Huse}},\ }\href@noop {} {\bibfield  {journal} {\bibinfo  {journal} {arXiv
  preprint arXiv:1905.05195}\ } (\bibinfo {year} {2019})}\BibitemShut {NoStop}%
\bibitem [{\citenamefont {Fan}\ \emph {et~al.}(2020)\citenamefont {Fan},
  \citenamefont {Vijay}, \citenamefont {Vishwanath},\ and\ \citenamefont
  {You}}]{fan2020self}%
  \BibitemOpen
  \bibfield  {author} {\bibinfo {author} {\bibfnamefont {R.}~\bibnamefont
  {Fan}}, \bibinfo {author} {\bibfnamefont {S.}~\bibnamefont {Vijay}}, \bibinfo
  {author} {\bibfnamefont {A.}~\bibnamefont {Vishwanath}}, \ and\ \bibinfo
  {author} {\bibfnamefont {Y.-Z.}\ \bibnamefont {You}},\ }\href@noop {}
  {\bibfield  {journal} {\bibinfo  {journal} {arXiv preprint arXiv:2002.12385}\
  } (\bibinfo {year} {2020})}\BibitemShut {NoStop}%
\bibitem [{\citenamefont {Ippoliti}\ \emph {et~al.}(2020)\citenamefont
  {Ippoliti}, \citenamefont {Gullans}, \citenamefont {Gopalakrishnan},
  \citenamefont {Huse},\ and\ \citenamefont
  {Khemani}}]{ippoliti2020entanglement}%
  \BibitemOpen
  \bibfield  {author} {\bibinfo {author} {\bibfnamefont {M.}~\bibnamefont
  {Ippoliti}}, \bibinfo {author} {\bibfnamefont {M.~J.}\ \bibnamefont
  {Gullans}}, \bibinfo {author} {\bibfnamefont {S.}~\bibnamefont
  {Gopalakrishnan}}, \bibinfo {author} {\bibfnamefont {D.~A.}\ \bibnamefont
  {Huse}}, \ and\ \bibinfo {author} {\bibfnamefont {V.}~\bibnamefont
  {Khemani}},\ }\href@noop {} {\bibfield  {journal} {\bibinfo  {journal} {arXiv
  preprint arXiv:2004.09560}\ } (\bibinfo {year} {2020})}\BibitemShut {NoStop}%
\bibitem [{\citenamefont {Anderson}(1972)}]{anderson1972more}%
  \BibitemOpen
  \bibfield  {author} {\bibinfo {author} {\bibfnamefont {P.~W.}\ \bibnamefont
  {Anderson}},\ }\href@noop {} {\bibfield  {journal} {\bibinfo  {journal}
  {Science}\ }\textbf {\bibinfo {volume} {177}},\ \bibinfo {pages} {393}
  (\bibinfo {year} {1972})}\BibitemShut {NoStop}%
\bibitem [{\citenamefont {Bao}\ \emph {et~al.}(2020)\citenamefont {Bao},
  \citenamefont {Choi},\ and\ \citenamefont {Altman}}]{bao2020theory}%
  \BibitemOpen
  \bibfield  {author} {\bibinfo {author} {\bibfnamefont {Y.}~\bibnamefont
  {Bao}}, \bibinfo {author} {\bibfnamefont {S.}~\bibnamefont {Choi}}, \ and\
  \bibinfo {author} {\bibfnamefont {E.}~\bibnamefont {Altman}},\ }\href@noop {}
  {\bibfield  {journal} {\bibinfo  {journal} {Physical Review B}\ }\textbf
  {\bibinfo {volume} {101}},\ \bibinfo {pages} {104301} (\bibinfo {year}
  {2020})}\BibitemShut {NoStop}%
\bibitem [{\citenamefont {Jian}\ \emph
  {et~al.}(2020{\natexlab{a}})\citenamefont {Jian}, \citenamefont {You},
  \citenamefont {Vasseur},\ and\ \citenamefont {Ludwig}}]{jian2020measurement}%
  \BibitemOpen
  \bibfield  {author} {\bibinfo {author} {\bibfnamefont {C.-M.}\ \bibnamefont
  {Jian}}, \bibinfo {author} {\bibfnamefont {Y.-Z.}\ \bibnamefont {You}},
  \bibinfo {author} {\bibfnamefont {R.}~\bibnamefont {Vasseur}}, \ and\
  \bibinfo {author} {\bibfnamefont {A.~W.}\ \bibnamefont {Ludwig}},\
  }\href@noop {} {\bibfield  {journal} {\bibinfo  {journal} {Physical Review
  B}\ }\textbf {\bibinfo {volume} {101}},\ \bibinfo {pages} {104302} (\bibinfo
  {year} {2020}{\natexlab{a}})}\BibitemShut {NoStop}%
\bibitem [{\citenamefont {Nahum}\ \emph {et~al.}(2021)\citenamefont {Nahum},
  \citenamefont {Roy}, \citenamefont {Skinner},\ and\ \citenamefont
  {Ruhman}}]{nahum2020measurement}%
  \BibitemOpen
  \bibfield  {author} {\bibinfo {author} {\bibfnamefont {A.}~\bibnamefont
  {Nahum}}, \bibinfo {author} {\bibfnamefont {S.}~\bibnamefont {Roy}}, \bibinfo
  {author} {\bibfnamefont {B.}~\bibnamefont {Skinner}}, \ and\ \bibinfo
  {author} {\bibfnamefont {J.}~\bibnamefont {Ruhman}},\ }\href@noop {}
  {\bibfield  {journal} {\bibinfo  {journal} {PRX Quantum}\ }\textbf {\bibinfo
  {volume} {2}},\ \bibinfo {pages} {010352} (\bibinfo {year}
  {2021})}\BibitemShut {NoStop}%
\bibitem [{\citenamefont {Hayden}\ \emph {et~al.}(2016)\citenamefont {Hayden},
  \citenamefont {Nezami}, \citenamefont {Qi}, \citenamefont {Thomas},
  \citenamefont {Walter},\ and\ \citenamefont {Yang}}]{hayden2016holographic}%
  \BibitemOpen
  \bibfield  {author} {\bibinfo {author} {\bibfnamefont {P.}~\bibnamefont
  {Hayden}}, \bibinfo {author} {\bibfnamefont {S.}~\bibnamefont {Nezami}},
  \bibinfo {author} {\bibfnamefont {X.-L.}\ \bibnamefont {Qi}}, \bibinfo
  {author} {\bibfnamefont {N.}~\bibnamefont {Thomas}}, \bibinfo {author}
  {\bibfnamefont {M.}~\bibnamefont {Walter}}, \ and\ \bibinfo {author}
  {\bibfnamefont {Z.}~\bibnamefont {Yang}},\ }\href@noop {} {\bibfield
  {journal} {\bibinfo  {journal} {Journal of High Energy Physics}\ }\textbf
  {\bibinfo {volume} {2016}},\ \bibinfo {pages} {9} (\bibinfo {year}
  {2016})}\BibitemShut {NoStop}%
\bibitem [{\citenamefont {Nahum}\ \emph {et~al.}(2018)\citenamefont {Nahum},
  \citenamefont {Vijay},\ and\ \citenamefont {Haah}}]{nahum2018operator}%
  \BibitemOpen
  \bibfield  {author} {\bibinfo {author} {\bibfnamefont {A.}~\bibnamefont
  {Nahum}}, \bibinfo {author} {\bibfnamefont {S.}~\bibnamefont {Vijay}}, \ and\
  \bibinfo {author} {\bibfnamefont {J.}~\bibnamefont {Haah}},\ }\href@noop {}
  {\bibfield  {journal} {\bibinfo  {journal} {Physical Review X}\ }\textbf
  {\bibinfo {volume} {8}},\ \bibinfo {pages} {021014} (\bibinfo {year}
  {2018})}\BibitemShut {NoStop}%
\bibitem [{\citenamefont {Vasseur}\ \emph {et~al.}(2018)\citenamefont
  {Vasseur}, \citenamefont {Potter}, \citenamefont {You},\ and\ \citenamefont
  {Ludwig}}]{vasseur2018entanglement}%
  \BibitemOpen
  \bibfield  {author} {\bibinfo {author} {\bibfnamefont {R.}~\bibnamefont
  {Vasseur}}, \bibinfo {author} {\bibfnamefont {A.~C.}\ \bibnamefont {Potter}},
  \bibinfo {author} {\bibfnamefont {Y.-Z.}\ \bibnamefont {You}}, \ and\
  \bibinfo {author} {\bibfnamefont {A.~W.}\ \bibnamefont {Ludwig}},\
  }\href@noop {} {\bibfield  {journal} {\bibinfo  {journal} {arXiv preprint
  arXiv:1807.07082}\ } (\bibinfo {year} {2018})}\BibitemShut {NoStop}%
\bibitem [{\citenamefont {Zhou}\ and\ \citenamefont
  {Nahum}(2019)}]{zhou2019emergent}%
  \BibitemOpen
  \bibfield  {author} {\bibinfo {author} {\bibfnamefont {T.}~\bibnamefont
  {Zhou}}\ and\ \bibinfo {author} {\bibfnamefont {A.}~\bibnamefont {Nahum}},\
  }\href@noop {} {\bibfield  {journal} {\bibinfo  {journal} {Physical Review
  B}\ }\textbf {\bibinfo {volume} {99}},\ \bibinfo {pages} {174205} (\bibinfo
  {year} {2019})}\BibitemShut {NoStop}%
\bibitem [{\citenamefont {Sang}\ and\ \citenamefont
  {Hsieh}(2020)}]{sang2020measurement}%
  \BibitemOpen
  \bibfield  {author} {\bibinfo {author} {\bibfnamefont {S.}~\bibnamefont
  {Sang}}\ and\ \bibinfo {author} {\bibfnamefont {T.~H.}\ \bibnamefont
  {Hsieh}},\ }\href@noop {} {\bibfield  {journal} {\bibinfo  {journal} {arXiv
  preprint arXiv:2004.09509}\ } (\bibinfo {year} {2020})}\BibitemShut {NoStop}%
\bibitem [{\citenamefont {Pollmann}\ \emph {et~al.}(2010)\citenamefont
  {Pollmann}, \citenamefont {Turner}, \citenamefont {Berg},\ and\ \citenamefont
  {Oshikawa}}]{pollmann2010entanglement}%
  \BibitemOpen
  \bibfield  {author} {\bibinfo {author} {\bibfnamefont {F.}~\bibnamefont
  {Pollmann}}, \bibinfo {author} {\bibfnamefont {A.~M.}\ \bibnamefont
  {Turner}}, \bibinfo {author} {\bibfnamefont {E.}~\bibnamefont {Berg}}, \ and\
  \bibinfo {author} {\bibfnamefont {M.}~\bibnamefont {Oshikawa}},\ }\href@noop
  {} {\bibfield  {journal} {\bibinfo  {journal} {Physical review b}\ }\textbf
  {\bibinfo {volume} {81}},\ \bibinfo {pages} {064439} (\bibinfo {year}
  {2010})}\BibitemShut {NoStop}%
\bibitem [{\citenamefont {Chen}\ \emph {et~al.}(2011)\citenamefont {Chen},
  \citenamefont {Gu},\ and\ \citenamefont {Wen}}]{chen2011classification}%
  \BibitemOpen
  \bibfield  {author} {\bibinfo {author} {\bibfnamefont {X.}~\bibnamefont
  {Chen}}, \bibinfo {author} {\bibfnamefont {Z.-C.}\ \bibnamefont {Gu}}, \ and\
  \bibinfo {author} {\bibfnamefont {X.-G.}\ \bibnamefont {Wen}},\ }\href@noop
  {} {\bibfield  {journal} {\bibinfo  {journal} {Physical review b}\ }\textbf
  {\bibinfo {volume} {83}},\ \bibinfo {pages} {035107} (\bibinfo {year}
  {2011})}\BibitemShut {NoStop}%
\bibitem [{\citenamefont {Edwards}\ and\ \citenamefont
  {Anderson}(1975)}]{edwards1975theory}%
  \BibitemOpen
  \bibfield  {author} {\bibinfo {author} {\bibfnamefont {S.~F.}\ \bibnamefont
  {Edwards}}\ and\ \bibinfo {author} {\bibfnamefont {P.~W.}\ \bibnamefont
  {Anderson}},\ }\href@noop {} {\bibfield  {journal} {\bibinfo  {journal}
  {Journal of Physics F: Metal Physics}\ }\textbf {\bibinfo {volume} {5}},\
  \bibinfo {pages} {965} (\bibinfo {year} {1975})}\BibitemShut {NoStop}%
\bibitem [{\citenamefont {Terhal}\ and\ \citenamefont
  {DiVincenzo}(2002)}]{terhal2002classical}%
  \BibitemOpen
  \bibfield  {author} {\bibinfo {author} {\bibfnamefont {B.~M.}\ \bibnamefont
  {Terhal}}\ and\ \bibinfo {author} {\bibfnamefont {D.~P.}\ \bibnamefont
  {DiVincenzo}},\ }\href@noop {} {\bibfield  {journal} {\bibinfo  {journal}
  {Physical Review A}\ }\textbf {\bibinfo {volume} {65}},\ \bibinfo {pages}
  {032325} (\bibinfo {year} {2002})}\BibitemShut {NoStop}%
\bibitem [{\citenamefont {Cao}\ \emph {et~al.}(2018)\citenamefont {Cao},
  \citenamefont {Tilloy},\ and\ \citenamefont {De~Luca}}]{cao2018entanglement}%
  \BibitemOpen
  \bibfield  {author} {\bibinfo {author} {\bibfnamefont {X.}~\bibnamefont
  {Cao}}, \bibinfo {author} {\bibfnamefont {A.}~\bibnamefont {Tilloy}}, \ and\
  \bibinfo {author} {\bibfnamefont {A.}~\bibnamefont {De~Luca}},\ }\href@noop
  {} {\bibfield  {journal} {\bibinfo  {journal} {arXiv preprint
  arXiv:1804.04638}\ } (\bibinfo {year} {2018})}\BibitemShut {NoStop}%
\bibitem [{\citenamefont {Fidkowski}\ \emph {et~al.}(2020)\citenamefont
  {Fidkowski}, \citenamefont {Haah},\ and\ \citenamefont
  {Hastings}}]{fidkowski2020dynamical}%
  \BibitemOpen
  \bibfield  {author} {\bibinfo {author} {\bibfnamefont {L.}~\bibnamefont
  {Fidkowski}}, \bibinfo {author} {\bibfnamefont {J.}~\bibnamefont {Haah}}, \
  and\ \bibinfo {author} {\bibfnamefont {M.~B.}\ \bibnamefont {Hastings}},\
  }\href@noop {} {\bibfield  {journal} {\bibinfo  {journal} {arXiv preprint
  arXiv:2008.10611}\ } (\bibinfo {year} {2020})}\BibitemShut {NoStop}%
\bibitem [{\citenamefont {Alberton}\ \emph {et~al.}(2020)\citenamefont
  {Alberton}, \citenamefont {Buchhold},\ and\ \citenamefont
  {Diehl}}]{alberton2020trajectory}%
  \BibitemOpen
  \bibfield  {author} {\bibinfo {author} {\bibfnamefont {O.}~\bibnamefont
  {Alberton}}, \bibinfo {author} {\bibfnamefont {M.}~\bibnamefont {Buchhold}},
  \ and\ \bibinfo {author} {\bibfnamefont {S.}~\bibnamefont {Diehl}},\
  }\href@noop {} {\bibfield  {journal} {\bibinfo  {journal} {arXiv preprint
  arXiv:2005.09722}\ } (\bibinfo {year} {2020})}\BibitemShut {NoStop}%
\bibitem [{\citenamefont {Sang}\ \emph {et~al.}(2021)\citenamefont {Sang},
  \citenamefont {Li}, \citenamefont {Zhou}, \citenamefont {Chen}, \citenamefont
  {Hsieh},\ and\ \citenamefont {Fisher}}]{sang2020entanglement}%
  \BibitemOpen
  \bibfield  {author} {\bibinfo {author} {\bibfnamefont {S.}~\bibnamefont
  {Sang}}, \bibinfo {author} {\bibfnamefont {Y.}~\bibnamefont {Li}}, \bibinfo
  {author} {\bibfnamefont {T.}~\bibnamefont {Zhou}}, \bibinfo {author}
  {\bibfnamefont {X.}~\bibnamefont {Chen}}, \bibinfo {author} {\bibfnamefont
  {T.~H.}\ \bibnamefont {Hsieh}}, \ and\ \bibinfo {author} {\bibfnamefont
  {M.~P.}\ \bibnamefont {Fisher}},\ }\href@noop {} {\bibfield  {journal}
  {\bibinfo  {journal} {PRX Quantum}\ }\textbf {\bibinfo {volume} {2}},\
  \bibinfo {pages} {030313} (\bibinfo {year} {2021})}\BibitemShut {NoStop}%
\bibitem [{\citenamefont {Chen}\ \emph {et~al.}(2020)\citenamefont {Chen},
  \citenamefont {Li}, \citenamefont {Fisher},\ and\ \citenamefont
  {Lucas}}]{chen2020emergent}%
  \BibitemOpen
  \bibfield  {author} {\bibinfo {author} {\bibfnamefont {X.}~\bibnamefont
  {Chen}}, \bibinfo {author} {\bibfnamefont {Y.}~\bibnamefont {Li}}, \bibinfo
  {author} {\bibfnamefont {M.}~\bibnamefont {Fisher}}, \ and\ \bibinfo {author}
  {\bibfnamefont {A.}~\bibnamefont {Lucas}},\ }\href@noop {} {\bibfield
  {journal} {\bibinfo  {journal} {arXiv preprint arXiv:2004.09577}\ } (\bibinfo
  {year} {2020})}\BibitemShut {NoStop}%
\bibitem [{\citenamefont {Liu}\ \emph {et~al.}(2020)\citenamefont {Liu},
  \citenamefont {Zhang},\ and\ \citenamefont {Chen}}]{liu2020non}%
  \BibitemOpen
  \bibfield  {author} {\bibinfo {author} {\bibfnamefont {C.}~\bibnamefont
  {Liu}}, \bibinfo {author} {\bibfnamefont {P.}~\bibnamefont {Zhang}}, \ and\
  \bibinfo {author} {\bibfnamefont {X.}~\bibnamefont {Chen}},\ }\href@noop {}
  {\bibfield  {journal} {\bibinfo  {journal} {measurement}\ } (\bibinfo {year}
  {2020})}\BibitemShut {NoStop}%
\bibitem [{\citenamefont {Jian}\ \emph
  {et~al.}(2020{\natexlab{b}})\citenamefont {Jian}, \citenamefont {Bauer},
  \citenamefont {Keselman},\ and\ \citenamefont
  {Ludwig}}]{jian2020criticality}%
  \BibitemOpen
  \bibfield  {author} {\bibinfo {author} {\bibfnamefont {C.-M.}\ \bibnamefont
  {Jian}}, \bibinfo {author} {\bibfnamefont {B.}~\bibnamefont {Bauer}},
  \bibinfo {author} {\bibfnamefont {A.}~\bibnamefont {Keselman}}, \ and\
  \bibinfo {author} {\bibfnamefont {A.~W.}\ \bibnamefont {Ludwig}},\
  }\href@noop {} {\bibfield  {journal} {\bibinfo  {journal} {arXiv preprint
  arXiv:2012.04666}\ } (\bibinfo {year} {2020}{\natexlab{b}})}\BibitemShut
  {NoStop}%
\bibitem [{\citenamefont {Nahum}\ and\ \citenamefont
  {Skinner}(2020)}]{nahum2020entanglement}%
  \BibitemOpen
  \bibfield  {author} {\bibinfo {author} {\bibfnamefont {A.}~\bibnamefont
  {Nahum}}\ and\ \bibinfo {author} {\bibfnamefont {B.}~\bibnamefont
  {Skinner}},\ }\href@noop {} {\bibfield  {journal} {\bibinfo  {journal}
  {Physical Review Research}\ }\textbf {\bibinfo {volume} {2}},\ \bibinfo
  {pages} {023288} (\bibinfo {year} {2020})}\BibitemShut {NoStop}%
\bibitem [{\citenamefont {Ryu}\ \emph {et~al.}(2007)\citenamefont {Ryu},
  \citenamefont {Mudry}, \citenamefont {Obuse},\ and\ \citenamefont
  {Furusaki}}]{Ryu2007}%
  \BibitemOpen
  \bibfield  {author} {\bibinfo {author} {\bibfnamefont {S.}~\bibnamefont
  {Ryu}}, \bibinfo {author} {\bibfnamefont {C.}~\bibnamefont {Mudry}}, \bibinfo
  {author} {\bibfnamefont {H.}~\bibnamefont {Obuse}}, \ and\ \bibinfo {author}
  {\bibfnamefont {A.}~\bibnamefont {Furusaki}},\ }\href {\doibase
  10.1103/PhysRevLett.99.116601} {\bibfield  {journal} {\bibinfo  {journal}
  {Phys. Rev. Lett.}\ }\textbf {\bibinfo {volume} {99}},\ \bibinfo {pages}
  {116601} (\bibinfo {year} {2007})}\BibitemShut {NoStop}%
\bibitem [{\citenamefont {Fu}\ and\ \citenamefont {Kane}(2012)}]{Fu2012}%
  \BibitemOpen
  \bibfield  {author} {\bibinfo {author} {\bibfnamefont {L.}~\bibnamefont
  {Fu}}\ and\ \bibinfo {author} {\bibfnamefont {C.~L.}\ \bibnamefont {Kane}},\
  }\href {\doibase 10.1103/PhysRevLett.109.246605} {\bibfield  {journal}
  {\bibinfo  {journal} {Phys. Rev. Lett.}\ }\textbf {\bibinfo {volume} {109}},\
  \bibinfo {pages} {246605} (\bibinfo {year} {2012})}\BibitemShut {NoStop}%
\bibitem [{\citenamefont {Lang}\ and\ \citenamefont
  {B{\"u}chler}(2020)}]{lang2020entanglement}%
  \BibitemOpen
  \bibfield  {author} {\bibinfo {author} {\bibfnamefont {N.}~\bibnamefont
  {Lang}}\ and\ \bibinfo {author} {\bibfnamefont {H.~P.}\ \bibnamefont
  {B{\"u}chler}},\ }\href@noop {} {\bibfield  {journal} {\bibinfo  {journal}
  {Physical Review B}\ }\textbf {\bibinfo {volume} {102}},\ \bibinfo {pages}
  {094204} (\bibinfo {year} {2020})}\BibitemShut {NoStop}%
\bibitem [{\citenamefont {Li}\ \emph {et~al.}(2019)\citenamefont {Li},
  \citenamefont {Chen},\ and\ \citenamefont {Fisher}}]{li2019measurement}%
  \BibitemOpen
  \bibfield  {author} {\bibinfo {author} {\bibfnamefont {Y.}~\bibnamefont
  {Li}}, \bibinfo {author} {\bibfnamefont {X.}~\bibnamefont {Chen}}, \ and\
  \bibinfo {author} {\bibfnamefont {M.}~\bibnamefont {Fisher}},\ }\href@noop {}
  {\bibfield  {journal} {\bibinfo  {journal} {arXiv preprint arXiv:1901.08092}\
  } (\bibinfo {year} {2019})}\BibitemShut {NoStop}%
\bibitem [{\citenamefont {Chen}\ \emph {et~al.}(2014)\citenamefont {Chen},
  \citenamefont {Lu},\ and\ \citenamefont {Vishwanath}}]{chen2014symmetry}%
  \BibitemOpen
  \bibfield  {author} {\bibinfo {author} {\bibfnamefont {X.}~\bibnamefont
  {Chen}}, \bibinfo {author} {\bibfnamefont {Y.-M.}\ \bibnamefont {Lu}}, \ and\
  \bibinfo {author} {\bibfnamefont {A.}~\bibnamefont {Vishwanath}},\
  }\href@noop {} {\bibfield  {journal} {\bibinfo  {journal} {Nature
  communications}\ }\textbf {\bibinfo {volume} {5}},\ \bibinfo {pages} {1}
  (\bibinfo {year} {2014})}\BibitemShut {NoStop}%
\bibitem [{\citenamefont {Lavasani}\ \emph {et~al.}(2021)\citenamefont
  {Lavasani}, \citenamefont {Alavirad},\ and\ \citenamefont
  {Barkeshli}}]{lavasani2021measurement}%
  \BibitemOpen
  \bibfield  {author} {\bibinfo {author} {\bibfnamefont {A.}~\bibnamefont
  {Lavasani}}, \bibinfo {author} {\bibfnamefont {Y.}~\bibnamefont {Alavirad}},
  \ and\ \bibinfo {author} {\bibfnamefont {M.}~\bibnamefont {Barkeshli}},\
  }\href@noop {} {\bibfield  {journal} {\bibinfo  {journal} {Nature Physics}\
  ,\ \bibinfo {pages} {1}} (\bibinfo {year} {2021})}\BibitemShut {NoStop}%
\bibitem [{\citenamefont {Aaronson}\ and\ \citenamefont
  {Gottesman}(2004)}]{aaronson2004improved}%
  \BibitemOpen
  \bibfield  {author} {\bibinfo {author} {\bibfnamefont {S.}~\bibnamefont
  {Aaronson}}\ and\ \bibinfo {author} {\bibfnamefont {D.}~\bibnamefont
  {Gottesman}},\ }\href@noop {} {\bibfield  {journal} {\bibinfo  {journal}
  {Physical Review A}\ }\textbf {\bibinfo {volume} {70}},\ \bibinfo {pages}
  {052328} (\bibinfo {year} {2004})}\BibitemShut {NoStop}%
\bibitem [{\citenamefont {Chandran}\ and\ \citenamefont
  {Laumann}(2015)}]{chandran2015semiclassical}%
  \BibitemOpen
  \bibfield  {author} {\bibinfo {author} {\bibfnamefont {A.}~\bibnamefont
  {Chandran}}\ and\ \bibinfo {author} {\bibfnamefont {C.}~\bibnamefont
  {Laumann}},\ }\href@noop {} {\bibfield  {journal} {\bibinfo  {journal}
  {Physical Review B}\ }\textbf {\bibinfo {volume} {92}},\ \bibinfo {pages}
  {024301} (\bibinfo {year} {2015})}\BibitemShut {NoStop}%
\bibitem [{\citenamefont {Webb}(2015)}]{webb2015clifford}%
  \BibitemOpen
  \bibfield  {author} {\bibinfo {author} {\bibfnamefont {Z.}~\bibnamefont
  {Webb}},\ }\href@noop {} {\bibfield  {journal} {\bibinfo  {journal} {arXiv
  preprint arXiv:1510.02769}\ } (\bibinfo {year} {2015})}\BibitemShut {NoStop}%
\bibitem [{\citenamefont {Nahum}\ \emph {et~al.}(2017)\citenamefont {Nahum},
  \citenamefont {Ruhman}, \citenamefont {Vijay},\ and\ \citenamefont
  {Haah}}]{nahum2017quantum}%
  \BibitemOpen
  \bibfield  {author} {\bibinfo {author} {\bibfnamefont {A.}~\bibnamefont
  {Nahum}}, \bibinfo {author} {\bibfnamefont {J.}~\bibnamefont {Ruhman}},
  \bibinfo {author} {\bibfnamefont {S.}~\bibnamefont {Vijay}}, \ and\ \bibinfo
  {author} {\bibfnamefont {J.}~\bibnamefont {Haah}},\ }\href@noop {} {\bibfield
   {journal} {\bibinfo  {journal} {Physical Review X}\ }\textbf {\bibinfo
  {volume} {7}},\ \bibinfo {pages} {031016} (\bibinfo {year}
  {2017})}\BibitemShut {NoStop}%
\bibitem [{\citenamefont {von Keyserlingk}\ \emph {et~al.}(2018)\citenamefont
  {von Keyserlingk}, \citenamefont {Rakovszky}, \citenamefont {Pollmann},\ and\
  \citenamefont {Sondhi}}]{von2018operator}%
  \BibitemOpen
  \bibfield  {author} {\bibinfo {author} {\bibfnamefont {C.}~\bibnamefont {von
  Keyserlingk}}, \bibinfo {author} {\bibfnamefont {T.}~\bibnamefont
  {Rakovszky}}, \bibinfo {author} {\bibfnamefont {F.}~\bibnamefont {Pollmann}},
  \ and\ \bibinfo {author} {\bibfnamefont {S.}~\bibnamefont {Sondhi}},\
  }\href@noop {} {\bibfield  {journal} {\bibinfo  {journal} {Physical Review
  X}\ }\textbf {\bibinfo {volume} {8}},\ \bibinfo {pages} {021013} (\bibinfo
  {year} {2018})}\BibitemShut {NoStop}%
\bibitem [{\citenamefont {Nahum}\ \emph {et~al.}(2013)\citenamefont {Nahum},
  \citenamefont {Serna}, \citenamefont {Somoza},\ and\ \citenamefont
  {Ortuno}}]{nahum2013loop}%
  \BibitemOpen
  \bibfield  {author} {\bibinfo {author} {\bibfnamefont {A.}~\bibnamefont
  {Nahum}}, \bibinfo {author} {\bibfnamefont {P.}~\bibnamefont {Serna}},
  \bibinfo {author} {\bibfnamefont {A.}~\bibnamefont {Somoza}}, \ and\ \bibinfo
  {author} {\bibfnamefont {M.}~\bibnamefont {Ortuno}},\ }\href@noop {}
  {\bibfield  {journal} {\bibinfo  {journal} {Physical Review B}\ }\textbf
  {\bibinfo {volume} {87}},\ \bibinfo {pages} {184204} (\bibinfo {year}
  {2013})}\BibitemShut {NoStop}%
\bibitem [{\citenamefont {Yang}\ and\ \citenamefont
  {Zhang}(1990)}]{yang1990so}%
  \BibitemOpen
  \bibfield  {author} {\bibinfo {author} {\bibfnamefont {C.~N.}\ \bibnamefont
  {Yang}}\ and\ \bibinfo {author} {\bibfnamefont {S.}~\bibnamefont {Zhang}},\
  }\href@noop {} {\bibfield  {journal} {\bibinfo  {journal} {Modern Physics
  Letters B}\ }\textbf {\bibinfo {volume} {4}},\ \bibinfo {pages} {759}
  (\bibinfo {year} {1990})}\BibitemShut {NoStop}%
\bibitem [{\citenamefont {Chen}\ \emph {et~al.}(2003)\citenamefont {Chen},
  \citenamefont {Hida},\ and\ \citenamefont {Sanctuary}}]{chen2003ground}%
  \BibitemOpen
  \bibfield  {author} {\bibinfo {author} {\bibfnamefont {W.}~\bibnamefont
  {Chen}}, \bibinfo {author} {\bibfnamefont {K.}~\bibnamefont {Hida}}, \ and\
  \bibinfo {author} {\bibfnamefont {B.}~\bibnamefont {Sanctuary}},\ }\href@noop
  {} {\bibfield  {journal} {\bibinfo  {journal} {Physical Review B}\ }\textbf
  {\bibinfo {volume} {67}},\ \bibinfo {pages} {104401} (\bibinfo {year}
  {2003})}\BibitemShut {NoStop}%
\bibitem [{\citenamefont {Schulz}(1986)}]{schulz1986phase}%
  \BibitemOpen
  \bibfield  {author} {\bibinfo {author} {\bibfnamefont {H.}~\bibnamefont
  {Schulz}},\ }\href@noop {} {\bibfield  {journal} {\bibinfo  {journal}
  {Physical Review B}\ }\textbf {\bibinfo {volume} {34}},\ \bibinfo {pages}
  {6372} (\bibinfo {year} {1986})}\BibitemShut {NoStop}%
\bibitem [{\citenamefont {Giamarchi}(2003)}]{giamarchi2003quantum}%
  \BibitemOpen
  \bibfield  {author} {\bibinfo {author} {\bibfnamefont {T.}~\bibnamefont
  {Giamarchi}},\ }\href@noop {} {\emph {\bibinfo {title} {Quantum physics in
  one dimension}}},\ Vol.\ \bibinfo {volume} {121}\ (\bibinfo  {publisher}
  {Clarendon press},\ \bibinfo {year} {2003})\BibitemShut {NoStop}%
\bibitem [{\citenamefont {Berg}\ \emph {et~al.}(2008)\citenamefont {Berg},
  \citenamefont {Dalla~Torre}, \citenamefont {Giamarchi},\ and\ \citenamefont
  {Altman}}]{berg2008rise}%
  \BibitemOpen
  \bibfield  {author} {\bibinfo {author} {\bibfnamefont {E.}~\bibnamefont
  {Berg}}, \bibinfo {author} {\bibfnamefont {E.~G.}\ \bibnamefont
  {Dalla~Torre}}, \bibinfo {author} {\bibfnamefont {T.}~\bibnamefont
  {Giamarchi}}, \ and\ \bibinfo {author} {\bibfnamefont {E.}~\bibnamefont
  {Altman}},\ }\href@noop {} {\bibfield  {journal} {\bibinfo  {journal}
  {Physical Review B}\ }\textbf {\bibinfo {volume} {77}},\ \bibinfo {pages}
  {245119} (\bibinfo {year} {2008})}\BibitemShut {NoStop}%
\bibitem [{\citenamefont {Sourlas}(1989)}]{sourlas1989spin}%
  \BibitemOpen
  \bibfield  {author} {\bibinfo {author} {\bibfnamefont {N.}~\bibnamefont
  {Sourlas}},\ }\href@noop {} {\bibfield  {journal} {\bibinfo  {journal}
  {Nature}\ }\textbf {\bibinfo {volume} {339}},\ \bibinfo {pages} {693}
  (\bibinfo {year} {1989})}\BibitemShut {NoStop}%
\bibitem [{\citenamefont {Nishimori}(2001)}]{nishimori2001statistical}%
  \BibitemOpen
  \bibfield  {author} {\bibinfo {author} {\bibfnamefont {H.}~\bibnamefont
  {Nishimori}},\ }\href@noop {} {\emph {\bibinfo {title} {Statistical physics
  of spin glasses and information processing: an introduction}}},\ \bibinfo
  {number} {111}\ (\bibinfo  {publisher} {Clarendon Press},\ \bibinfo {year}
  {2001})\BibitemShut {NoStop}%
\end{thebibliography}%

\appendix

\section{Local symmetries}\label{app:local_symmetry}

\subsection{Qubit circuit}
Here we show that, for the circuits considered in Section~\ref{sec:framework} and \ref{sec:Z2model}, the average hybrid dynamics in $\mathcal{H}^{(n)}$ exhibits local symmetries generated by $\mathcal{X}_j = \prod_{a = 1}^n (X_{a}X_{\bar{a}})_j$, $\mathcal{Y}_j = \prod_{a = 1}^n (Y_{a}Y_{\bar{a}})_j$, and $\mathcal{Z}_j = \prod_{a = 1}^n (Z_{a}Z_{\bar{a}})_j$.
Our proof relies on the facts: (i) unitary gates $U = e^{-\ri \theta O_{P}}$ are generated by Pauli string operators $O_{P}$, e.g. $O_{P} = \tau_{i}^\alpha\tau_{j}^{\beta}$; (ii) the random couplings $\theta$ are drawn from a symmetric distribution to zero; (iii) measurement operators are Pauli strings.
We note that, as a special case discussed in Section~\ref{sec:framework}, the average hybrid dynamics in $\mathcal{H}^{(2)}$ generated by uniformly random single-qubit phase rotations $e^{-\ri\theta_iZ_i}$ exhibits these local symmetries. 
In the following, we demonstrate the symmetries for the average dynamics in $\mathcal{H}^{(n)}$ generated by unitary gates and projective measurements separately.

Without lose of generality, we consider a unitary gate $U_i = e^{-\ri\theta^{\alpha\beta}_i\tau^\alpha_i\tau^\beta_{i+1}}$. $U_i$ acts on $\mathcal{H}^{(n)}$ after averaging over the random coupling as
\begin{align}
    \mathcal{U}_i^{(n)} = \overline{e^{-\ri\theta_i^{\alpha\beta}H_i^{(n)}}} = \sum_{k = 0}^{\infty} \frac{(-1)^k}{(2k)!}\overline{\left(\theta_i^{\alpha\beta}\right)^{2k}}\left(H_i^{(n)}\right)^{2k},
\end{align}
where $H_i^{(n)} = \sum_{a = 1}^n \tau_{i,a}^\alpha\tau_{i+1,a}^\beta - \tau_{i,\bar{a}}^\alpha\tau_{i+1,\bar{a}}^\beta$.
Here, we use the fact that $\theta_{i}^{\alpha\beta}$ is drawn from a symmetric distribution to zero, and odd powers of $H_i^{(n)}$ in the expansion of $\mathcal{U}_i^{(n)}$ vanish.
A special case of $\mathcal{U}_i^{(n)}$ with $n = 2$ and Gaussian random $\theta_i^{\alpha\beta}$ is given in Eq.~\eqref{eq:Uavg}.
To demonstrate the local symmetries in the averaged dynamics $\mathcal{U}_i^{(n)}$, it suffices to show
\begin{align}
    \left[\left(H_i^{(n)}\right)^2, \mathcal{R}_j\right] = 0
\end{align}
for any $i, j$ and $\mathcal{R}_j \in \{\mathcal{X}_j, \mathcal{Y}_j, \mathcal{Z}_j\}$.
Each nontrivial term in $(H_i^{(n)})^2$ contains a product of Pauli matrices $\tau^\alpha_{i,a}\tau^\alpha_{i,a'}$ on site $i$ and $\tau^\beta_{i+1,a}\tau^\beta_{i+1,a'}$ on site $i+1$, which commutes with $\mathcal{R}_j$.
Hence, $(H_i^{(n)})^2$ commutes with $\mathcal{R}_j$, and $\mathcal{R}_j$ are local symmetries of $\mathcal{U}_i^{(n)}$.
The generalization to other unitary gates generated by Pauli string operators is straightforward.

The probabilistic application of a projective measurement averaged over measurement results acts on $\mathcal{H}^{(n)}$ as
\begin{align}
    \mathcal{M}_\nu^{(n)} = (1 - \Gamma_\nu\delta t)\mathds{1}^{\otimes 2n} + \Gamma_\nu\delta t\sum_{m_\nu = \pm} P_{m_\nu}^{\otimes 2n}.
\end{align}
A special case of $\mathcal{M}_\nu^{(2)}$ is given in Eq.~\eqref{eq:Mavg}.
Here, $P_{m_\nu} = (1 \pm M_\nu)/2$ are the projection operators associated with the measurements on the Pauli string $M_\nu$.
We note that measurement results for any Pauli string can only be $m_\nu = \pm 1$.
To demonstrate the local symmetries of $\mathcal{M}_\nu^{(n)}$, it suffices to show
\begin{align}
\label{eqn_app:Mavg_commute}
    [P_{+}^{\otimes 2n} + P_{-}^{\otimes 2n}, \mathcal{R}_j] = 0
\end{align}
for any $j$ and $\mathcal{R}_j \in \{\mathcal{X}_j, \mathcal{Y}_j, \mathcal{Z}_j\}$.
To verify that Eq.~\eqref{eqn_app:Mavg_commute} is satisfied, we expand $P_{+}^{\otimes 2n} + P_{-}^{\otimes 2n}$ according to the power of operator $M_\nu$.
Each nonvanishing term contains a product of even number of operator $M_\nu$ (acting on different branches) and therefore commute with $\mathcal{X}_j, \mathcal{Y}_j$, and $\mathcal{Z}_j$.

\subsection{Fermionic circuit}
For the Gaussian fermionic circuits considered in Section~\ref{sec:fermion}, the averaged dynamics preserves local symmetries generated by $\mathcal{R}_\ell = \prod_{\alpha = 1}^n \ri\gamma_{\ell,\alpha\ua}\gamma_{\ell,\alpha\da}$ at every Majorana site $\ell$.
The local symmetry $\mathcal{R}_\ell$ commutes with any Majorana bilinear defined on the same site, i.e. $[\mathcal{R}_\ell, \gamma_{\ell',\alpha\sigma}\gamma_{\ell',\beta\sigma'}] = 0$.
Using this property, we can verify, for any replica index $n$, both the averaged unitary gates $\mathcal{U}_{s/b,j}^{(n)}$ and the averaged measurements $\mathcal{M}_{s/b,j}^{(n)}$ commute with $\mathcal{R}_\ell$.

\section{Effective quantum Hamiltonian for $\mathbb{Z}_2$ symmetric quantum circuits}\label{app:Hq_Z2}

To demonstrate our method, we present an explicit formula of $H_{\text{eff}}$ for a simple paradigmatic $1$D circuit model with $\mathbb{Z}_2$ symmetry that conserves the total parity of Pauli-Z operator.
Our model follows the structure depicted in Fig.~\ref{fig:qubit_model}.
The inter-layer hybrid dynamics involves the following unitary gate and measurements: 
(i) random $\theta_{ij}^{xx} X_i X_j$ interactions with Gaussian random $\theta_{ij}^{xx}$ of zero mean,
(ii) random $\theta_{ij}^{zz} Z_i Z_j$ interactions with Gaussian random $\theta_{ij}^{xx}$ of zero mean,
(iii) measurements of $Z_j$ with probability $\Gamma^z_{j}\delta t$, and 
(iv) measurements of $X_i X_{j}$ with probability $\Gamma^{xx}_{ij}\delta t$.
All unitary couplings and measurement projections in this model respect the $\mathbb{Z}_2$-parity symmetry generated by $\hat{\pi} = \prod_j Z_j$.
As explained in Appendix~\ref{app:local_symmetry}, $\mathcal{X}_j$ are local conserved quantities, and $s_j$ variables denote local integrals of motion.

Using the framework developed in Section~\ref{sec:framework}, we can show the effective Hamiltonian $H_{\text{eff},+}$ for the dynamics of two copies of density matrices in the even sector of local parities is of the form
\begin{align}
    H_{\text{eff},+} = H_{JXX} + H_{JZZ} + H_{PZ}+ H_{PXX},
    \label{eq:Z2_Heff}
\end{align}
where $H_{JXX}$ and $H_{JZZ}$ arise from unitary gates with random couplings $\theta^{xx}_{ij}$ and $\theta^{zz}_{ij}$, and $H_{PZ}$ and $H_{PXX}$ arise from $Z_j$ and $X_i X_j$ measurements.
The Hamiltonian terms can be written in terms of spin-1 operators
\begin{align}
    H_{JZZ} = \sum_{i < j}& J^{zz}_{ij}\left(M_{i,+} M_{j,-} + M_{i,-} M_{j, +}\right), \\
    H_{JXX} = \sum_{i < j}& -J^{xx}_{ij}\Big(S_{i}^xS_{j}^x + S_{i}^y e^{\ri\pi S_{i}^x} e^{\ri\pi S_{j}^x} S_{j}^y \nonumber \\
    &\quad\quad\quad - L_{i} L_{j}\Big), \\
    H_{PXX} = \sum_{i < j}& -\frac{\Gamma^{xx}_{ij}}{4}\Big(S_{i}^xS_{j}^x + S_{i}^y e^{\ri\pi S_{i}^x} e^{\ri\pi S_{j}^x} S_{j}^y \nonumber \\
    &\quad\quad\quad + L_{i} L_{j}\Big),  \\
    H_{PZ} = \sum_{j}& \Gamma_j^{z} \left(S_{j}^z\right)^2,
\end{align}
where the coupling $J^{zz}_{ij} \delta t = 8\overline{(\theta^{zz}_{ij})^2}$, $J^{xx}_{ij} \delta t = 2\overline{(\theta^{xx}_{ij})^2}$, the operator $L_{i} = [(S_{i}^+)^2 + (S_{i}^-)^2]/2$ with $S_{i}^{\pm}$ being the spin ladder operators, and $M_{j,\pm} = \kket{m = \pm 1}\bbra{m = \pm 1}_j$ are projectors to the state $\kket{m = \pm 1}$ at site $j$.

The effective Hamiltonian exhibits a $D_4 \times \mathbb{Z}_2^\mathbb{H}$ symmetry, where $D_4 = (\mathbb{Z}_2^{\Pi_L} \times \mathbb{Z}_2^{\Pi_1})\rtimes \mathcal{S}_2$.
In the spin-1 Hilbert space, the symmetry generators are given by $\hat{\Pi}_L = \prod_j\mathrm{diag}(-1,1,-1)_j$, $\hat{\Pi}_1 = \prod_j \mathrm{diag}(1,1,-1)_j$, and $\mathcal{C}_\ell = \prod_j -\exp(\ri\pi \sum_j S_j^x)$.
The Hermitian conjugate $\mathbb{H}$ manifests as the complex conjugation $\mathcal{K}$.
This $D_4 \times \mathbb{Z}_2^\mathbb{H}$ symmetry agrees with the analysis in Section~\ref{sec:z2_phases}.

\section{Exchange symmetry and hermiticity symmetry in the $n = 2$ qubit model}\label{app:symmetry}

In this section, we derive the exchange operation $\mathcal{C}_\ell\mathcal{C}_r$ and Hermitian conjugation $\mathbb{H}$ in the representation of $\mathcal{G}^{(2)}_{\text{eff}}$ on the even parity sector of the $n = 2$ qubit model.
In particular, we show the exchange operation acts trivially as an identity matrix in the even parity sector and therefore cannot be broken.
This proof only relies on the presence of local conserved quantities $\mathcal{X}_j$, $\mathcal{Y}_j$, and $\mathcal{Z}_j = 1$ in the qubit model and does not require the physical circuit symmetry.

To start with, the exchange operation $\mathcal{C}_\ell\mathcal{C}_r$ is given by a product of single-site exchange operations $\mathcal{C}_{\ell,i}\mathcal{C}_{r,i}$.
The single-site left and right permutation can be written in terms of Pauli operators as
\begin{align}
    \mathcal{C}_{\ell,i} &= (IIII + XIXI + YIYI + ZIZI)_i/2, \\
    \mathcal{C}_{r,i} &= (IIII + IXIX + IYIY + IZIZ)_i/2.
\end{align}
In the even parity sector, one can show the product $\mathcal{C}_{\ell,i}\mathcal{C}_{r,i}$ acts as an identity matrix regardless of the physical symmetry, and so is $\mathcal{C}_\ell\mathcal{C}_r$. Hence, $\mathcal{S}_2^X$ is in the kernel of the representation and cannot be broken.

The hermiticity symmetry can be written as a combination of transpose $\mathcal{T}$ and complex conjugation $\mathcal{K}$.
The transpose operation $\mathcal{T} = \mathcal{T}_1\mathcal{T}_2$ is a product of transpose for the first and second copy.
$\mathcal{T}_{1}$ and $\mathcal{T}_2$ can be further written respectively as a product of single-site transpose operations $\mathcal{T}_{1,i}$ and $\mathcal{T}_{2,i}$ which take the form
\begin{align}
    \mathcal{T}_{1,i} &= (IIII + XXII + YYII + ZZII)_i/2, \\
    \mathcal{T}_{2,i} &= (IIII + IIXX + IIYY + IIZZ)_i/2.
\end{align}
We can verify the product $\mathcal{T}_{1,i}\mathcal{T}_{2,i}$ is identity in the even parity sector, and so is $\mathcal{T} = \prod_i \mathcal{T}_{1,i}\mathcal{T}_{2,i}$.
Hence, the Hermitian conjugate $\mathbb{H} = \mathcal{K}$ in the even parity sector.

\section{Higher replicas ($n\geq 3$)}\label{app:higher_replicas}
Having shown the existence of a broad array of symmetry enriched phases in the case of two replica copies $n = 2$, 
we here discuss the correspondence of each phase in the higher replica models ($n \geq 3$).
We verify that two area-law and three volume-law phases found in the two-chain model in Section~\ref{sec:z2_numerics} have natural generalizations in the higher replica models with identical physical signatures.
This provides evidence that these phases can be extrapolated to the replica limit $n \to 1$, predicting distinct phases of quantum circuits.

Among the rest five (two area-law and three volume-law) phases which haven't been realized in a concrete model, we show that only the composite volume-law phase has a higher replica generalization.
The coexistence of EA order and parity variance is generically not allowed in the higher replica models; four coexistence phases (two area-law and two volume-law phases) do not have natural higher replica generalizations.
However, we note this does not exclude the possibility of the realizing these phases in the replica limit. 
Whether there is fundamental physical obstruction to realizing these phases in a concrete quantum circuit remains an open question.

According to the analysis in Section~\ref{sec:symmetry}, the dynamics of $n$ replicas exhibits an enlarged symmetry $\mathcal{G}^{(n)} = (B_n \times B_n) \rtimes \mathbb{Z}_2^{\mathbb{H}}$, where the hyperoctahedral group $B_n = \mathbb{Z}_2 \wr \mathcal{S}_n = \mathbb{Z}_2^{\otimes n} \rtimes \mathcal{S}_n$ is the symmetry group of an $n$-dimensional hypercube.
The dynamics also conserves local parities $\mathcal{X}_j$, $\mathcal{Y}_j$, and $\mathcal{Z}_j$ (see Appendix~\ref{app:local_symmetry}).

Similar to the case of two replicas $n = 2$, the effective symmetry that determines the ground state phases is reduced due to the conservation of local parities. 
We are interested in the ground state in the same sector of local parities as the reference state, i.e. $\mathcal{X}_j = \mathcal{Z}_j = +1$ and $\mathcal{Y}_j = (-1)^n$.
In this sector, the nonvanishing expectation values of local parities necessitate the breaking of single-branch symmetries.
Hence, we are left with an effective global symmetry $\mathcal{G}^{(n)}_{\text{eff}}$ generated by the elements of the left and right permutation in $\mathcal{S}_n$, $\mathbb{H}$ together with $\hat{\Pi}_1\equiv \prod_j (Z_{1}Z_{\bar{1}})_j$ and $\hat{\Pi}_{12}\equiv \prod_j(Z_{1}Z_{2})_j$.
Distinct phases of quantum circuits are characterized by the residual symmetry which is a subgroup of $\mathcal{G}^{(n)}_{\text{eff}}$.

We note that, in contrast to the case of two replicas ($n = 2$), the generators of the $\mathcal{S}_n^X$ exchange symmetry, i.e. $\mathcal{C}_{\ell,\xi} \mathcal{C}_{r,\xi}$ for $\xi \in \mathcal{S}_n$, act nontrivially in the even parity sector; $\mathcal{S}_n^X$ can in principle be broken for $n \geq 3$.

\emph{Area-law phases}.---
The area-law phase is again characterized by the unbroken $\mathcal{S}_n$ symmetry or the composite symmetry of the cyclic permutation $\mathcal{C}_\ell$ and $g_\mathcal{I}$ present in the reference state $\kket{\mathcal{I}}$.
The presence of the hermiticity symmetry necessitates the dynamical symmetry $(\mathcal{S}_n \times \mathcal{S}_n) \rtimes \mathbb{Z}_2^\mathbb{H}$.

Starting from the fully symmetric area-law phase with residual symmetry $\mathcal{G}^{(n)}_{\text{eff}}$, we can condense the charge $\mathcal{Q}_1\equiv X_1X_{\bar{1}}$ together with all operators related to it by the $(\mathcal{S}_n\times \mathcal{S}_n)\rtimes \mathbb{Z}_2^\mathbb{H}$ symmetry. This leads to a broken symmetry area-law phase with the residual symmetry $(\mathcal{S}_n\times \mathcal{S}_n)\rtimes \mathbb{Z}_2^\mathbb{H}$. It is easily verified that both the symmetric and the broken symmetry state have exactly the same physical signatures as their corresponding state in the $n=2$ model.

The coexistence of EA order and parity variance is generically not allowed in the model of higher replicas.
To obtain a nonvanishing EA order, one needs to condense $\mathcal{Q}_{ab} \equiv X_a X_b$ for $a,b=1,2,\cdots,n$.
Such a condensate breaks not only $\hat{\Pi}_{a} \equiv \prod_j (Z_aZ_{\bar{a}})_j$ but also $\hat{\Pi}_{ab} \equiv \prod_j (Z_aZ_b)_j$, indicating an exponentially decaying parity variance.
One can also verify that condensing $X_aZ_bY_c$ and its related charges by $(\mathcal{S}_n \times \mathcal{S}_n)\rtimes \mathbb{Z}_2^\mathbb{H}$ in the composite area-law phase breaks both $\hat{\Pi}_a$ and $\hat{\Pi}_{ab}$, giving a vanishing parity variance.
Hence, two area-law phases with coexisting orders do not have a higher replica generlation.

The symmetric SPT phase, generalized to $n\geq 3$, is found starting from the broken symmetry area-law phase. We can restore the full symmetry by condensing domain walls of $\hat{\Pi}_{12}$ and $\hat{\Pi}_1$ as well as all domain walls related to them by the symmetry $(\mathcal{S}_n\times \mathcal{S}_n)\rtimes \mathbb{Z}_2^\mathbb{H}$. The symmetric SPT phase is obtained by condensing $\hat{\Pi}_{ab}$ and $\hat{\Pi}_a$ domain walls bound to charges of the $\mathcal{S}_n$ symmetries. The dual picture of the same phase is a condensate of the $\mathcal{S}_n$ domain walls bound to the charge $\mathcal{Q}_1$ and symmetry related operators. 

\emph{Volume-law phases}.---
We now turn to the volume-law regime. One correspondence of the symmetric volume-law phase in the higher replicas is the phase with fully broken left and right $\mathcal{S}_n$, while retaining the symmetry $\hat{\Pi}_a \equiv \prod_j (Z_a Z_{\bar{a}})_j$, $\hat{\Pi}_{ab} \equiv \prod_j (Z_a Z_b)_j$, and $\mathbb{Z}_2^\mathbb{H}$.\footnote{We note that the volume-law phase in the replica limit is believed to be a condensate of elementary domain walls~\cite{jian2020measurement}, which breaks the full $\mathcal{S}_n$ symmetry. Hence, in higher replica models, we only consider the volume-law phase with fully broken $\mathcal{S}_n$.}
One can check that the physical quantities in this phase behave the same as in its $n = 2$ correspondence.

Starting from the symmetric volume-law phase, we can condense $\mathcal{Q}_a = X_a X_{\bar{a}}$, leading to a featureless phase preserving symmetry $\hat{\Pi}_a$, while breaking $\hat{\Pi}_{ab} = Z_a Z_b$.
We can further condense $\mathcal{Q}_{ab}$ in the featureless phase to break the symmetry $\hat{\Pi}_{a}$, leading to the broken symmetry volume-law phase with $\mathbb{Z}_2^\mathbb{H}$ symmetry.
Two coexistence volume-law phases do not have natural generalizations in the higher replicas as condensing either $\mathcal{Q}_{ab}$ or $\mathcal{Q}_{a\bar{b}} \equiv X_a X_{\bar{b}}$ for all $a,b=1,2,\ldots,n$ breaks $\hat{\Pi}_{a'b'}$ leading to vanishing parity variance.

A correspondence of the composite volume-law phase can be shown to exist in the model of even number of replicas ($n$ being even). 
This phase features a residual symmetry group $(\mathbb{Z}_n \times \mathbb{Z}_n) \rtimes \mathbb{Z}_2^\mathbb{H}$.
Two $\mathbb{Z}_n$ subgroups are generated respectively by the composite symmetry $\mathcal{C}_\ell \hat{\Pi}_L$ and $\mathcal{C}_\ell\mathcal{C}_r$, where $\hat{\Pi}_L \equiv \prod_j (\prod_{a =1}^n Z_a)_j$.
Starting from the symmetric area-law phase, we can obtain this phase by condensing $(X_1Y_{\bar{1}}Z_2)_j$ symmetrized under the residual symmetry.
We note that the symmetrized operator is nonvanishing for the even replicas.

\section{Fermionic states and operators in the duplicated Hilbert space}\label{app:correspondence}

Here we detail the correspondence between second quantized fermionic operators $f_j$, acting in the Hilbert space of a single copy of quantum circuit, and the $f_{j,\alpha,\sigma}$ acting in the duplicated Hilbert space of the effective Hamiltonian. 

We start from the Fock states $\ket{\vec{n}}=\prod_j (f^\dagger_{j})^{n_{j}}\ket{0}$ in the Hilbert space of quantum circuit. The space of double density matrices is accordingly spanned by the states:
$\ket{\vec{n}_1}\bra{\vec{m}_1}\otimes\ket{\vec{n}_2}\bra{\vec{m}_2}$.
To define a proper fermionic Fock space for the duplicated system, we make the correspondence:
\begin{align}
&\ket{\vec{n}_1}\bra{\vec{m}_1}\otimes\ket{\vec{n}_2}\bra{\vec{m}_2}\leftrightarrow \label{eq:state_corr}\\
& \prod_i (f^\dagger_{i,1,\ua})^{n_{1i}} \prod_j (f^\dagger_{j,1,\da})^{m_{1j}}\prod_k(f^\dagger_{k,2,\ua})^{n_{2k}} \prod_l (f^\dagger_{l,2,\da})^{m_{2l}}\kket{\text{vac}}_f\nonumber\\
&\equiv \kket{\vec{n}_1,\vec{m}_1,\vec{n}_2,\vec{m}_2}.\nonumber
\end{align}
Here, the second quantized operators $f^\dagger_{i,\alpha,\sigma}$ create a fermion in the $\sigma$ branch (ket or bra) of copy  $\alpha$. 

The above correspondence between states gives a simple representation of the boundary state
$\bbra{\mathcal{I}}$, which implements the doubled trace operation: 
\be
\bbra{\mathcal{I}}=\sum_{\vec{n}_1,\vec{n}_2} \bbra{\vec{n}_1,\vec{n}_1,\vec{n}_2,\vec{n}_2}.
\label{eq:I}
\ee
Indeed we can check that $\bbrakket{\mathcal{I}}{\rho^{(2)}}=\tr\rho^{(2)}$.

The above correspondence between states implies a correspondence between operators \begin{subequations}
\begin{align}
&f^\dagger_j\otimes \mathds{1}\otimes \mathds{1}\otimes \mathds{1}\leftrightarrow f^\dagger_{j,1,\ua},\\
&\mathds{1}\otimes f_j\otimes \mathds{1}\otimes \mathds{1}\leftrightarrow f^\dagger_{j,1,\da}(-1)^{N_{1\ua}},\\
&\mathds{1}\otimes \mathds{1}\otimes f^\dagger_j\otimes \mathds{1}\leftrightarrow f^\dagger_{j,2,\ua}(-1)^{N_{1\ua}+N_{1\da}},\\
&\mathds{1}\otimes \mathds{1}\otimes \mathds{1}\otimes f_j\leftrightarrow f^\dagger_{j,2,\da}(-1)^{N_{1\ua}+N_{1\da}+N_{2\ua}}.
\end{align}
\end{subequations}
The sign factors on the right are needed to ensure that operators on the two sides of the correspondence have identical action on the respective states. $N_{\alpha\sigma}$ denote the total number in copy $\alpha$ and ``branch'' $\sigma$ (ket $\ua$ or bra $\da$).
We make the following remarks.
First, with this correspondence, all the fermion parity conserving local operators, acting on a single circuit state (in any copy or branch) remain local and the sign factors cancel out of them. The extension to any number of copies is obvious.
Second, due to the additional sign factors, the hermiticity symmetry transformation appears to be nonlocal in the duplicated Hilbert space, namely $\mathbb{H}f^\dagger_{j,\alpha,\ua}\mathbb{H}^{-1} = (-1)^{N_{\alpha,\ua}} f^\dagger_{j,\alpha,\da}$.
However, $\mathbb{H}$ is local for any bosonic operator made of fermionic bilinears and therefore becomes a local symmetry in the effective Hilbert space.

Having defined the operators in the duplicated Hilbert space $\mathcal{H}^{(2)}$, we can write the effective Hamiltonian in a Fock basis using any ordering convention of the fermions. In particular, the form of the effective spin-1 Hamiltonian in Section~\ref{sec:fermi_mapping} implicitly assumes that it operates between Fock states with a site-local ordering convention:
\begin{align}
&\bigotimes_{j}\kket{n_{j,1,\uparrow}, n_{j,1,\downarrow},n_{j,2,\uparrow},n_{j,2,\downarrow}} \equiv \\
&\prod_j\Big[(f^\dagger_{j,1,\ua})^{n_{j,1,\ua}} (f^\dagger_{j,1,\da})^{n_{j,1,\da}}(f^\dagger_{j,2,\ua})^{n_{j,2,\ua}}(f^\dagger_{j,2,\da})^{n_{j,2,\da}}\Big]\kket{\text{vac}}_f.\nonumber
\end{align}
The same Hamiltonian will take a different form if written in terms of Fock states with copy-local ordering $\kket{\vec{n}_{1\ua},\vec{n}_{1\da},\vec{n}_{2\ua},\vec{n}_{2\da}}$. In particular, the $U(1)$ symmetries will become nonlocal and not be explicitly apparent because they are generated by conservation of the  $c^\dagger_{\ell,\sigma}$, which are not local to one copy.

The price we pay for working with the site-local Fock states $\kket{{\bf{n}}}$ in which the effective spin Hamiltonian is simple and the $U(1)$ symmetry apparent is that we also need to rewrite the boundary state $\bbra{\mathcal{I}}$ with the same Fock states. This in order to compute its overlap with the state generated by the effective imaginary time evolution (together with any boundary operators we want to measure). 
We have seen that $\bbra{\mathcal{I}}$ has a simple representation in terms of the copy-local Fock states in Eq.~\eqref{eq:I}. The Fock states in the two ordering conventions are related to each other by a sign factor
which is a function of the occupation numbers in the state. These factors take a simple form for the basis states included in the boundary state, giving:
\begin{align}
\bbra{\mathcal{I}}=\sum_{\vec{n}_1,\vec{n}_2} g(N_1, N_2) \bigotimes_{j} \bbra{n_{j,1},n_{j,1},n_{j,2},n_{j,2}},
\label{eq:I_site_local}
\end{align}
where the factor $g(N_1, N_2) = (-1)^{{1\over 2}N_1(N_1-1)+{1\over 2}N_2(N_2-1)}$.

\section{$U(1)$ symmetry breaking in $\kket{\mathcal{I}}$}\label{app:boundary}
In this section, we show the boundary state $\kket{\mathcal{I}}$ breaks both $U(1)$ symmetries corresponding to the conservation of $\Sigma^z$ and $\eta^z$. We demonstrate the long-range order in $\kket{\mathcal{I}}$ manifested by nondecaying correlation functions of order parameters $\bbra{\mathcal{I}}\eta_i^y \eta_j^y\kket{\mathcal{I}}/4^L = \bbra{\mathcal{I}}\Sigma_i^y \Sigma_j^y\kket{\mathcal{I}}/4^L = 1/2$.

To evaluate the correlation of order parameters in $\kket{\mathcal{I}}$, we write the boundary state $\kket{\mathcal{I}}$ [Eq.~\eqref{eq:I}] in terms of second quantized fermionic operators $f_{j,\alpha,\sigma}$ acting on the Hilbert space of branch $\sigma$ of copy $\alpha$
\begin{align}
    \kket{\mathcal{I}} = &\prod_{i = 1}^N \left[ 1 + f_{i,1,\uparrow}^{\dagger} f_{i,1,\downarrow}^{\dagger} e^{\ri \pi \sum_{i' = i+1}^N n_{1i'}}\right] \nonumber \\
    &\left[ 1 + f_{i,2,\uparrow}^{\dagger} f_{i,2,\downarrow}^{\dagger} e^{\ri \pi \sum_{i' = i+1}^N n_{2i'}} \right] \kket{\text{vac}}_f.
\end{align}
Here, the state $\kket{\mathcal{I}}$ has a definite $f$-fermion parity on every site in the Hilbert space of forward and backward branch of each copy, i.e. $\bbra{\mathcal{I}}\hat{\Pi}_{j,\alpha,\ua}\hat{\Pi}_{j,\alpha,\da}\kket{\mathcal{I}} = 1$, where $\hat{\Pi}_{j,\alpha,\sigma} = 1 - 2f_{j,\alpha,\sigma}^\dagger f_{j,\alpha,\sigma}$ is the $f$-fermion parity.

In terms of $f$-fermion operators, the order parameters $\Sigma^\pm_j$ and $\eta^\pm_j$ take the form
\begin{align}
    \Sigma_j^+ = & -\ri\mathcal{O}_{j,1\uparrow,1\downarrow} - \mathcal{O}_{j,1\uparrow,2\downarrow} -  \mathcal{O}_{j, 2\uparrow,1\downarrow} +\ri \mathcal{O}_{j,2\uparrow,2\downarrow},\\
    \Sigma_j^- = & \ri\mathcal{O}_{j,1\uparrow,1\downarrow} - \mathcal{O}_{j,1\uparrow,2\downarrow} - \mathcal{O}_{j, 2\uparrow,1\downarrow} -\ri \mathcal{O}_{j,2\uparrow,2\downarrow},\\
    \eta_j^+ =& \ri\mathcal{O}_{j,1\uparrow,1\downarrow} - \mathcal{O}_{j,1\uparrow,2\downarrow} + \mathcal{O}_{j, 2\uparrow,1\downarrow} +\ri \mathcal{O}_{j,2\uparrow,2\downarrow},\\
    \eta_j^- =& -\ri\mathcal{O}_{j,1\uparrow,1\downarrow} - \mathcal{O}_{j,1\uparrow,2\downarrow} +  \mathcal{O}_{j, 2\uparrow,1\downarrow} -\ri \mathcal{O}_{j,2\uparrow,2\downarrow},
\end{align}
where $\mathcal{O}_{j,\alpha\sigma, \beta\sigma'}$ is the fermion bilinear defined as
\begin{align}
    \mathcal{O}_{j, \alpha\sigma, \beta\sigma'} = \frac{\ri}{2}\left(f_{j,\alpha,\sigma}^\dagger f_{j,\beta,\sigma'}^\dagger - f_{j,\beta,\sigma'}f_{j,\alpha,\sigma}\right).
\end{align}

Operator $\mathcal{O}_{j,\alpha\sigma,\beta\sigma'}$ flips the local $f$-fermion parity in Hilbert space of copy $\alpha$ and $\beta$ on site $j$. Therefore, the correlation function of $\mathcal{O}_{j,\alpha\sigma,\beta\sigma'}$ in the boundary state $\kket{\mathcal{I}}$ is non-vanishing only when $\alpha = \beta$. We can further show the only nonvanishing two-point correlations are given by
\begin{align}
    \frac{1}{4^L}\bbra{\mathcal{I}}\mathcal{O}_{i,\alpha\uparrow,\alpha\downarrow} \mathcal{O}_{j,\alpha\uparrow,\alpha\downarrow} \kket{ \mathcal{I}} = \frac{1}{4},
    \label{eq:Ocorr}
\end{align}
where $1/4^L$ is the normalization of $\kket{\mathcal{I}}$, i.e. $\bbrakket{\mathcal{I}}{\mathcal{I}} = 4^L$.
Using Eq.~\eqref{eq:Ocorr}, we can show the order parameter correlation functions in the boundary state $\kket{\mathcal{I}}$ is nondecaying
\begin{align}
    \frac{1}{4^L}\bbra{\mathcal{I}}\Sigma_i^y \Sigma_j^y\kket{\mathcal{I}} = \frac{1}{4^L}\bbra{\mathcal{I}}\eta_i^y \eta_j^y\kket{\mathcal{I}} = \frac{1}{2}.
\end{align}
Hence, the boundary state exhibits long-range orders and breaks both the $U(1)$ symmetries generated by $\eta^z$ and $\Sigma^z$.
Furthermore, we note that the expectation value of order parameters in $\kket{\mathcal{I}}$ is vanishing, i.e. $\bbra{\mathcal{I}}\Sigma^\pm_j\kket{\mathcal{I}} = \bbra{\mathcal{I}}\eta^\pm_j\kket{\mathcal{I}} = 0$.
For this reason, the boundary state $\kket{\mathcal{I}}$ can be understood as a superposition of $U(1)$ symmetry breaking states with the ordering of $U(1)$ phase in different directions.

\section{Swap operator in fermionic duplicated Hilbert space}\label{app:swap}

The swap operator $\mathcal{C}_{\ell,A}$ is in general nonlocal in the fermionic duplicated Hilbert space $\mathcal{H}^{(2)}$.
In this section, we show that $\mathcal{C}_{\ell,A}$ bears a simple local description when acting on the boundary state $\kket{\mathcal{I}}$.
This allows a long wavelength description for the swap boundary state $\kket{\mathcal{C}_{\ell,A}} \equiv \mathcal{C}_{\ell,A} \kket{\mathcal{I}}$ in Section~\ref{sec:fermi_phases}.

To start with, we consider the swap operator in the Hilbert space of two copies of quantum circuits
\begin{align}
    \mathcal{C}_{A} = \sum_{\vec{n}_{1}, \vec{n}_2} \ket{\vec{n}_{2A}\vec{n}_{1B}} \bra{\vec{n}_{1A}\vec{n}_{1B}} \otimes \ket{\vec{n}_{1A}\vec{n}_{2B}} \bra{\vec{n}_{2A}\vec{n}_{2B}},
\end{align}
where $\vec{n}_\alpha = (\vec{n}_{\alpha, A}, \vec{n}_{\alpha,B})$ is the occupation of fermionic modes in the subsystem $A$ and its complement $B$. 
The purity of subsystem can be obtained by applying the permutation operator to the double density matrix either from left and then taking the trace, i.e.  $\tr\rho^2_A = \tr(\mathcal{C}_{A}\, \rho\otimes\rho)$.

Using the correspondence established in Appendix~\ref{app:correspondence}, we can formulate the purity as an overlap between the swap boundary state $\bbra{\mathcal{C}_{\ell,A}}$ and $\kket{\rho}$, where $\kket{\mathcal{C}_{\ell,A}}$ takes the form
\begin{align}
    \kket{\mathcal{C}_{\ell,A}} = \sum_{\vec{n}_1,\vec{n}_2} \kket{\vec{n}_{2A}\vec{n}_{1B}, \vec{n}_{1A}\vec{n}_{1B}, \vec{n}_{1A}\vec{n}_{2B},\vec{n}_{2A}\vec{n}_{2B}}.
\end{align}
Alternatively, we can write $\kket{\mathcal{C}_{\ell,A}}$ in the site-local ordering convention as
\begin{align}
    \kket{\mathcal{C}_{\ell,A}} &= \sum_{\vec{n}_1,\vec{n}_2} g(N_{1},N_{2}) \label{eq:swap_state} \\
    &\bigotimes_{j\in A} \kket{n_{j,2},n_{j,1},n_{j,1},n_{j,2}} \bigotimes_{j \in B}\kket{n_{j,1},n_{j,1},n_{j,2},n_{j,2}}. \nonumber
\end{align}

We note that the boundary state $\kket{\mathcal{C}_{\ell,A}}$ [Eq.~\eqref{eq:swap_state}] and $\kket{\mathcal{I}}$ [Eq.~\eqref{eq:I_site_local}] in the site-local convention only differ by a local exchange of the occupation in the forward branches of copy 1 and 2 in subsystem $A$.
This allows one to define an effective swap operator $\mathcal{\widetilde C}_{\ell,A} = \prod_{j\in A} \mathcal{\widetilde C}_{\ell,j}$ such that $\mathcal{\widetilde C}_{\ell,A}\kket{\mathcal{I}} = \kket{\mathcal{C}_{\ell,A}}$, where $\mathcal{\widetilde C}_{\ell,j}$ is the single-site swap operator satisfying
\begin{align}
    \mathcal{\widetilde C}_{\ell,j} \kket{n_{j,1}, n_{j,1}, n_{j,2}, n_{j,2}} = \kket{n_{j,2}, n_{j,1}, n_{j,1}, n_{j,2}}.
\end{align}
The single-site swap operator $\mathcal{\widetilde C}_{\ell,j}$ can be written in terms of $f$-fermionic operators and further in terms of $U(1)$ symmetry generators
\begin{align}
    \widetilde{\mathcal{C}}_{\ell,j} =& f^\dagger_{j,1,\uparrow} f_{j, 2, \uparrow} + f_{j,1,\uparrow} f_{j,2,\uparrow}^\dagger + \frac{1}{2}\left(1 + \hat{\Pi}_{j,1,\ua}\hat{\Pi}_{j,2,\ua}\right) \nonumber \\
    =& e^{-\ri\frac{\pi}{2} \left(\Sigma_j^z + \eta^z_j\right)},
\end{align}
where $\hat{\Pi}_{j, \alpha, \ua} = 1 - 2f_{j,\alpha,\ua}^\dagger f_{j,\alpha,\ua}$ is the fermion parity in the forward branch of copy $\alpha$ at site $j$.
Therefore, the swap operator $\mathcal{C}_{\ell,A}$, when acting on the boundary state $\kket{\mathcal{I}}$, is equivalent to $\mathcal{\widetilde C}_{\ell,A}$, which rotates both $U(1)$ order parameters ($\eta^+_j$ and $\Sigma^+_j$) counterclockwise by an angle $\pi/2$ in region $A$.
In the long-wavelength description developed in Section~\ref{sec:fermi_phases}, $\mathcal{\widetilde C}_{\ell,A}$ inserts a pair of half vortices at the edges of region $A$ on the boundary of (1+1)d XY model.
Here, we note again that $\mathcal{\widetilde C}_{\ell,A}$ is not the swap operator when acting on a general state and $\mathcal{\widetilde C}_{\ell,A}^2 \neq \mathds{1}$.
We also note that the swap operator breaks the time-reversal symmetry for the spin-$1/2$ fermions as it only operates on the forward branches ($\sigma = \uparrow$).

\section{Details of the numerical simulation in Gaussian fermionic circuits}\label{app:numerics}

\subsection{Power-law decay exponents of correlations in the critical phase}\label{app:n_power_law}

\begin{figure*}[t!]
    \centering
    \includegraphics[width=\textwidth]{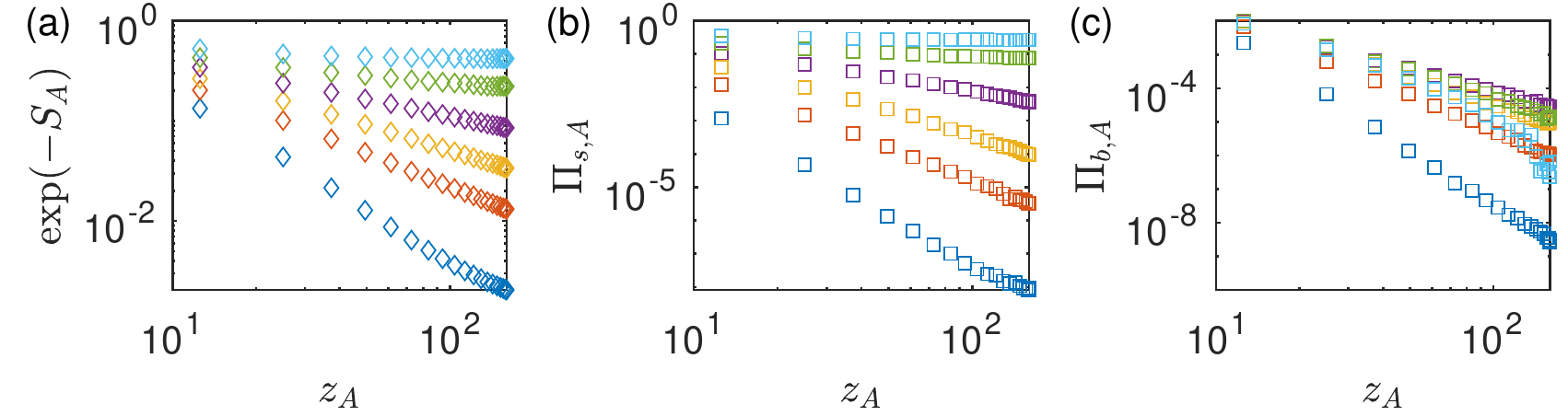}
    \caption{Phase transition in Gaussian fermionic circuits along Cut 1 in Fig.~\ref{fig:fermi_introductory}. Entanglement entropy $\exp(-S_A)$ [panel (a)], parity variances $\Pi_{s,A}$ [panel (b)] and $\Pi_{b,A}$ [panel (c)] as function of the conformal coordinate $z_A$ for various measurement probabilities $p_s$. Data points for the same $p_s$ are marked in the same color in different panels. $p_s$ ranges from $0.02$ to $0.25$ and increases from the bottom to the top in panel (a). The numerical results are obtained with system size $L = 160$, averaged over $400$ realizations, and plotted in a log-log scale.}
    \label{fig:app_power_law}
\end{figure*}

In this section, we present the numerical results of $e^{-S_A}$, $\Pi_{s,A}$ and $\Pi_{b,A}$ in Gaussian fermionic circuits along cut 1. 
We use the numerical results to extract power-law exponents presented in Fig.~\ref{fig:phase_transition}(a-c).

Figure~\ref{fig:app_power_law} presents $e^{-S_A}$ [panel(a)], $\Pi_{s,A}$ [panel(b)] and $\Pi_{b,A}$ [panel(c)] as a function of the conformal coordinate $z_A$ for various measurement probabilities $p_s$.
We show that these quantities exhibit power-law decay in the critical phase (when $p_s < p_{s,c}$ on cut 1).
The scaling deviates from power laws when increasing the measurement probability $p_s$.
Numerical results show an area-law $S_A$, a constant $\Pi_{s,A}$, and an exponentially decaying $\Pi_{b,A}$ in the trivial area-law phase (when $p_s > p_{s,c}$ on cut 1), which verifies our theoretical predictions.
Our theoretical understanding also explains the non-monotonic behavior in the bond parity variance $\Pi_{b,A}$ as a function of $p_s$.
In the critical phase, $\Pi_{b,A}$ increases with $p_s$ owing to the decreasing Luttinger parameter $K$ towards the critical point.
On the other hand, in the trivial area-law phase, $\Pi_{b,A}$ decreases with $p_s$ due to the decreasing correlation length.
We note that the values of $\Pi_{b,A}$ are small along cut 1 owing to the vanishing bond measurement probability, i.e. $p_b = 0$.

\subsection{Correlation length in the area-law phase of Gaussian fermionic circuit}\label{app:n_corr_len}

\begin{figure}[b!]
    \centering
    \includegraphics[width=0.48\textwidth]{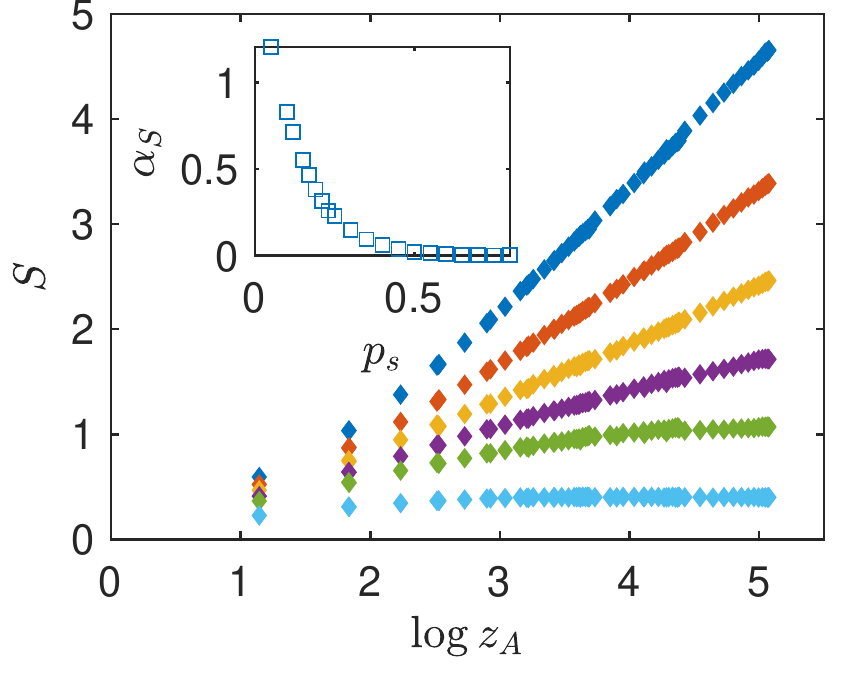}
    \caption{Subsystem entanglement entropy as a function of conformal coordinate $z_A = L\cos(\pi L_A/L)$ for various different measurement probabilities $p_s$ along cut 1. $p_s$ ranges from $0.05$ to $0.35$ and increases in the data points from the top to the bottom. (Inset) Logarithmic coefficient $\alpha(p_s)$ of the entanglement entropy as a function of $p_s$.}
    \label{fig:app_corr_len}
\end{figure}

We present the details on numerically extracting correlation length in the area-law phase of the Gaussian fermionic circuit. 
Specifically, we show the entanglement entropy data used to extract the correlation length and provide the procedure to determine the logarithmic coefficient $\alpha_S(p_s)$ and constant $b(p_s)$ in Eq.~\eqref{eq:entropy_critical}.

To start with, we show in Fig.~\ref{fig:app_corr_len} the entanglement entropy along cut 1 exhibits distinct behaviors across the phase transition.
When measurement rate $p_s$ is small, the entanglement exhibits a critical scaling as a function of the subsystem size $L_A$ (linear in $\log z_A$).
Increasing $p_s$ along cut 1, the circuit enters an area-law phase, which is characterized by the saturation of entanglement entropy as a function of $L_A$.

Here, we focus on the area-law phase and use the entanglement data to extract the finite correlation length $\xi$.
The entanglement exhibits distinct scaling for different subsystem sizes $L_A$.
When $L_A \ll \xi$, the entanglement exhibits critical scaling as Eq.~\eqref{eq:entropy_critical}.
Increasing the subsystem size $L_A$, the entanglement scaling changes to Eq.~\eqref{eq:entropy_conj}.
We extract $\xi$ in two steps: (1) extract $\alpha(p_s)$ and $b(p_s)$ using the entanglement data for small subsystem sizes, i.e. $L_A \ll \xi$; (2) extract $\xi$ using the entanglement data for larger subsystem sizes by fitting to Eq.~\eqref{eq:entropy_conj}.

To extract $\alpha(p_s)$ and $b(p_s)$, we use the data for various $L_A$ and $L$ in the following way: (1) For a given $p_s$, we plot $S(z_A, p_s, L)$ as a function of $\log z_A$ for various $L$; (2) We choose at least seven consecutive data points of $z_A$ and perform the linear regression according to Eq.~\eqref{eq:entropy_critical}. We optimize the $r^2$ of the linear regression over all possible choices to obtain the best $\alpha(p_s)$ and $b(p_s)$.
The optimized $\alpha(p_s)$ is shown in the inset of Fig.~\ref{fig:app_corr_len}.

Using the optimized $\alpha(p_s)$ and $b(p_s)$, we perform the least square fitting for $S(z_A, p_s, L)$ as a function of $z_A$ according to Eq.~\eqref{eq:entropy_conj} for every $p_s$ and $L$ to extract $\xi(p_s, L)$.
The result is presented in Fig.~\ref{fig:corr_len}.

\end{document}